\documentclass[preprint,12pt,3p]{elsarticle}
\usepackage{amssymb}
\usepackage{graphicx}
\usepackage{xcolor}
\usepackage{caption}
\usepackage{subcaption}
\usepackage{wrapfig}
\usepackage{nomencl}
\usepackage{macros_ad}
\usepackage{macros-rohan-vlastni}
\usepackage{amsfonts}
\usepackage{animate}
\usepackage{subcaption}
\usepackage{amsmath,array}
\usepackage{amsmath}
\def\figPath{img}
\def\figPathMacI{img/case_VIa}
\def\figPathMac{img/case_VI}

\newtheorem{remark}{Remark}

\newcommand\cf{{\textit{cf.~}}}

\begin{document}
	
	\title{Two-scale model of quasi-steady flow of electrolyte in weakly piezoelectric porous media}
	%\tnotetext[t1]{The research was supported in part by project GA16-03823S and NT 13326 of the Ministry of Health of the Czech Republic and by the European Regional Development Fund (ERDF), project “NTIS - New Technologies for Information Society”, European Centre of Excellence, CZ.1.05/1.1.00/02.0090. Jana Turjanicov\'{a} is also grateful for the support of her work by project SGS-2016-059.}
	
	\author[1]{Jana Camprov\'a Turjanicov\'a\corref{cor1}}
	\ead{turjani@ntis.zcu.cz}
	
	\author[1]{Eduard Rohan}
	\ead{rohan@kme.zcu.cz}
	
%	\author[1]{Vladim\'{\i}r Luke\v{s}}
%	\ead{vlukes@kme.zcu.cz}
	
	\address[1]{European Centre of Excellence, NTIS -- New Technologies for
		Information Society, Faculty of Applied Sciences, University of West Bohemia,
		Univerzitni\'{\i} 8, 30614 Pilsen, Czech Republic}

	\cortext[cor1]{Corresponding author}

        \begin{abstract}
This paper presents a new homogenized model of two-component electrolyte transport through a weakly piezoelectric porous medium. The model relevant to the microscopic scale describes quasi-stationary states of the medium while reflecting essential physical phenomena, such as electrochemical interactions in a dilute Newtonian solvent under assumptions of slow flow. The dimensional analysis of the mathematical model introduces scaling of the viscosity, electric permittivity, piezoelectric coupling, and dielectric tensor. The micromodel is linearized around the reference state represented by the thermodynamic equilibrium and further upscaled through the asymptotic homogenization method. Due to the scaling of parameters, the derived limit model retains the characteristic length associated with the pore size and the electric double-layer thickness. The upscaling procedure gives a fully coupled macroscopic model describing the electrolyte flow in terms of a global pressure and streaming potentials of the two ionic species in the weakly piezoelectric matrix. By virtue of the characteristic responses, quantities of interest are reconstructed at the microscopic scale using the resolved macroscopic fields. The coupling between electrochemical and mechanical phenomena influenced by the skeleton piezoelectricity is illustrated using numerical examples. The model, which was motivated by the cortical bone porous tissue, is widely applicable in designing new biomaterials involving piezoelectric stimulation.

\end{abstract}

\maketitle

\section{Introduction}
Modeling the transport of electrolyte through a porous medium (TEMP) is a complex nonlinear multiphysics problem with a wide range of applications in geosciences, environmental engineering, physiology, tissue biomechanics, and material sciences. In the technological-oriented areas of engineering, computational quantitative analysis and optimization are needed to design such materials while respecting the link between microstructure features and effective macroscopic properties. The multi-scale modeling based on the asymptotic homogenization can provide a very efficient analytical and computational tool, although the relevance of such an approximation requires the scale separation assumption to be satisfied. This homogenization approach provides clear relationships between the effective material constitutive laws pronounced by virtue of the homogenized coefficients and the physical interactions in microstructures, \ie at the heterogeneity level. This remarkable advantage is appreciated especially in a case of problems and models featured by multiphysical interactions, such as those treated in this paper dealing with the influence of the skeleton piezoelectricity on the electrolyte flow.

\par The complexity of TEMP models originates from the fluid-structure interactions (FSI) on the interface between fluid-filled pores and the solid phase due to the physical phenomena associated with electric charges and related material properties. %%%%
%% The material properties of the matrix influence the complexity of the model significantly.
A significant contribution to the research made in the area has been made while assuming a rigid porous medium with a given static charge on the pore surfaces, see \cite{looker2006homogenization,allaire2013asymptotic,allaire2013ion}, thus, reducing the modelling problem only to the fluid phase. Regarding deformable media, besides a few papers respecting the skeleton compliance or evolving porous structures, see \eg \cite{ray2012multiscale,allaire2015ion}, several works devoted to the mechanical part of the FSI without any electroosmotic or electromechanical coupling, see  \eg \cite{andreasen2013topology,rohan-etal-CMAT2015,Sandstrom-Larssen-CMAME2016,Rohan-AMC}, have established useful platforms to extend those particular models to account for the phenomena featuring the transport of electrolytes.

\par Important applications lie in the field of geosciences; quite recognized are works \cite{moyne2002electro,moyne2006two}, which deal with the modeling of expansive clays composed of the charged solid skeleton saturated by an electrolyte solution via homogenization. The microscopic model includes equations describing electro-hydrodynamics coupled with the equation governing the flow of the electrolyte solution, ion electrodiffusion, and electric potential distribution. The derived two-scale model describes electroosmotic phenomena occurring in TEPM, such as the electroosmotic flow driven by the streaming potential gradient, the electrophoretic motion of mobile charges, and the osmosis-induced swelling.

The coupling between fluid motion and deformation of the solid matrix was also explored in the work \cite{mikelic2012interface} and pursued further in the study motivated by nuclear waste storing, \cite{allaire2015ion}. This work extends the known Poisson-Nernst-Planck system of equations by the elasticity of the solid part. Therein, the authors show that by a suitable choice of time scale, the deformation of the porous medium becomes only weakly coupled to the electrokinetic system, which is advantageous for the model implementation and numerical simulations.

\par The electrodiffusion phenomenon deserves a particular interest in tissue biomechanics. Besides the cell physiology featured by the complex processes on the membrane presenting a specific porous structure, the homogenization-based upscaling of the electrodiffusion phenomena was considered in the series of papers  \cite{lemaire2006multiscale,lemaire2010modelling,lemaire2010multiphysical} devoted to the modelling of bone fluid flow at two porosity levels in the cortical bone tissue and discussing its possible application in the research of mechanosensing, bone growth, and remodeling, \cf \cite{nguyen2009numerical}.

%%%
However, we are interested in modeling ion transport in piezoelectric porous media. In this respect, only a few works deal with this topic, although it enjoys steadily increasing attention, namely in the context of the bone tissue and biomaterial engineering \cite{Silva-2022,Zhang2023b}, \cf \cite{miara2005piezomaterials} devoted to the homogenization of periodic structures formed by simplified osteocytes in piezoelectric scaffolds. Most of the works concern the theoretical issues of mathematical modeling without numerical simulations. The influence of the skeleton piezoelectricity on the electroosmotic effect in cortical bone tissue has been studied in \cite{lemaire2011multiscale} in the homogenization framework and under a priori assumptions based on the theoretical investigation of electric measurements in the bone.

\paragraph{The main contribution of presented paper}{In this paper, we extend the model presented previously in \cite{turjani2018} in several important aspects which open new avenues of the homogenization-based modelling for material research of electroactive porous media. Specifically, %% To this aim,
  the present study concerns  nonstationary processes and introduces phenomena related to the a weakly piezoelectric skeleton inducing a stronger electro-chemo-mechanical coupling in the fluid-structure interactions. Also, the boundary condition considered in the micromodel now respect the ionic exchange between phases at the pore surface.
%%  we present
These ingredients provide  a sound mathematically rigorous basis necessary to explore the influenece of the scaffold geometry in the conext of nonstationary interacions of the electrolyte with deforming electroactive (piezoelectric) solid -- theses phenomena are not often resepcted in otherwise very detailed research addressing the chemical processes,  \cf \cite{Ferranndez-2021,Lasia-2023}.
  %% previously presented model of stationary ionic transport in deformable porous media and its implementation, \cite{turjani2018}. We model the solid skeleton as a weakly piezoelectric medium, which leads to stronger electro-chemo-mechanical coupling and fluid-structure interactions. Further, the flow of the electrolyte is now assumed to be quasi-static.
  %% The final addition is the ionic exchange between phases at the pore surface, represented by the Neumann-type boundary condition. 
	The other related works \cite{lemaire2011multiscale,moyne2006two} and also the most of the papers cited above concern the theoretical issues of mathematical modeling without numerical simulations. On the contrary, results reported in this paper  for  3D microstructures are obtained using the implemented homogenized model, whereby also the ``downscaling'' procedure is considerd to recover the macroscopic response at the local level of the heterogeneous medium.  To our knowledge, this has not been pursued so far in the published literature. We aim to demonstrate the relevance of such a model for practical applications, such as the modeling of cortical bone porous structure. 

        \paragraph{Outline of the paper}{The paper is orgabized, as follows. \sref{sec:1} provides a complete mathematical description of the electro-chemo-mechanical interactions between the fluid considered as a two-component electrolyte, and solid phases; for the latter, due to the scaling of the piezoelectric and dielectric coeffinient, the concept of a weakly piezoelectric skeleton is introduced. Then the dimensional analysis and linearization issues are reporten in \sref{sec:2}. The two-scale model is derived using the homogenization in \sref{sec_homog}, where the principal results are presented, namely the local problems for the so-called corrector functions, formulae for computing the homogenized equations, and, finally, the effective macroscopic model. \sref{sec_numeric} deals with the implementation of numerical simulation and presents the numerical solution of the non-stationary macroscopic model as well as its comparison to the model of ionic transport through deforming porous media. An application of the derived model to the modeling of cortical bone tissue follows in \sref{sec:app}. \sref{sec_conclusion} summarizes Results and further outlooks are summarized in the Conclusion, \ref{sec_conclusion}

\paragraph{Basic notations} We shall adhere to the following notation throughout the paper. The position $x$ in the medium is specified through the coordinates
$(x_1,x_2,x_3)$ with respect to a Cartesian reference frame. We shall use the microscopic (dilated) Cartesian reference system of coordinates $(y_1,y_2,y_3)$. The gradients are employed, $\nabla_x = ( \pd / \pd x_i)$ and
$\nabla_y = (\pd / \pd y_i)$  alternatively. Time derivatives will be denoted by the symbol $\pd{_t}=(\pd / \pd t)$.
As usual, bold letters denote the vectors; for instance,
$\ub(t,x)$ denotes the solid matrix displacement vector field depending on the
spatial variable $x$ and time variable $t$.  Moreover, the components of this vector will be denoted by $u_i$ for $i=1,...,3$  thus $\ub = (u_i)$.  The Einstein summation
convention is used, which stipulates implicitly that repeated indices are summed
over. 
By a common consensus, $\Ib$ denotes the identity matrix, and $\delta_{ij}$ denotes Kronecker's delta. The symbol $\otimes$ denotes the dyadic product of two vectors, which results in the second-order tensor. Two sets of indexes appear throughout the text. The indexes denoted by letters of the Latin alphabet $(i,j,k,l,\dots)$ usually refer to the problem space dimension. On the other hand, indexes $\alpha$ and $\beta$ denote the $\alpha-$th or $\beta-$th ionic species.
The symbol $\RR$ denotes the set of real numbers.
The differential volume and surface elements are
denoted by $\dV$ and $\dS$, respectively. 
Functional spaces are introduced subsequently in the text. The symbols denoting the most frequently used quantities are presented in Tabs.~\ref{tab:nomencl} and ~\ref{tab:nomencl2}.

\begin{table}[!p]\centering
	\begin{tabular}{ll}
		\hline
		Symbol	& Quantity \\
		\hline
		{$d$}{}&{dimension, } \\
		{$\veps$}&{scale parameter, $0<\varepsilon<< 1$} \\
		{$L_{c}$}&{characteristic dimension of macroscopic domain}\\
		{$\ell_\mic$}&{characteristic dimension dimension of microscopic periodic structure}\\
		{$\sigmabf_s$}{}&{stress tensor of piezoelectric solid phase}\\
		{$\vec{D}_s$}{}&{electric displacement of solid}\\
		{$\Psi_s$}&{potential of solid phase}\\
		{$q_s$}&{electric volume charge of solid}\\
		{$\ub$}&{displacement}\\
%		{$\boldsymbol{\sigma}_s$}&{stress tensor of linear elastic solid phase}\\
		{$\str\ub$}&{elastic strain tensor,$\str\ub=\frac{1}{2}\left(\nabla\ub+(\nabla\ub)^{T}\right)$}\\
		{$\vec{D}_f$}{}&{electric displacement of fluid phase}\\
		{$\sigmabf_f$}{}&{stress tensor of the fluid phase}\\
		{$\boldsymbol{\tau}_M$}&{Maxwell 2nd order tensor}\\
		{$p$}&{fluid pressure}\\
		{$P$}&{global pressure}\\
		{$\fb=\{f_k\}$}&{external body force}\\
		{$\vb_f$}&{fluid velocity}\\
		{$\jb_\alpha$}&{migration-diffusion flux of the $\alpha$-th ionic species}\\
		{$\wb$}&{convective velocity}\\
		{$\mu_\alpha^0$}&{standard electrochemical potential expressed at infinite dilution}\\
		{$\mu_\alpha$}&{electrochemical potential of  $\alpha$-th ionic species}\\
		{$c_\alpha$}&{concentration of $\alpha$-th ionic species}\\
		{$z_\alpha$}&{valence of $\alpha$-th ionic species}\\
		{$\Phi_\alpha$}&{ionic potential related to of  $\alpha$-th ionic species}\\
		{$\Psi_f$}&{electrostatic potential of a fluid phase}\\
		{$q_f$}&{electric volume charge of fluid}\\
		{$\kappa$}&{dielectric constant, $\kappa=\mathcal{E}_r\mathcal{E}_0$}\\
		{$\vec{E}=\{\mathrm{E}_k\}$}&{electric field, $\vec{E}=-\nabla\Psi_f$}\\
		{$\Sigma$}&{surface charge at solid-fluid interface}\\
		{$\sqcup^\veps$}&{quantity/domain is related to the scale parameter $\veps$ (also quantity is dimensionless)}\\
		{$\sqcup^0$}&{quantity is related to the "slow" macroscopic scale}\\
		{$\sqcup^1$}&{quantity is related to the "fast" microscopic scale}\\
		{$\sqcup^\textrm{eq}$}&{quantity in equilibrium}\\
		{$\sqcup^\textrm{per}$}&{pertubed quantity}\\
		{$\sqcup^\textrm{mic}$}&{fluctutions part of quantity reconstructed at microscale}\\
		{$\sqcup^\textrm{rec}$}&{quantity recovered at microscopic scale}\\
		{$\sqcup^\textrm{eff}$}&{quantity is solution to dimensionalized macroscopic problem}\\
		\hline
	\end{tabular}
	\caption{The nomenclature: frequently used symbols and superscripts}	
	\label{tab:nomencl}
\end{table}
\begin{table}[!h]\centering
	\begin{tabular}{llll}
		\hline
		Symbols	& Quantity 	Symbols	& Quantity\\
		\hline
		{$\Ab=\{A_{ijkl}\}$}&{elasticity tensor}&{$T$}&{absolute temperature}\\
		{$\db=\{d_{ij}\}$}&{dielectric tensor}&{$k_B$}&{Boltzmann constant}\\
		{$\gb=\{g_{kij}\}$}&{piezoelectric coupling tensor}&{$e$}&{elementary charge}\\
		{$\eta_f$}&{dynamic viscosity}&{$\lambda_D$}&{Debye length parameter}\\
		{$D_\alpha^0$}&{diffusivity of  $\alpha$-th ion. spec.}&{$\gamma$}&{parameter $\gamma=(\ell_\mic/\Lambda_D)^2$}\\
		\hline
	\end{tabular}
	\caption{The nomenclature: material parameters and electrochemical constants used through the text}	
	\label{tab:nomencl2}
\end{table}
\nomenclature[O]{$x$}{coordinate system on macroscopic scale ("slow" scale)}
\nomenclature[O]{$y$}{coordinate system on representative periodic cell ("fast" scale)}
\nomenclature[O]{$\veps$}{scale parameter, $0<\varepsilon<< 1$}
\nomenclature[C]{$L_{c}$}{characteristic dimension of macroscopic domain}
\nomenclature[C]{$l$}{characteristic dimension dimension of microscopic periodic structure}
\nomenclature[O]{$\sigmabf_s^p$}{stress tensor of piezoelectric solid phase}
\nomenclature[O]{$\vec{D}_s$}{electric displacement}
\nomenclature[O]{$q_s$}{electric volume charge of solid}
\nomenclature[C]{$\db=\{d_{ij}\}$}{dielectric tensor}
\nomenclature[C]{$\gb=\{g_{kij}\}$}{piezoelectric coupling tensor}
\nomenclature[F]{$\ub$}{displacement}
\nomenclature[O]{$\boldsymbol{\sigma}_s$}{stress tensor of linear elastic solid phase}
\nomenclature[O]{$\str\ub$}{elastic strain tensor,$\str\ub=\frac{1}{2}\left(\nabla\ub+(\nabla\ub)^{T}\right)$}
\nomenclature[C]{$\Ab=\{A_{ijkl}\}$}{elasticity tensor}
\nomenclature[O]{$\boldsymbol{\sigma}_f$}{stress tensor of the fluid phase}
\nomenclature[O]{$\boldsymbol{\tau}_M$}{Maxwell 2nd order tensor}
\nomenclature[F]{$p$}{fluid pressure}
\nomenclature[F]{$\Psi_s$}{potential of solid phase}
\nomenclature[O]{$q_s$}{electric volume charge of solid}
\nomenclature[C]{$\eta_f$}{dynamic viscosity}
\nomenclature[O]{$\fb=\{f_k\}$}{external body force}
\nomenclature[F]{$\vb_f$}{fluid velocity}
\nomenclature[O]{$\jb_\alpha$}{migration-diffusion flux of the $\alpha$-th ionic species}
\nomenclature[C]{$D_\alpha^0$}{diffusivity of the $\alpha$-th ionic species}
\nomenclature[F]{$\wb$}{convective velocity}
\nomenclature[C]{$T$}{absolute temperature}
\nomenclature[C]{$k_B$}{Boltzmann constant}
\nomenclature[C]{$\mu_\alpha^0$}{standard electrochemical potential expressed at infinite dilution}
\nomenclature[O]{$\mu_\alpha$}{electrochemical potential of  $\alpha$-th ionic species}
\nomenclature[F]{$c_\alpha$}{concentration of $\alpha$-th ionic species}
\nomenclature[C]{$z_\alpha$}{valence of $\alpha$-th ionic species}
\nomenclature[C]{$\lambda_D$}{Debye length parameter}
\nomenclature[C]{$e$}{elementary charge} 
\nomenclature[F]{$\Psi$}{electrostatic potential related to EDL}
\nomenclature[O]{$\nb$}{unit exterior normal}
\nomenclature[F]{$\Psi_f$}{electrostatic potential of a fluid phase}
\nomenclature[C]{$\kappa$}{dielectric constant, $\kappa=\mathcal{E}_r\mathcal{E}_0$}
\nomenclature[O]{$\mathbf{E}=\{E_k\}$}{electric field, $\mathbf{E}=-\nabla\Psi_f$}
\nomenclature[C]{$\Sigma$}{surface charge at solid-fluid interface}
\nomenclature[S]{$\sqcup^\veps$}{quantity/domain is related to the scale parameter $\veps$ (also quantity is dimensionless)}
\nomenclature[S]{$\sqcup^0$}{quantity is related to the "slow" macroscopic scale}
\nomenclature[S]{$\sqcup^1$}{quantity is related to the "fast" microscopic scale}
\nomenclature[S]{$\sqcup^\textrm{eq}$}{quantity in equilibrium}
\nomenclature[S]{$\sqcup^\textrm{per}$}{pertubed quantity}
\nomenclature[S]{$\sqcup^\textrm{mic}$}{fluctutions part of quantity reconstructed at microscale}
\nomenclature[S]{$\sqcup^\textrm{rec}$}{quantity recovered at microscopic scale}
\nomenclature[S]{$\sqcup^\textrm{eff}$}{quantity is solution to dimensionalized macroscopic problem}

%\begin{multicols}{2}
%	\printnomenclature
%\end{multicols}

%\input{references.tex}

%\end{document}

\section{Mathematical model}\label{sec:1}
The porous medium occupies an open bounded domain $\Om \subset \R^d$,  where the dimension is {$d=3$}{}. Without loss of generality, we may assume that the domain represents a block specimen $\Om=\prod_{i=1}^d]0, L_i[$, which will enable us to impose periodic boundary conditions on its boundary $\pd\Om$. Domain $\Om$ consists of the fluid $\Om_f$ and solid $\Omega_s$ parts, whereby both $\Om_f$ and $\Omega_s$ are connected domains. We refer to the solid-fluid interface by $\Gamma=\pd\Om_s\cap\pd\Om_f$, and $\nb$ designates the unit normal vector on $\Gamma$, being outward to $\Om_f$. The subscripts $\sqcup_s$ and $\sqcup_f$ will also be used throughout the text to denote the constants and variables of the respective phases. 

\begin{figure}[t]\centering
	\includegraphics[width=0.6\linewidth]{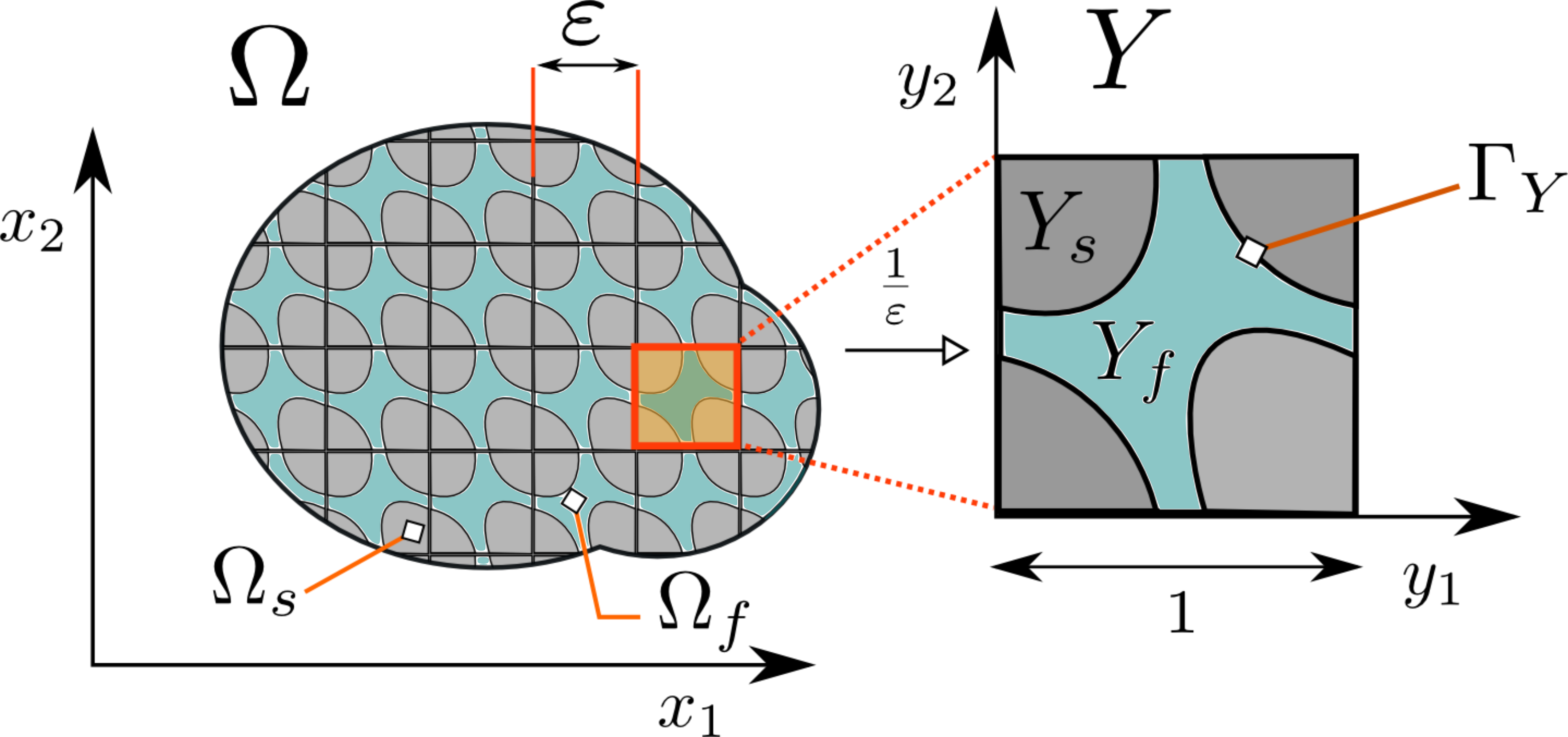}
	\caption{Decomposition of porous body $\Om$ into solid and fluid phase, $\Omega_s$ and $\Omega_f$, that have designated parts of the external boundary, $\partial_{\ext}\Omega_s$ and $\partial_{\ext}\Omega_f$, and interface $\Gamma$.}
\end{figure}

\subsection{Processes in solid phase}\label{sec:1-2}
This section will deal with the mathematical description of processes happening in the solid porous matrix, which is assumed to be piezoelectric. Piezoelectric materials produce an electrical charge in response to applied mechanical stress and \textit{vice versa}. Properties of such material are described by an elasticity tensor $\Ab=\{A_{ijkl}\}$, a dielectric tensor $\db=\{d_{ij}\}$ and a  piezoelectric coupling tensor $\gb=\{g_{kij}\}$. The symmetries of $\Ab$ and $\db$ are obvious, \ie $A_{ijkl} = A_{klij} = A_{jikl}$ and $d_{ij} = d_{ij}$. The third-order $g_{kij}$ adheres the symmetry $g_{kji}=g_{kij}$ asociated with the strain tensor symmetry, whereby the nonzero entries depend on the material crystal structure;
piezoelectric materials are classified into symmetry groups, characterized by specific sparsity and symmetry patterns of the tensor $\gb$, \cite{zou2013symmetry}.

%%In technical practice,  the symmetric indices $ij$ are usually replaced by a single index $l$ as follows
%% \begin{equation}
%% 11\rightarrow 1,\quad 22\rightarrow 2,\quad 33\rightarrow 3,\quad 23\rightarrow 4,\quad 31\rightarrow 5,\quad 12\rightarrow 6.
%% \end{equation}
%% It has a maximum of 18 distinct components, but this number can be significantly reduced if certain symmetries exist in the material crystal structure.

The linear constitutive equations for piezoelectric material  are 
\begin{equation}
\begin{split}\label{eq_piezo_const}
\sigmabf_s&=\Ab\str\ub-\gb^T\nabla\Psi_s,  \\
\vec{D}_s&=\gb\str\ub+\db\nabla\Psi_s,
\end{split}
\end{equation}
where  $\sigmabf_s$ is the stress tensor and $\vec{D}_s$ is the electric displacement.
$\Psi_s$ is the electric potential in the solid in response to applied mechanical stress according to the following relations
\begin{align}
-\diver\sigmabf_s&=\fb\inom s\label{eq_piezo1}   \\
-\diver\vec{D}_s&=q_s\inom s,\label{eq_piezo2}
\end{align}
where $q_s$ is the electric volume charge. \Eq{eq_piezo1} is referred to as piezoelectricity equation and \Eq{eq_piezo2} as balance of charge equation.

\subsection{Processes in fluid phase}\label{sec:1-1}
The fluid is a two-component electrolyte solution consisting of the solvent and two ionic species with valencies $z_\alpha, \alpha=1,2$, $z_1=-1, z_2=+1$. Each ionic species (labeled by $\alpha$) dissolved in the electrolyte is associated with the electrochemical potential $\mu_\alpha$, which depends on the concentration $c_\alpha$, the electrostatic potential $\Psi_f$ and the absolute temperature $T$, such that, \cite{hunter2001foundations},
\begin{equation}\label{eq_elchem_potential}
\mu_\alpha=\mu_\alpha^0+k_B T\ln c_\alpha+ez_\alpha\Psi_f\;,
\end{equation}
where $\mu_\alpha^0$ is the standard electrochemical potential expressed at infinite dilution, $k_B$ is Boltzmann constant, and  $e$ is the elementary charge.

Three phenomena drive the ionic transport in the fluid: 1) convection of the solvent, which is determined by a convective velocity $\wb$,  2) diffusion of the $\alpha$-th ionic species in the solvent characterized by the diffusivity $D_\alpha$ of the $\alpha$-th ionic species, and 3) motion of ions due to the electrical field. 

Considering the mechanical interaction on  solid-fluid interface $\Gamma$, the convective velocity $\wb$ takes into account the extension of the solid deformation velocity to the fluid part
\begin{equation}\label{eq_w}
\wb=\vb_f-\frac{\pd\ub}{\pd t},
\end{equation}
where $\vb_f$ denotes fluid velocity. 
To describe these processes, we introduce a set of constitutive equations that defines the fluid electric displacement $\vec{D}_f$, the fluid stress tensor $\sigmabf_f$ and the diffusive fluxes $\jb_\alpha, \alpha=1,2$
\begin{align}
\vec{D}_f&=\kappa\nabla\Psi_f,\label{eq_diel}\\
\sigmabf_f&=-p\Ib+2\eta_f\str{\vb_f}+\boldsymbol{\tau}_M,\label{eq_stress_fl}\\
\jb_\alpha&=-\frac{c_\alpha D_\alpha}{\textrm{k}_B\textrm{T}}\nabla\mu_\alpha=-\frac{c_\alpha D_\alpha}{\textrm{k}_B\textrm{T}}\left(\frac{\textrm{k}_B\textrm{T}}{c_\alpha }\nabla c_\alpha  + ez_\alpha\nabla\Psi_f\right),\qquad \alpha=1,2. \label{eq_j}
\end{align}
The full fluid stress tensor $\boldsymbol{\sigma}_f$ consist of the fluid stress involving the fluid velocity $\vb_f$, hydrostatic pressure $p$ and dynamic viscosity $\eta_f$ and is extended by the Maxwell 2nd order stress tensor $\boldsymbol{\tau}_M=\kappa\left(\vec{E}\otimes\vec{E}-\frac{1}{2}|\vec{E}|^2\Ib\right).$ The symbol $\vec{E}$ refers to the  electric field and $\str{\vb_f}=1/2\left(\nabla\vb_f+(\nabla\vb_f)^{\textrm{T}}\right)$.

Using these constitutive relations, the processes in fluid phase $\Om_f$ are described by the following system of equations
\begin{align}\label{eq_transport}
\frac{\pd c_\alpha}{\pd t} + \diver \left(\jb_\alpha+\wb c_\alpha\right)&=0&\textrm{ in }\Om_f, \alpha=1,2,\\
-\diver \vec{D}_f&=q_f&\textrm{ in }\Om_f,\label{eq_GP}\\
-\diver\sigmabf_f&=\fb&\textrm{ in }\Om_f,\label{eq_NS}\\
\diver(\wb+\pdt\ub)&=0&\textrm{ in }\Om_f,\label{eq_incompress}
\end{align}
where net charge density $q_f$ is by definition $q_f\equiv e\sum\limits_{\beta=1}^{2}z_\beta c_\beta$.
The transport equation is given by the Eulerian mass conservation law \Eq{eq_transport}, where the effects of diffusion and migration caused by an external electric field are expressed by the migration-diffusion flux $\jb_\alpha$ given by \Eq{eq_j}. The electrokinetics is described by the Gauss-Poisson problem \Eq{eq_GP}, where $\kappa$ refers to the dielectric coefficient of the solvent (assumed to be constant). The flow of the solvent is given by the Navier-Stokes equation \Eq{eq_NS}, where $\fb$ is the external body force. We consider the quasi-steady flow of the solvent only; thus, the convective term, as well as the inertia term, can be neglected. The system of equations is completed by incompressibility condition \Eq{eq_incompress}.  

In contrast with reference \cite{allaire2015ion}, we do not consider any external electrical field taking effect.
%\subsection{Fluid solid interaction}\label{sec:1-2}
%The porous medium is composed of a solid skeleton, which we consider to be an elastic conducting material. Therefore, we assume that the electrostatic potential $\Phi_s$ is constant in the whole sub-domain $\Omega_s$, whereas a constant surface charge $\Sigma$ \opr{exhibits} on its surface $\pd \Om_s$. 

\subsection{Solid-fluid interactions}\label{sec:1-3}
The processes occurring in both phases are closely coupled through the interaction on the solid-fluid interface. Thus, to complete the model given above, we must apply the interface condition on $\Gamma$.

Through equation \Eq{eq_piezo2}, we introduced the electrokinetic variable $\Psi_s$ into the system, which is separated from the potential of fluid phase $\Psi_f$. This separation of variables is possible due to their different origin. However, to preserve their continuity, we must consider the conditions on the solid-fluid interface, \cite{lemaire2011multiscale}.

The solid skeleton exhibits a surface charge $\Sigma$ on the pore surface $\pd \Om_s$. Due to the presence of the Stern layer, the electrical jump { $\Sigma_{fs}=[\vec{D}_s\cdot\nb]_s^f$}{} between surface charges of fluid and solid phases is sometimes considered. However, we follow the approach from \cite{lemaire2011multiscale} and assume that the electrical jump remains small when compared with the surface charge densities of fluid and solid phases, so it can be neglected, \ie  $\Sigma_{fs}\equiv 0$ on the interface. Thus, the condition of surface charge continuity holds
´\begin{equation}\label{eq_piezo_bc1}
\vec{D}_s\cdot\nb=\Sigma=\vec{D}_f\cdot\nb \ongamma,\\
\end{equation}
In addition, we apply the following interface conditions to ensure continuity of potentials, normal stresses, and velocities on the solid-fluid interface $\Gamma$
\begin{align}\begin{aligned}
\Psi_s&=\Psi_f &\ongamma,\\
\pdt{\ub}&=\vb_f&\textrm{ on }\Gamma,\\
\sigmabf_s\cdot\nb&=\sigmabf_f\cdot\nb&\textrm{ on }\Gamma.
\end{aligned}
\label{eq_elasticity_bc}
\end{align}

In addition and contrarily to our previous work \cite{turjani2018}, we consider the existence of the ionic exchange between phases on the solid-fluid interface $\Gamma$; thus, interface condition on migration-diffusion fluxes reads
\begin{equation}\label{eq_j_bc}
\jb_\alpha \cdot\nb=k^\alpha\frac{\pd c_\alpha}{\pd t}\qquad\textrm{ on }\Gamma, \alpha=1,2,
\end{equation}
where $k^\alpha$ is coefficients quantifying ionic surface exchanges, see \cite{lemaire2010modelling}.

To simplify the derivation of the two-scale model, we will assume that the problem is $L$-periodic, where $L$ is the characteristic size of the macroscopic domain $\Om$.  
\begin{remark}[$L$-periodicity]\label{rem:L-per}
	Let us assume that domain $\Omega$ is defined as an N-dimensional block with the length of side $L>0$, so that $\Om=\prod_{d=1}^{N}(0, L).$
	Any smooth variable $\phi$ is considered $L$-periodic if its value and partial derivatives are the same on the opposing sides of the block $\Om$.
\end{remark}
Following this remark, we assume $\Psi_f, \Psi_s, c_\alpha, \ub, \vb$ and $p$ to be $L$-periodic, unless stated otherwise.

The assumption of $L$-periodicity proves helpful during the upscaling process because it eliminates the need to deal with various conditions at the outer boundary. At the same time, it can be limiting for the simulation of real processes with conditions other than periodic at the outer boundary. Dealing with this limitation will be shown later.

\section{Linearization and decomposition into subproblems}\label{sec:2}
In the following section, we linearize the nonlinear model \Eq{eq_transport}-\Eq{eq_elasticity_bc} using the classical linearization approach as introduced by \cite{obrien1978electrophoretic}. The linearization is enabled by introducing the reference equilibrium state and establishing a linearized problem only for perturbed fields.

\subsection{Periodic structure of the porous medium}

The porous medium is characterized by the characteristic size of its micropores $\ell_\mic$ and its macroscopic characteristic length $L_c$. The ratio between these two parameters defines the scale parameter $\veps=\ell_\mic/L_c, 0<\varepsilon\ll 1$  and represents the smallest zoom, by which the microstructure becomes visible from the macroscopic point of view. Then, we can generate the porous medium as a periodic lattice by repeating the representative volume element (RVE) occupying domain $Y^\veps=\veps Y$. The RVE cell $Y = \Pi_{i=1}^3]0,\bar y_i[ \subset \RR^3$ consists of the solid part
occupying domain $Y_s$ and the complementary fluid part $Y_f$, thus
\begin{equation}\label{eq-mi6}
\begin{split}
Y   = Y_s  \cup Y_f  \cup \Gamma_Y \;,\quad
Y_s   = Y \setminus \ol{Y_f}  \;,\quad
\Gamma_Y   = \ol{Y_s } \cap \ol{Y_f }\;.
\end{split}
\end{equation}
The relation between the micro- and macroscopic coordinates is approximately $y=x/\veps$; see the precise decomposition ansatz in \Eq{eq:3a}.
For a given scale $\veps > 0$, $\ell_i = \veps \bar y_i$ is the characteristic size associated with the $i$-th coordinate direction, whereby
also $\veps \approx \ell_i / L_c$, hence $\ell_i \approx \ell_\mic$ (for all $i=1,2,3$) specifies the microscopic characteristic length $\ell_\mic$. %% for a given macroscopic characteristic length $L$.
Below, we introduce two-scale functions depending on $x \in \Om$ and $y \in Y$ using the unfolding operator.

\subsection{Linearized problem}\label{sec:dimless}
This section will summarize the necessary steps to obtain a non-dimensional linearized form of the equations from \sref{sec:1-1}-\ref{sec:1-3}. We report the main adimensional choices and results in the following text and refer to our previous paper \cite{turjani2018} for more insight into the procedure. The non-dimensional problem is obtained through the process of scale analysis. There are some significant benefits to the scaling of mathematical models. The scale analysis of a mathematical model serves to identify small parameters, such as the scale parameter $\veps$ described above. Another benefit of scaling is related to running numerical simulations since scaling simplifies the choice of values for the input data greatly and makes the simulation results more widely applicable.

The macroscopic coordinate is rescaled by characteristic length $L_c$, so $x^\prime=x/L_c$. Thus, the dimensionless operator $\nabla^\prime$ can be defined by $\nabla^\prime=L_c\nabla,  \nabla=(\partial_x)$. Similarly, the time variable is rescaled by $t^\prime=t/t_c$ where we take the diffusion time $t_c=L_c^2/D_\alpha$ as the characteristic time scale. The dimensionless operator $\pdT^\prime$ is defined by $\pdT^\prime=t_c\pdT.$

The dimensionless variables are expressed in terms of their respective characteristic quantities denoted by subscript $\sqcup_c$. They are denoted by superscript $\sqcup^\veps$ to symbolize their relation to small parameter $\veps$. 
\begin{equation}
p^\veps=\frac{p}{p_c}\;,\quad \vb_f^\veps=\frac{\vb_f}{v_c}\;,\quad \wb^\veps=\frac{\wb}{v_c}\;,\quad \Psi_f^\veps=\frac{\Psi_f}{\Psi_c}\;,\quad \Psi_s^\veps=\frac{\Psi_s}{\Psi_c}\;,\quad c_\alpha^\veps=\frac{c_\alpha}{c_c}\;,\quad \ub^\veps=\frac{\ub}{u_c}\;,
\end{equation}
where $p_c$ is the characteristic pressure, $v_c$ the characteristic velocity, $c_c$ characteristic concentration, $\Psi_c$ the characteristic potential and $u_c$ the characteristic displacement. The characteristic pressure $p_c$ is expressed using the ideal gas law, $p_c=c_ck_BT$ and we use the approximation of $\zeta$-potential as a value of the characteristic potential $\Psi_c$, \ie $\Psi_c=\frac{k_BT}{e}$.

According to
\cite{lemaire2011multiscale},  the ratio between the velocity and pressure magnitudes
$\lambda_c:=\frac{v_c\eta_f}{L_cp_c}$ is obtained by the dimensional analysis of the Darcy law, which can also represent the viscous flow in pores. This yields 
\begin{equation*}\label{darcy_ap}
v_c=\frac{kp_c}{\eta_fL_c},
\end{equation*}
where $k$ denotes the intrinsic permeability (units [m$^2$]) depending only on the size of the micropores,  $k\sim l^2$, hence holds and
\begin{equation}\label{eq-B3_ap}
\lambda_c=\frac{v_c\eta_f}{L_c p_c}=\frac{k}{L_c^2 }\sim\frac{l^2}{L_c^2 }=\veps^2.
\end{equation}
We choose the dimensionless forcing terms in the same manner and denote them by superscript $\sqcup^\prime$
\begin{equation}\label{dimless_f}
\Sigma^\prime=\frac{\Sigma}{\Sigma_c}\;,\quad\fb^\prime=\frac{\fb L_c}{p_c}\;{q_s}^\prime=\frac{q_s\Sigma_cE_c}{p_cg_c}.
\end{equation}
The effect of electric double layer (EDL) is represented by the so-called Debye length $\lambda_D$
\begin{equation}\label{debye}
\lambda_D=\sqrt{\kappa\frac{\Psi_c}{e c_c}\left(\sum\limits_{\beta=1}^{2}z_\beta^2\right)^{-1}}
=\sqrt{\kappa\textrm{k}_B\textrm{T}\left(e^2c_c\sum\limits_{\beta=1}^{2}z_\beta^2\right)^{-1}}  ,
\end{equation}
where the last expression is due to the special type of the electrolyte with $z_1=-1,$ $ z_2=+1$. Apart of this parameter,  other dimensionless parameters, such as the Peclect number $ \Pec_\alpha$, the ratio between electrical and thermal energy $\textrm{N}_\sigma$, Damkohler number $\textrm{Da}_\alpha$,  and parameter $\textrm{U}_L$ (see \cite{moyne2006two}) are employed in the the dimensionless form of the system \Eq{eq_transport}-\Eq{eq_elasticity_bc}; 
\begin{equation}\label{param}
\begin{aligned}
\Pec_\alpha=&\frac{\ell_\mic^2\textrm{k}_B\textrm{T}c_c}{\eta_f D_\alpha},\quad &\textrm{N}_\sigma=\frac{e\ell_\mic\Sigma_c}{\kappa\textrm{k}_B\textrm{T}}, \quad
\textrm{Da}_\alpha=&\frac{k_cL_c}{t_cD_\alpha}, \quad
&\textrm{U}_L=&\frac{\eta_fD_\alpha}{ E \ell_\mic^2},\\
\textrm{M}_g=&\frac{\textrm{k}_B\textrm{T}g_c}{\ell_\mic p_ce},\quad &\textrm{M}_\Psi=\frac{\epsilon_s\textrm{k}_B\textrm{T}E_c}{g_c\ell_\mic p_ce},\quad \textrm{C}_p=&\frac{\Sigma_cE_c}{p_cg_c},\qquad &\gamma=&\frac{\ell_\mic^2}{\lambda_D^2},
\end{aligned}
\end{equation}
where  $k_c$ is the characteristic coefficient of ionic exchanges, such that $\hat{k}_\alpha=k_\alpha/k_c$ and $g_c$ is the characteristic value of piezoelectric coupling such that dimensionless tensor of piezoelectric coupling is given by $\gb^\prime=\gb/g_c$. It can be shown that $k_c\equiv \ell_\mic$, see \cite{lemaire2010modelling}. Further, we introduce the dimensionless elasticity tensor $\Ab^\prime=\Ab E_c^{-1}$, where $ E_c$ is the characteristic size of elastic moduli.

\begin{remark}
	Through the following text, we will drop the prime $^\prime$ denoting non-dimensionalized operators to refer to the rescaled macroscopic coordinates and rescaled time by $x$ and $t$, respectively. Similarly, we drop the prime $^\prime$ denoting non-dimensionalized forcing terms and quantities.
\end{remark}

The dimensionless form of the problem \Eq{eq_piezo_const}-\Eq{eq_elasticity_bc} is subsequently linearized. The linearization is performed under the assumptions of sufficiently small fields $\fb$ and $\Psi^{\ext}$. Then, the state variables are only slightly perturbed from equilibrium. Following the linearization procedure in \cite{allaire2013asymptotic}, any unknown $a^\veps$ can be split into its equilibrium part $a^{eq,\veps}$ and its perturbation $\delta a^\veps$, as follows
\begin{equation}
\label{eq_decomposition1}
a^\veps(t,x)=a^{\eq,\veps}(x)+ a^{\perturb,\veps}(t,x),
\end{equation}
where superscript $\sqcup^\textrm{eq}$ indicates equilibrium quantities, which corresponds to solution of system dimensionless form of system given in \sref{sec:1-1} and \sref{sec:1-2} for $\fb=0$, $\wb^{\eq,\veps}=0$ and, by the consequence, zero diffusive fluxes $\jb^\veps_\alpha=0$.

\subsubsection{Equilibrium state quantities}
The equilibrium solution serves as the reference state of the electrolyte such that the linearized state problem governs the perturbation. We define it as a solution of system given in \sref{sec:1-1} and \sref{sec:1-2} for $\fb=0$, $\wb^{\eq,\veps}=0$ and, by the consequence, zero diffusive fluxes $\jb^\veps_\alpha=0$.  The proof of the existence of equilibrium solution $(c_\alpha^{\eq,\veps}(x),\Psi^{\eq,\veps}(x),p^{\eq,\veps}(x),\ub^{\eq,\veps}(x))$  was given elsewhere, \cite{allaire2013asymptotic}, therefore we only provide the results for the sake of completeness.

%Although we assume $\Sigma$ to be a constant defined on interface $\Gamma^\veps$, even for a periodic distribution of charges $\Tuf{\Sigma}=\tilde\Sigma(y)$, $ y\in \Gamma_Y$, the Poisson-Boltzmann problem in equilibrium yields $\veps Y$-periodic solutions $\Psi^{\eq,\veps}$ in $\Om_f^\veps$, recalling the ``macroscopic'' $L$-periodicity on $\pd_\ext\Om_f^\veps$. 

{For a periodic distribution of surface charges $\Tuf{\Sigma}=\tilde\Sigma(y)$, $ y\in \Gamma_Y$, the Poisson-Boltzmann problem in equilibrium yields $\veps Y$-periodic solutions $\Psi^{\eq,\veps}$ in $\Om_f^\veps$, recalling the ``macroscopic'' $L$-periodicity on $\pd_\ext\Om_f^\veps$. This property allows us to consider only the local problem in the zoomed RVE represented by cell $Y_f$.}{} 
Then,
\begin{equation}
\Psi_f^{\eq,\veps}(x)=\Psi_f^{\eq}(y), \quad  c_\alpha^{\eq,\veps}(x)=c_\alpha^{\eq}(y),
\end{equation}
where concentrations $c_\alpha^{\eq}(y), \alpha=1,2$ obey the form of the Boltzmann distribution 
\begin{equation}\label{eq_equilibrium_conc}
c_\alpha^{\eq}(y)=c_\alpha^{b}\exp(-z_\alpha\Psi_f^{\eq}(y)),
\end{equation}
depending on the characteristic concentration in the bulk $c_\alpha^b$, representing the concentration of the $\alpha$-th ionic species in an infinite pore.
The potential $\Psi^{\eq}(y)\in \Vb^1(Y_f)$ is a solution of the Poisson-Boltzmann equation imposed in  $Y_f$, in particular 
\begin{equation}
\begin{aligned}
\nabla_y^2\Psi_f^{\eq}=&\gamma\sum_{\beta=1}^{2}z_\beta c_\beta^b\exp(-z_\beta\Psi_f^{\eq})&\textrm{in }Y_f,\\
\nabla_y\Psi_f^\eq\cdot\nb&=-N_\sigma\Sigma=0&\textrm{on }\Gamma_Y.
\end{aligned}\label{eq_adim_PB0}
\end{equation}
Existence of equilibrium solution  $\Psi^{\eq,\veps}$ for zero Neumann condition is guaranteed under the assumption of $L-$periodicity of $\Psi^{\eq,\veps}$ and providing classic electroneutrality condition $ \sum_{\beta=1}^{2}z_\beta c_\beta^b=0$, which ensures that $\Psi^{\eq,\veps}$  vanishes for the zero surface charge, {\cite{allaire2015ion}}{}.
The equilibrium pressure $p^{\eq,\veps}(x)$ depends on the equilibrium potential $\Psi^{\eq,\veps}$ and equilibrium concentrations $c_\alpha^{eq,\veps}, \alpha=1,2,$ as follows
\begin{equation}
p^{\eq,\veps}=\sum_{\beta=1}^{2} c_\beta^{\eq,\veps}\exp(-z_\beta\Psi_f^{\eq,\veps}).
\end{equation}
Since the convective velocity vanishes at the equilibrium ($\wb^{\eq,\veps}=0$), the convective velocity $\wb^\veps(x)$ is defined by its perturbation part only, \ie  $\wb^\veps(x)=\wb^{\perturb,\veps}(x)$.

Similarly,the equilibrium displacement $\ub^{\eq,\veps}(x)$ and potential $\Psi_s^{\eq,\veps}(x)$ are obtained as a solution of piezoelectric problem in $Y_s$, see \ref{sec_equilib}.

To conclude, by virtue of \Eq{eq_equilibrium_conc}, the unfolded  equilibrium concentrations $\Tuf{c^{\eq,\veps}_\alpha}$ and the unfolded pressure field $\Tuf{p^{\eq,\veps}}$ are $Y$-periodic functions. Moreover,  the unfolded displacements are also $Y$-periodic functions by the consequence, whereby the macroscopic strains vanish. Therefore, we neglect any influence of the equilibrium displacements field on the reference configuration associated with the linearization procedure considered in what follows. The equilibrium pore geometry might be perturbed due to the local strains in $Y_s$.   

\subsubsection{Perturbed state quantities and change of variables}
The work \cite{hunter2001foundations} states that quantities such as $c^\veps_\alpha$ vary strongly across the fluid domain. That suggests that ion concentrations are not adequate for the homogenization procedure. Thus, following the approach suggested in \cite{moyne2003} and \cite{obrien1978electrophoretic}, we adopt the change in variables from ion concentration to so-called ionic potentials $\Phi^\veps_\alpha$. These are referred to as streaming potentials generated by the movement of ions. The change of variables is achieved by expressing the ionic concentrations in the context of Boltzmann distribution,
\begin{equation}\label{eq_ionic_pot}
c^\veps_\alpha(x)=c^b_\alpha\exp(-z_\alpha\left(\Psi_f^\eps(x)+\Phi^\veps_\alpha(x)\right)).
\end{equation}

To summarize,  the potentials $\Psi_f^\veps$ and $\Phi^\veps_\alpha$ that are present in the Boltzmann distribution \Eq{eq_ionic_pot} are defined as follows
\begin{itemize}
	\item Potential $\Psi_f^\veps$ reflects only the effects of the EDL on the ion distribution;  $\Psi_f^\veps$ is given by Poisson-Boltzmann problem \Eq{eq_adim_PB0} in the equilibrium state.
	\item Ionic potentials $\Phi_\alpha^\veps$ (often referred to as the streaming potential) represent the electric field produced by the motion of $\alpha-$th ionic species. In the equilibrium, $\Phi_\alpha^\veps$ vanishes since the convection $\wb^\veps$ and the ionic flux $\jb^\veps_\alpha$ vanish. Thus, the ionic potential is identified by its perturbation only, $\Phi^\veps_\alpha(x)=\dPhie$.
\end{itemize}

\begin{remark}
	The choice of boundary conditions does not influence the mobility of particles, see \cite{obrien1978electrophoretic}; thus, the introduction of ionic potentials eliminates the boundary condition \Eq{eq_piezo_bc1}$_2$ from the system. However, the boundary condition \Eq{eq_piezo_bc1}$_2$ will apply to problem \Eq{eq_adim_PB0}, which determines the potential in equilibrium $\Psi_f^{\eq,\veps}$ on RVE Y. 
\end{remark}

Further, we employ the first-order Taylor expansion to \Eq{eq_ionic_pot} to express the linearized perturbed concentrations $c_\alpha^{\perturb,\veps}$ as follows
\begin{equation}\label{eq_delta_c}
c_\alpha^{\perturb,\veps}(x)=-c_\alpha^{\eq,\veps}(x)z_\alpha\left(\dPsie(x)+\Phi^\veps_\alpha(x)\right).
\end{equation}
Next, we follow the approach from \cite{moyne2003,lemaire2010multiphysical} and introduce global pressure as a combination of hydrodynamic and Donnan pressure, 
\begin{equation}\label{eq_glob_P}
P^\veps= p^{\perturb,\eps}+\sum\limits_{\beta=1}^{2}c^{\eq,\veps} z_\beta \left(\dPsie+\Phi_\beta^{\veps}\right).
\end{equation}

The decomposition of unknown fields \Eq{eq_decomposition1} is substituted into the dimensionless form of the problem \Eq{eq_piezo_const}-\Eq{eq_elasticity_bc}, where fluid velocity is expressed by \Eq{eq_w} and \eqref{eq_delta_c} is employed to express $c^{\perturb,\veps}_\alpha(x)$.
Note that products of the small quantities, \eg $(c^{\perturb,\veps}_\alpha)^2$, are neglected.
Due to the linearization and the introduction of the global pressure \Eq{eq_glob_P}, the non-dimensionalized problem splits into three subproblems that can be solved subsequently. While using global pressure $P^\veps$ given by \Eq{eq_glob_P}, the linearized stress in fluid $\sigmabf_f^\veps$  becomes
\begin{multline}
\sigmabf_f^\veps=-P^\veps\Ib+2\veps^2\str{\wb^\veps}+2\textrm{U}_L\veps^2\str{\pdt{\dube}}+\sum\limits_{\beta=1}^{2}z_\beta c^{\eq,\veps}(\dPhije+\dPsie)\Ib+\\
+\frac{\veps^2}{\gamma}\left(\nabla\Psi^{\eq,\veps}\otimes\nabla\dPsie+\nabla\dPsie\otimes\nabla\Psi^{\eq,\veps}-\nabla\Psi^{\eq,\veps}\cdot\nabla\dPsie\Ib\right)\label{eq_fluid_stress_lin0},
\end{multline}
and the linearized migration-diffusion fluxes $\jb_\alpha^{\perturb,\veps}$ read
\begin{equation}\label{eq_fluid_j_lin}
\jb_\alpha^{\perturb,\veps}=c_\alpha^{\eq,\veps}\nabla\dPhie.
\end{equation}

\paragraph{Dimensionless linearized piezoelectric system} The dimensionless linearized problem of perturbed quantities reads: Find $(\dwbe, \dPe, \dPhie, \dPsie, \dPsis, \dube)$, such that following equations hold
\begin{equation}
\begin{aligned}\label{eq_lin_el}
\veps^2\Delta\dPsie + \gamma\sum_{\beta=1}^{2}z_\beta^2c^{\eq,\veps} \dPsie+\gamma\sum_{\beta=1}^{2}z_\beta^2c^{\eq,\veps} \dPhije&=0&\textrm{in }\Omega^\veps_f,\\
c_\alpha^{\eq,\veps}\pdt{}(\dPhije+\dPsie)+c_\alpha^{\eq,\veps}\diver(\nabla\dPhie)+\frac{\Pe_\alpha}{z_\alpha}\diver(\dwbe c_\alpha^{eq,\veps})&=0&\textrm{in }\Omega^\veps_f,\\
\nabla P^\veps-\veps^2\Delta(\dwbe+\textrm{U}_L\pdt{\dube})-\sum_{\beta=1}^{2}z_\beta c^{\eq,\veps} (\nabla\dPhije+\nabla\Psi^{\ext})&=\fb&\textrm{in }\Omega^\veps_f,\\
\diver(\dwbe+\textrm{U}_L\pdt{\dube})&=0&\textrm{in }\Omega^\veps_f,\\
-\diver(\Ab\str{\dube}-\veps \textrm{M}_g\gb^{T}\nabla\dPsis)&=\fb&\textrm{in }\Omega^\veps_s,\\
-\diver(\veps\gb\str{\dube}+\veps^{2}\textrm{M}_\Psi\db\nabla\dPsis)&=\textrm{C}_p\qb&\textrm{in }\Omega^\veps_s,
\end{aligned}
\end{equation}
with the interface conditions being
\begin{equation}\label{eq_lin_el_PB}
\begin{aligned}
\veps\nabla(\dPsie)\cdot\nb=&0,&\textrm{ on }\Gamma^\veps,\\
c_\alpha^{\eq,\veps}\nabla\dPhie\cdot\nb=&-\varepsilon k_\alpha z_\alpha c_\alpha^{\eq,\veps}\pdt{}(\dPhie+\dPsie),&\textrm{ on }\Gamma^\veps,\\
(\Ab\str{\dube}-\veps \textrm{M}_g\gb^{T}\nabla\dPsis)\cdot\nb=&\sigmabf_f^\veps\cdot\nb, &\textrm{ on }\Gamma^\veps,\\
(\gb\str{\dube}+\veps \textrm{M}_\Psi\db\nabla\dPsis)\cdot\nb=&0, &\textrm{ on }\Gamma^\veps,
\end{aligned}
\end{equation}
and $\dube, \dPsie,\dPhije$ and $\dwbe$ being $Y$-periodic. 

Introduction of the parameter $\veps$ into \Eq{eq_lin_el} - \Eq{eq_lin_el_PB} is the natural consequence of the adimensional choices and dimensional analysis of the system, see \cite{turjani2018}. The scaling of piezoelectric coupling $\gb$ and dielectric tensor $\db$ is consistent with a weakly piezoelectric medium, see  \ie  \cite{Rohan-2018}.

\section{Homogenization}\label{sec_homog}
This section aims to provide an outlook of the upscaling procedure and the structure of the homogenized model, namely to define local problems for so-called characteristic responses on the cell $Y$, give formulae for computing effective coefficients, and formulate the macroscopic homogenized model.  The upsacaling is achieved by the unfolding homogenization method (UFM), which was introduced in \cite{cioranescu2008periodic}.  The UFM uses the properties of the so-called unfolding operator $\Tuft$, which is in more detail discussed in \ref{sec_A}.

\subsection{Weak formulation}
One of the prerequisites of the UFM is to work with weak forms of equations \Eq{eq_lin_el}--\eqref{eq_fluid_j_lin}. 

We assume that $\fb \in \Lb^2(\Om^\veps_f), \qb\in\Lb^2(\Om^\veps_s)$ and $\vec{E} \in \RR^3$ are given. The following functional spaces are employed.
\begin{align}
\Vb^1(\Om_p^\veps)=&\left\{\phibf\in H^1(\Om_p^\veps)^d,  \ L-\textrm{periodic in }x \right\},\nonumber\\
\Vb^1_0(\Om_p^\veps)=&\left\{\phibf\in H^1(\Om_p^\veps)^d,  \phibf=0\qquad\textrm{ on }\Gamma^\veps,\ L-\textrm{periodic in }x \right\},\nonumber\\
V^1(\Om_p^\veps)=&\left\{\psi\in H^1(\Om_p^\veps),\ L-\textrm{periodic in }x\right\},\nonumber
\end{align}
where $H^1(\Om_p^\veps)$ is the Sobolev space $W^{1,2}(\Om_p^\veps)$, subscript $p=s,f$.

Then, using standard per-partes integration, \Eq{eq_lin_el}-\Eq{eq_fluid_j_lin} are transformed into their weak form given below. 

The weak form of problem \Eq{eq_lin_el}--\eqref{eq_fluid_j_lin} reads: Find $(\dwbe,\Pe,\dPhie,\dPsie)\in \Hb^1_{\# 0}(\Om^\veps_f)\times [H^1_\#(\Om^\veps_f)]^3\ \alpha=1,2$, and $(\dPsis,\dube)\in \Hb^1_{\# 0}(\Om^\veps_s)\times [H^1_\#(\Om^\veps_s)]^3$ such that the following equations hold
%\\ \textit{Processes in solid phase:}
\begin{equation}\label{eq_weak_piezo1}
\begin{aligned}
\int\limits_{\Om_s}&\left[\Ab\str{\dube}-\veps \textrm{M}_g\bar\gb^T\nabla\dPsis\right]:\str{\testu}\dV - \int\limits_{\Gamma}\sigmabf_s^\veps\cdot\nb\testu\dS=\int\limits_{\Om_s}\fb\cdot\testu\dV,\\
\int\limits_{\Om_s}&\left[\veps\bar\gb\str{\dube}+\veps^{2}\textrm{M}_\Psi\bar\db\nabla\dPsis\right]\cdot\nabla\testpsis\dV =\int\limits_{\Om_s}\textrm{C}_p\bar\qb\testpsis\dV,
\end{aligned}
\end{equation}
for all $ \testpsis \in H^1_\#(\Om_s^\veps)$ and $\testu \in\Hb^1_\#(\Om_s^\veps)$,
%\\ \textit{Stokes flow:}
\begin{align}\label{eq_weak_piezo2}
\begin{aligned}
\int\limits_{\Omega_f^\veps}\testp&\diver(\dwbe+\textrm{U}_L\pdt{\dube})=0,\\
\veps^2\int\limits_{\Omega_f^\veps}&\nabla(\dwbe+\textrm{U}_L\pdt{\dube}):\nabla\testw  \dx+\int\limits_{\Omega_f^\veps}\Pe\nabla\testw  \dx=\\
& \qquad\qquad\qquad=\int\limits_{\Om_f^\veps}\testw\cdot\fb \dx+\sum_{\beta=1}^{2}z_\beta\int\limits_{\Omega_f^\veps} c^{\eq,\veps}_\beta\testw\cdot\nabla\dPhije \dx,
\end{aligned}
\end{align}
for all $\testw \in \Hb^1_{\# 0}(\Om_f^\veps), \testp \in L^2(\Om_f^\veps)$,
%\\ \textit{Ionic transport:}
\begin{multline}
\int\limits_{\Omega_f^\veps} c^{\eq,\veps}_\alpha(\pdt{\dPsie}+\pdt{\dPhie})\testphi_\alpha  \dx -\int\limits_{\Omega_f^\veps} c^{\eq,\veps}_\alpha\frac{\Pec_\alpha}{z_\alpha}(\dwbe)\cdot\nabla\testphi_\alpha  \dx=\\
=\int\limits_{\Omega_f^\veps} c^{\eq,\veps}_\alpha(\nabla\dPsie+\vec{E})\nabla\testphi_\alpha \dx-\int\limits_{\Gamma^\veps}\veps\testphi_\alpha\nb\cdot k^\alpha z_\alpha c^{\eq,\veps}_\alpha(\pdt{\dPsie}+\pdt{\dPhie}),
\end{multline}
for $\alpha=1,2$ and for all $ \testphi_\alpha \in H^1_\#(\Om_f^\veps)^2$,
%\\ \textit{Poisson-Boltzmann equation:}
\begin{multline}
\veps^2\int\limits_{\Omega^\veps_f}\nabla\dPsie\cdot\nabla\testpsi \dx + \gamma \int\limits_{\Omega^\veps_f}\left(\sum_{\beta=1}^{2}z_\beta^2 c^{\eq,\veps}_\beta\right)\dPsie\testpsi \dx =\\
=-\gamma\sum_{\beta=1}^{2}z_\beta^2\int\limits_{\Omega^\veps_f} c^{\eq,\veps}_\beta\dPhije\testpsi\dx,\label{eq_weak_piezo3}
%\end{aligned}
\end{multline}
for all $  \testpsi \in H^1_\#(\Om_f^\veps)$.

%\subsubsection{Dealing with Neumann condition in piezoelasticity equation}

The second term in \Eq{eq_weak_piezo1}$_1$ can be rewritten by using the continuity of stresses and Stokes divergence theorem, becoming
\begin{equation}\label{eq_stress_f}
\int\limits_{\Gamma}\sigmabf_s^\veps\cdot\nb\testu\dS=\int\limits_{\Gamma}\sigmabf_f^\veps\cdot\nb\testu\dS=-\int\limits_{\Omega_f}\diver{(\sigmabf_f^\veps\testu)}\dS=\int\limits_{\Om_f}\fb\cdot\testu\dV-\int\limits_{\Omega_f}\sigmabf_f^\veps:\nabla\testu\dV,
\end{equation}
where we substitute $\diver{(\sigmabf_f^\veps\testu)}$ by  \Eq{eq_lin_el}$_3$ and where the fluid stress $\sigmabf^\veps_f$ is given by \Eq{eq_fluid_stress_lin0}. Next, we deal with the \rhs forcing terms that appear in both \Eq{eq_weak_piezo1}$_1$ and \Eq{eq_stress_f}  with regards to the fluid and the solid part. Since we consider the boundary conditions to be periodic and continuous stresses at the solid-fluid interface, according to \cite{allaire2015ion}, the volume force $\fb$ must satisfy the compatibility condition 
\begin{equation}
\int\limits_{\Om}\fb\dV=0.
\end{equation}
Considering this force compatibility condition and the \eqref{eq_stress_f}, the piezoelasticity equation \Eq{eq_weak_piezo1}$_1$ becomes
\begin{equation}\label{eq_weak_piezoel_2}
\int\limits_{\Om_s}\left[\Ab\str{\dube}-\veps \textrm{M}_g\bar\gb\nabla\dPsies\right]:\str{\testu}\dV + \int\limits_{\Om_f}\sigmabf_f^\veps:\nabla\testu\dV=\int\limits_{\Om}\fb\cdot\testu\dV,
\end{equation}
for all test functions $\testu \in\Hb_\#^1(\Om_s^\veps)$ and where $\sigmabf_f^\veps$ is given by \Eq{eq_fluid_stress_lin0}.

\subsection{Convergences and recovery sequences}
The two-scale limit problem can be obtained due to the weak convergences in the unfolded domain $\Om\times Y$. More information about the two-scale convergence method provides \cite{allaire2010homogenizationA} and \cite{allaire2015ion}. 
In our case, the unfolded equations of the weak formulation are obtained using the unfolding operator $\mathcal{T}_\veps$ defined in  \ref{sec_A}, see \cite{cioranescu2008periodic}.
{Due to the a~priori estimates on the solutions \Eq{eq_weak_piezo1}-\Eq{eq_weak_piezo3}, it can be proved that there exist limit fields $\wb^0, P^0, P^1, \Psi^0, \ub^0, \ub^1$ and $\left\{\Phi^0_\alpha,\Phi^1_\alpha\right\}_{\alpha=1,2}, $ for $\veps\rightarrow 0$ and any fixed time $t>0$, see \ref{appendix_conv}. We only explain them briefly as these convergence results do not differ from the previously explored steady-state case presented in our work \cite{turjani2018}.}{}

As a consequence of the convergences, see \ref{appendix_conv}, truncated asymptotic expansions of the unfolded unknown fields can be introduced, which satisfy the same convergence result. These form the recovery sequences  in subdomains $\Om_f^\veps$ and $\Om_s^\veps$ 
%(see Remark~\ref{r_rec_sec})
\begin{align}
%	\begin{split}
\Tuf{\chi_f^\veps\dwbe}&\approx\wb^0(t,x,y),\qquad
\Tuf{\chi_f^\veps \Pe}\approx P^0(t,x)+\veps P^1(t,x,y),\nonumber\\
\Tuf{\chi_f^\veps\dPsief}&\approx\Psi^0_f(t,x,y),\qquad
\Tuf{\chi_s^\veps\dube}\approx\ub^0(t,x)+\veps \ub^1(t,x,y),\label{eq_seq_piezo}\\
\Tuf{\chi_s^\veps\dPsies}&\approx\Psi^0_s(t,x,y),\qquad
\Tuf{\chi_f^\veps \dPhie}\approx\Phi_\alpha^0(t,x)+\veps \Phi_\alpha^1(t,x,y),\quad \alpha=1,2,  \nonumber
%	\end{split}
\end{align}
where $\chi^\veps_s$ and $\chi^\veps_f$ are characteristic functions of subdomains $\Om_f^\veps$ and $\Om_s^\veps$, respectively.
Furthermore, $( \wb^0, P^0, P^1, \{\Phi^0_\alpha, \Phi^1_\alpha\})\alpha=1,2,$ $\Psi_f^0, \Psi_s^0$ and $( \ub^0, \ub^1)$ are the unique solutions to the corresponding two-scale limit problems, which are obtained by substituting the recovery sequences \Eq{eq_seq_piezo} into the weak form of the problem \Eq{eq_weak_piezo1}--\eqref{eq_weak_piezo3}. All these functions depend on time $t$, however, we only treat the quasi-static problem where the time dependence is not involved significantly, following the approach of \cite{rohan2019biot}. Analogous approximations of recovery sequences are considered for the test functions $(\testw,\testp,\testphi_\alpha,\testpsi,\testpsis,\testu)$, which are associated with the unknowns $( \wb^\veps, P^\veps,  \Phi_\alpha^\veps, \Psi^\veps, \Psi_s^\veps, \ub^\veps)$. 

\subsubsection{The two-scale problem}
{The equations of the two-scale limit problem are obtained by substituting the recovery sequences \Eq{eq_seq_piezo} into the weak form of the problem \Eq{eq_weak_piezo1}--\eqref{eq_weak_piezo3} and read as follows: Find $( \ub^0, \ub^1,\Psi_s^0),( \wb^0, P^0,\{\Phi^0_\alpha, \Phi^1_\alpha\})\alpha=1,2,$ and $\Psi_f^0$  such that}{}
\begin{multline}
\intOmYi{s}\left[\Ab\left(\strx{\ub^0}+\stry{\ub^1}\right)-\textrm{M}_g\bar\gb\nablay{\Psi}_s^0\right]:\left(\strx{\testu^0}+\stry{\testu^1}\right)=\\
=-\intOmYi{f}\vec{\sigmabf}^1_f:\left(\nablax{\testu^0}+\nablay{\testu^1}\right)+\intOmYi{s}\fb\cdot\testu^0,\label{eq_piezo_limit_1}
\end{multline}
where limit $\vec{\sigmabf}^1_f$ is given by 
\begin{align}
\vec{\sigmabf}^1_f=&-P^0(x)\Ib+\sum\limits_{\beta=1}^{N=2}z_\beta c_\beta^\eqe(y)(\Psio(x,y)+\Phijo(x))\Ib\nonumber\\
+&\gamma^{-1}\left(\nablay\Psieq(y)\otimes\nablay\Psio(x,y)+\nablay\Psio(y)\otimes\nablay\Psieq(y)-\nablay\Psieq(y)\cdot\nablay\Psio(y)\Ib\right),\label{eq_sigma1}
\end{align}
%for all $\testu^0 \in\Hb^1_\#(\Om_s^\veps)$.
\begin{flalign}
\intOmYi{s}\bar\gb:\left[\strx{\ub^0}+\stry{\ub^1}\right]\cdot\nabla_y\testpsis^0+\intOmYi{s}\textrm{M}_\Psi\nablay{\Psi}_s^0\cdot \nablay\testpsis^0&=\intOmYi{s}\textrm{C}_p\qb\testpsis^0,&
\end{flalign}
%for all $ \testpsis^0 \in H^1_\#(\Om_s^\veps)$.

%Find $(\delta\wb^0,P^0,\{\Phi^0_\beta, \Phi^1_\beta\})$:
\begin{flalign}
\intOmYi{f}&\testp^0\nablay\cdot(\wb^0+\textrm{U}_L\pdT\ub^0)=0,&
\end{flalign}
\begin{multline}
\intOmYi{f}\nablay(\wb^0):\nablay\testw^0 + \intOmYi{f}P^1\nablay\testw^0-\intOmYi{f}\testw^0\cdot(\fbh+\nablax P^0)=\\
= \sum\limits_{\beta=1}^{N=2}z_\beta\intOmYi{f}c_\beta^\eq\left[\testw^0\left(\nablax\Phi_\beta^0+\nablay\Phi_\beta^1+\vec{E}\right)\right],
\end{multline}

%for all $\testw^0 \in \Hb^1_{\# 0}(\Om_f^\veps)$, $\testp^0 \in L^2(\Om_f^\veps)$, $ \testpsis^0 \in H^1_\#(\Om_s^\veps)$, $\testu^0 \in\Hb^1_\#(\Om_s^\veps)$.

\begin{multline}
\intOmYi{f}c_\alpha^\eq(\pdT\Psio+\pdT\Phio)\testphi_\alpha^0-\intOmYi{f}c_\alpha^\eq(\nablax\Phi_\alpha^0+\nablay\Phi_\alpha^1+\vec{E})\cdot(\nablax\testphi_\alpha^0+\nablax\testphi_\alpha^1)+\\
+\intOmYi{f}c_\alpha^\eq\frac{\Pec_\alpha}{z_\alpha}\left[(\nablax\testphi_\alpha^0+\nablay\testphi_\alpha^1)\cdot\wb^0\right]=0,
\end{multline}
%for all $\testphi^0_\alpha \in H^1_\#(\Om_f^\veps)$.
%\textit{ Poisson-Boltzmann:}
%Find $\Psio(x,y)$ such that
\begin{flalign}
\intOmYi{f}\nablay{\Psi^0}\nablay\testpsi^0+\gamma\intOmYi{f}\left(\sum\limits_{\beta=1}^{N=2}z_\beta^2 c_\beta^\eq\right)\Psi^0\testpsi^0&=-\gamma\sum\limits_{\beta=1}^{N=2}z_\beta^2\intOmYi{f}c_\beta^\eq\Phi_\beta^0)\testpsi^0,&\label{eq_piezo_limit_2}
\end{flalign}
for all $(\testpsi^0,\testphi^0_\alpha)\in H^1_\#(\Om_f^\veps)^2$,  $\testw^0 \in \Hb^1_{\# 0}(\Om_f^\veps)$, $\testp^0 \in L^2(\Om_f^\veps)$, $ \testpsis^0 \in H^1_\#(\Om_s^\veps)$, $\testu^0 \in\Hb^1_\#(\Om_s^\veps)$.

\subsubsection{Scale separation formulae}
To separate the fast and the slow variables, we introduce the so-called scale-separation formulae, which allow us to establish local problems for characteristic responses. Therefore, scale decomposition formulae of the limits $\wb^0,P^1,\Phi_\alpha^1,\alpha=1,2,$ are introduced as
\begin{align}\label{eq_split_piezo1}
\begin{split}
\wb^0(x,y,t)=  \sum\limits_{\beta=1}^{2}\ombf^{\beta,k}(y)\left(\frac{\pd \Phi_\beta^0}{\pd x_k} +\mathrm{E}_k\right)(x)+\ombf^{P,k}(y)\left(f_k-\frac{\pd P^0}{\pd x_k} \right)(x),\\
\Phi_\alpha^1(x,y,t)=  \sum\limits_{\beta=1}^{2}\theta^{\beta,k}(y)\left(\frac{\pd \Phi_\beta^0}{\pd x_k} +\mathrm{E}_k\right)(x)+\theta^{P,k}(y)\left(f_k-\frac{\pd P^0}{\pd x_k} \right)(x),\\
P^1(x,y,t)= \sum\limits_{\beta=1}^{2}\pi^{\beta,k}(y)\left(\frac{\pd \Phi_\beta^0}{\pd x_k} +\mathrm{E}_k\right)(x)+\pi^{P,k}(y)\left(f_k-\frac{\pd P^0}{\pd x_k} \right)(x),
\end{split}
\end{align}
where two families of corrector base functions $(\ombf^{P,k}, \pi^{P,k},\theta_\beta^{P,k})$, $(\ombf^{\alpha,k}, \pi^{\alpha,k},\theta_\beta^{\alpha,k}), \alpha=1,2,$  were introduced, indexed by $k\in\{1,\dots,n\}$, where $n$ is spatial dimension of the problem. The standard Einstein summation convention holds.
%\par We denote $\mathbf{e}^k$ canonical basis of $\RR^d$, $k=1,...,d$ and $\delta_{ij}$ the Kronecker symbol. 

Similarly, the scale decomposition formulae of the limits $ \Psi^0_f, \Psi^0_s$ and $\ub^1$ read
\begin{align}\label{eq_split_piezo2}
\begin{split}
\Psi_f^0(t,x,y)&=\sum\limits_{\beta=1}^{2}\varpi^\beta(y)\left(\Phi_\beta^0(x) {\ext}(x)\right),\\
\Psi_s^0(t,x,y)&=\eta^{ij}(y)e_{ij}\left(\ub^0(x)\right)-P^0(x)\eta^P(y)+\sum\limits_{\beta=1}^{2}\eta^\beta(y)z_\beta\Phi_\beta^0(x) ,\\
\ub^1(t,x,y)&= \wb^{ij}(y)e_{ij}\left(\ub^0(x)\right)-P^0(x)\wb^P(y)+\sum\limits_{\beta=1}^{2}\wb^\beta(y)z_\beta\Phi_\beta^0(x) ,
\end{split}
\end{align}
where $\varpi^\beta,\wb^{ij},\wb^P, \wb^{\alpha}, \eta^{ij},\eta^P, \eta^{\alpha}, \alpha=1,2,$ are corrector base functions that are obtained as the solutions of the local cell problems introduced below. 
\subsubsection{Cell problems}
Corrector base functions introduced in the previous section are a solution to the local cell problems defined on the RPC $Y$. For better orientation, they can be divided into the following three groups:

\begin{itemize}
	\item three cell problems related to electrokinetic system with solutions {$(\ombf^{P,k}, \pi^{P,k},\theta_\beta^{P,k})$}{},  $(\ombf^{\alpha,k}, \pi^{\alpha,k},\theta_\beta^{\alpha,k})$ and $(\ombf^{u,k}, \pi^{u,k},\theta_\beta^{u,k})$ for $\alpha=1,2$ and  $\beta=1,2$,
	\item auxiliary cell problem with solution $\varpi^\beta$ for $\beta=1,2$,
	\item three cell problems related to poroelasticity with solutions $,\wb^{ij},\wb^P$ and $\wb^{\alpha}$ for $\alpha=1,2$.
\end{itemize}

All seven cell problems are introduced below. In their description, we use the following bilinear forms for the sake of brevity,
\begin{equation}
\begin{aligned}
\aYi{\ombf}{\phibf}{s}&=\intYi{s}\Ab\stry\ombf:\stry\phibf \dV,&
\gYi{\ombf}{\varphi}{s}&=\intYi{s}\left[\gb:\stry\ombf\right] \cdot \nablay\phi \dV,\\
\ghYi{\ombf}{\varphi}{s}&=\intYi{s}\textrm{M}_g\gb^{T}\nablay\varphi:\stry\ombf \dV,&
\dYi{\psi}{\varphi}{s}&=\intYi{s} \textrm{M}_\Psi \nablay\psi\cdot\nablay\varphi\dV,\\
\bYi{\pi}{\phibf}{f}&=\intYi{f}\pi\nabla\cdot\phibf\dV,&
\cYi{\underline{\pi}}{\phibf}{f}&=\sum\limits_{\beta=1}^{2}\intYi{f} z_\beta c^\eq_\beta(y)\nabla\pi_\beta\cdot\phibf \dV,\label{eq_bilin_6}\\
\hYib{\psi}{\pi}{f}&=\intYi{f}  c^\eq_\beta(y)\nabla\psi\cdot\nabla\pi\dV,&
\pYib{\pi}{\ombf}{f}&=\intYi{f}  c^\eq_\beta(y)\Pec_\beta z_\beta^{-1}  \pi\nabla\cdot\ombf\dV,\\
\eYi{\ombf}{\phibf}{s}&=\intYi{s}\stry\ombf:\stry\phibf \dV,& \left<\psi,\pi\right>_{Y_f}&=\intYi{f}\psi\cdot\pi\dV,
\end{aligned}
\end{equation}
where $\underline{\pi}\equiv\{\pi_\alpha\}=\{\pi_1,\pi_2\}$.
\par The first group of local problems consists of the two autonomous problems solved in $Y_f$. The local problem that gives the response to the macroscopic pressure gradient reads: Find $(\ombf^{P,k}, \pi^{P,k},\theta_\alpha^{P,k})\in \Hb^1_{\#0}(Y_f)$,  $k=1,2,3,\ \alpha=1,2$: 
\begin{equation}
\begin{split}
\label{eq_lp_el0}
\eYi{\ombf^{P,k}(y)}{\testw}{f} + \bYi{\pi^{P,k}(y)}{\testw}{f}-\cYi{\underline{\theta}^{P,k}}{\testw}{f}&=\left<y_k,\testw\right>_{Y_f},\\
\hYib{\testphi}{\theta_\beta^{P,k}}{f}+\pYib{\testphi}{\ombf^{P,k}(y)}{f}&=0,\\
\bYi{\testp}{\ombf^{P,k}(y)}{f}&=0,
\end{split}
\end{equation}
for any test functions $\testw\in \Vb^1_0(Y_f)$, $\testp\in L^2(Y_f)$, $\testphi\in V^1(Y_f)$ and for $\underline{y}^1\equiv\{y_k,0\}$, $\underline{y}^2\equiv\{0,y_k\}$.

\par The weak form of the second autonomous cell problem that returns the response to the macroscopic diffusive flux reads: Find $(\ombf^{\alpha,k}, \pi^{\alpha,k},\theta_\beta^{\alpha,k})\in \Hb^1_{\#0}(Y_f)$, $k=1,2,3,\ \alpha=1,2, \ \beta=1,2$:
\begin{equation}
\begin{split}
\label{eq_lp_ela}
\eYi{\ombf^{\alpha,k}(y)}{\testw}{f} + \bYi{\pi^{\alpha,k}(y)}{\testw}{f}-\cYi{\underline{\theta}^{\alpha,k}}{\testw}{f}&=\cYi{\delta_{\alpha\beta}y_k}{\testw}{f},\\
\hYib{\testphi}{\theta_\beta^{\alpha,k}}{f}+\pYib{\testphi}{\ombf^{\alpha,k}(y)}{f}&=-\hYib{\testphi}{\delta_{\alpha\beta}y_k}{f},\\
\bYi{\testp}{\ombf^{\alpha,k}(y)}{f}&=0,
\end{split}
\end{equation}
for any test functions $\testw\in \Vb^1_0(Y_f)$, $\testp\in L^2(Y_f)$, $\testphi\in V^1(Y_f)$ and for $\underline{y}^1\equiv\{y_k,0\}$, $\underline{y}^2\equiv\{0,y_k\}$.

%The linear forms $f_{Y_f}(\testw), g_{Y_f}(\testphi)$ and $h_{Y_f}(\testp)$ on the r.h.s. of \Eq{eq_lp_el0} are  defined for respective cell problems with solutions $(\ombf^{0,k}, \pi^{0,k},\theta_\beta^{0,k})$,  $(\ombf^{\alpha,k}, \pi^{\alpha,k},\theta_\beta^{\alpha,k})$ and $(\ombf^{u,k}, \pi^{u,k},\theta_\beta^{u,k}), \alpha=1,2,$  as follows:
%\begin{eqnarray}
%f_{Y_f}(\testw)=&\left\{\begin{array}{ll}
%\intYi{f}y_k\cdot\testw\dV, &\textrm{ for }	(\ombf^{0,k}, \pi^{0,k},\theta_\beta^{0,k}),\\
%\cYi{\delta_{\alpha\beta}y_k}{\testw}{f} ,	 &\textrm{ for }(\ombf^{\alpha,k}, \pi^{\alpha,k},\theta_\beta^{\alpha,k}),\\
%-\aYi{\textrm{U}_L\underline{y}^\alpha}{\testw}{f}, &\textrm{ for } (\ombf^{u,k}, \pi^{u,k},\theta_\beta^{u,k}).
%	\end{array}\right. \\
%g_{Y_f}(\testphi)=& \left\{\begin{array}{ll}
%0, &\textrm{ for }	(\ombf^{0,k}, \pi^{0,k},\theta_\beta^{0,k}),\\
%-\dYi{\testphi}{\delta_{\alpha\beta}y_k}{f},	 &\textrm{ for }(\ombf^{\alpha,k}, \pi^{\alpha,k},\theta_\beta^{\alpha,k}),\\
%0, &\textrm{ for } (\ombf^{u,k}, \pi^{u,k},\theta_\beta^{u,k}).
%\end{array}\right. \\
%h_{Y_f}(\testp)=&\left\{\begin{array}{ll}
%0, &\textrm{ for }	(\ombf^{0,k}, \pi^{0,k},\theta_\beta^{0,k}),\\
%0, &\textrm{ for }	(\ombf^{\alpha,k}, \pi^{\alpha,k},\theta_\beta^{\alpha,k}),\\
%-\intYi{f}\textrm{U}_L\testp\nabla\cdot(y_k)\dV, &\textrm{ for } (\ombf^{u,k}, \pi^{u,k},\theta_\beta^{u,k}).
%\end{array}\right.
%\end{eqnarray}
%for any test functions $\testw\in \Vb^1_0(Y_f)$, $\testp\in L^2(Y_f)$, $\testphi\in V^1(Y_f)$ and for $\underline{y}^1\equiv\{y_k,0\}$, $\underline{y}^2\equiv\{0,y_k\}$.

\par The auxiliary cell problem related to potential perturbation reads:  
Find corrector base functions $\varpi^\alpha\in V^1(Y_f),\alpha=1,2$, such that
\begin{equation}\label{eq_lp_pot}
\intYi{f} \nabla\varpi^\alpha\cdot\nabla\testpsi\dV+\gamma\intYi{f}\sum\limits_{\beta=1}^{2}\left(z_\beta^2c^\eq_\beta(y)\right)\varpi^\alpha\testpsi\dV=-\intYi{f} \gamma z_\alpha^2  c^\eq_\alpha(y)\testpsi\dV,
\end{equation}
for any test functions $\testpsi\in V^1(Y_f)$.

Finally, we introduce the three local problems that constitute the third group. They read as follows:
\begin{itemize}
	\item Find $\wb^{ij}\in\Hb^1_\#(Y_s), \intYi{s} \wb^{ij}\dV=0, \eta^{ij}\in H^1_\#(\Om_s^\veps)$ such that
	\begin{equation}
	\begin{split}\label{eq_lp_disp_p1}
	\aYi{\wb^{ij}+\Pibf^{ij}}{\testu}{s}-\ghYi{\testu}{\eta^{ij}}{s}&=0\\
	\gYi{\wb^{ij}+\Pibf^{ij}}{\testpsis}{s}+\dYi{\eta^{ij}}{\testpsis}{s}&=0,
	\end{split}
	\end{equation}
	for any test functions $\testu\in\Hb^1_\#(Y_s)$ and $\testpsis\in H^1_\#(\Om_s^\veps)$. {The transformation tensor $\Pibf^{ij}=(\Pi^{ij}_k), i,j,k=1,\dots,n$ enable to establish local displacements defined in Y
		generated by affine transformation of the macroscopic strains $\eb_x(\ub_0)$ defined in $\Om$;}{} it holds
	that $\eb_y(\Pibf_{ij}e^x_{ij}(\ub_0))=\eb_x(\ub_0)$, where $\Pi^{ij}_k=y_j\delta_{ik}$.
	
	\item Find $\wb^P\in\Hb^1_\#(Y_s), \intYi{s} \wb^P \dV=0, \eta^{P}\in H^1_\#(\Om_s^\veps)$ such that
	\begin{equation}
	\begin{split}\label{eq_lp_disp_p2}
	\aYi{\wb^{P}}{\testu}{s}-\ghYi{\testu}{\eta^{P}}{s}&=-\intYG{\Gamma_Y}\testu\cdot\nb \dSy,\\
	\gYi{\wb^{P}}{\testpsis}{s}+\dYi{\eta^{P}}{\testpsis}{s}&=0,
	\end{split}		
	\end{equation}
	for any test functions $\testu\in\Hb^1_\#(Y_s)$ and $\testpsis\in H^1_\#(\Om_s^\veps)$.
	\item Find $\wb^{\alpha}\in\Hb^1_\#(Y_s), \intYi{s}\wb^{\alpha}\dV=0, \eta^{\alpha}\in H^1_\#(\Om_s^\veps)$ such that
	\begin{equation}
	\begin{split}\label{eq_lp_disp_p3}
	&\aYi{\wb^{\alpha}}{\testu}{s}-\ghYi{\testu}{\eta^{\alpha}}{s}=
	-\sum\limits_{\beta=1}^2\intYG{\Gamma_Y}\testu\cdot \left( z_\beta c^\eq_\beta\left(\delta_{\alpha\beta}+\varpi^\alpha\right)\Ib\right)\cdot\nb \dSy-\\
	&\quad-\intYG{\Gamma_Y}\testu\cdot\left( \frac{1}{\gamma}\left(\nabla_y\Psi^\eq\otimes\nabla_y\varpi^\alpha+\nabla_y\varpi^\alpha\otimes\nabla_y\Psi^\eq-\nabla_y\Psi^\eq\cdot\nabla_y\varpi^\alpha\Ib\right)\right)\cdot\nb \dSy,\\
	&\gYi{\wb^{\alpha}}{\testpsis}{s}+\dYi{\eta^{\alpha}}{\testpsis}{s}=0,
	\end{split}		
	\end{equation}
	for any test functions $\testu\in\Hb^1_\#(Y_s)$ and $\testpsis\in H^1_\#(\Om_s^\veps)$.
\end{itemize}
The cell problems corresponding to the electrokinetic system can be treated separately. In contrast, the cell problems related to poroelasticity are coupled through the \rhs of \Eq{eq_lp_disp_p3} with the solutions $\varpi^\alpha, \alpha=1,2,$ of the auxiliary cell problems \Eq{eq_lp_pot}.

\subsubsection{Homogenized coefficients}
Below, we list expressions of homogenized coefficients that arise from the upscaling of  \Eq{eq_weak_piezo1}--\eqref{eq_weak_piezo3}. These coefficients constitute the effective material parameters of the upscaled porous medium and are expressed in terms of the corrector basis functions. We may distinguish two groups of coefficients according to which corrector functions they are associated.

The first group of the corrector functions defines the following homogenized coefficients
\begin{equation}\label{eq_ec_el_p1}
\begin{split}
\Jcal^\alpha_{lk}&=\left<\ombf^{\alpha,k}(y),\mathbf{e}^l\right>_{Y_f},\quad\\
\Kcal_{lk}&=\left<\ombf^{P,k}(y),\mathbf{e}^l\right>_{Y_f},\\
\Dcal^{\alpha\beta}_{lk}&=\left<c ^\eq_\beta(y)\left(\Pec_\beta z^{-1}_\beta\ombf^{\alpha,k}(y)+\left(\mathbf{e}^k\delta_{\alpha\beta}+\nabla_y\theta_\beta^{\alpha,k}(y)\right)\right),\mathbf{e}^l\right>_{Y_f},\\
\Lcal^{\alpha}_{lk}&=\left<c ^\eq_\alpha(y)\left(\Pec_\alpha z^{-1}_\alpha\ombf^{P,k}(y)+\nabla_y\theta_\alpha^{P,k}(y)\right),\mathbf{e}^l\right>_{Y_f}, 
\end{split}
\end{equation}
which are identical to the effective coefficients describing the model of ionic transport through linear elastic porous media, see \cite{turjani2018}. These are the permeability tensor $\Kcalbf=(\mathcal{K}_{lk})$, the diffusivity tensors $\Dcalbf^{\alpha\beta}=(\mathcal{D}^{\alpha\beta}_{lk})$, the tensors $\Jcalbf^\alpha=(\mathcal{J}^{\alpha}_{lk})$ related to the flow driven by electric fields and coupling tensors $\Lcalbf^\alpha=(\mathcal{L}^{\alpha}_{lk})$.

The following four effective coefficients are new additions. They arise from the homogenization of quasi-steady problems and read
\begin{equation}\label{eq_ec_el_p2}
\begin{aligned}
\Qcalbf^{\alpha\beta}&=\left< c^\eq_\alpha(y),\varpi^{\beta}(y)+\delta_{\alpha\beta}\right>_{Y_f}, \quad
&\hat{\Qcalbf}^{\alpha\beta}&=\left< c^\eq_\alpha(y),\varpi^{\beta}(y)\right>_{Y_f}\\
\Scalbf^{\alpha\beta}&=\intYi \Gamma -z_\alpha c^\eq_\alpha(y)\hat{k}_\alpha(\varpi^{\beta}(y)+\delta_{\alpha\beta})\cdot\nb\dSy,\quad
&\hat{\Scalbf}^{\alpha\beta}&=\intYi \Gamma -z_\alpha c^\eq_\alpha(y)\hat{k}_\alpha\varpi^{\beta}(y)\cdot\nb\dSy.
\end{aligned}
\end{equation}
These coefficients are connected to the terms, which include time derivatives $\pd{_t}$ and serve to couple the electrokinetics, ionic transport, and mechanics. The coefficients $\Qcalbf^{\alpha\beta} $ and $\hat{\Qcalbf}^{\alpha\beta} $ characterize the evolution of ionic potentials. The coefficients $\Scalbf^{\alpha\beta}$ and $\hat{\Scalbf}^{\alpha\beta} $ bring ionic exchanges up to the macroscopic scale.  It can be shown that these coefficients meet Onsager's reciprocity conditions, \cite{allaire2010homogenizationA}.

The second group is relevant to the piezoelectric behavior of the macroscopic body and consists of the following coefficients,

\begin{equation}\label{eq_ec_piezo_disp_1}
\begin{split}
\Acal_{ijkl}&= 	\aYi{\wb^{ij}+\Pibf^{ij}}{\wb^{kl}+\Pibf^{kl}}{s}+\dYi{\eta^{ij}}{\eta^{kl}}{s}=\\
&=\aYi{\wb^{kl}+\Pibf^{kl}}{\Pibf^{ij}}{s}-\ghYi{\Pibf^{ij}}{\eta^{kl}}{s},\\
\Bcal_{ij}&=\phi_f\delta_{ij}+\aYi{\wb^{P}}{\Pibf^{ij}}{s}-\ghYi{\Pibf^{ij}}{\eta^{P}}{s},\\
\Ccal^\alpha_{ij}&=\aYi{\wb^{\alpha}}{\Pibf^{ij}}{s}-\ghYi{\Pibf^{ij}}{\eta^{\alpha}}{s}+\sum\limits_{\beta=1}^{2}z_\beta\Ib\intYi{f} c_\beta^0(y)\left(\varpi^\alpha(y)+\delta_{\alpha\beta}\right)\dV+\\
&+\intYi{f}\gamma^{-1}\left(\nabla_y\Psi^\eq\otimes\nabla_y\varpi^\alpha+\nabla_y\varpi^\alpha\otimes\nabla_y\Psi^\eq-\nabla_y\Psi^\eq\cdot\nabla_y\varpi^\alpha\Ib\right)\dV,\\
\Mcal&=\intYi{\Gamma}\wb^P(y)\cdot\nb\dSy=\aYi{\wb^P}{\wb^P}{s}+\dYi{\eta^P}{\eta^P}{s},\\
\Ncal^{\alpha}&=\intYi{\Gamma}\wb^\alpha(y)\cdot\nb\dSy,
\end{split}
\end{equation}
where $\Acalbf=(\mathcal{A}_{ijkl})$ is the fourth-order symmetric positive definite effective tensor of a piezoelectric drained skeleton,  $\Bcalbf=(\Bcal_{ij})$ is the coupling tensor related to the pressure. These effective coefficients are similar to Biot's poroelasticity coefficients as they were derived in work \cite{turjani2018}. However, they are extended by the piezoelectric terms. The coefficients $\Mcal$ and $\Ncal^\alpha$ are new additions related to the evolution of pressure and potentials, respectively.

The coefficient $\Ccalbf^\alpha=(\mathcal{C}^\alpha_{ij})$ is the tensor responsible for the coupling between the electrokinetic and poropiezoelectric system.

The two sets of effective coefficients that are defined by \Eq{eq_ec_el_p1}--\eqref{eq_ec_piezo_disp_1} characterize the macroscopic behavior of the PZPM saturated by an electrolyte solution and occur in the definition of the macroscopic homogenized problem.

\subsubsection{Homogenized model}

The homogenized model describing macroscopic behavior is derived from the two-scale limit problems 
%\Eq{eq_piezo_limit_1}-\Eq{eq_piezo_limit_2}
by letting all the test function components irrelevant to the macroscopic scale vanish. We present the macroscopic model in its nondimensional form:  Find $P^0\in L^2(\Om),\ub^0\in \Hb^1_{\# 0}(\Om)$ and $\Phi_\alpha^0\in H_\#^1(\Om),\alpha=1,2,$ such that
\begin{align}
\begin{split}
\label{eq_macro_dimless}
\int_{\Om}&\Kcalbf (\fb-\nabla_x P^0)\nabla_x q \dV
+\int_{\Om}\Mcal \pdt P^0 q \dV-\int_{\Om}q\Bcalbf : \strx{\pdt\ub^0}+\\
+&\sum\limits_{\beta=1}^{2}\left[\int_{\Om} \Jcalbf^\beta  (\nabla_x\Phi_\beta^\eff+\vec{E})\nabla_xq \dV-\int_{\Om}\Ncal^\beta (\pdt{\Phi_\beta^0}-\pdt\Psi^\ext) q \dV\right]=0\\
\sum\limits_{\beta=1}^{2}& \left[\int_{\Om}({\Qcalbf}^{\alpha\beta}+{\Scalbf}^{\alpha\beta})\pdt{\Phi_\beta^0} \varphi\dV+\int_{\Om}(\hat{\Qcalbf}^{\alpha\beta}+\hat{\Scalbf}^{\alpha\beta})\pdt{\Psi^\ext} \varphi\dV\right]+\\
+&\int_{\Om}\Lcalbf^\alpha (\fb-\nabla_xP^0)\nabla_x\varphi \dV 
+\sum\limits_{\beta=1}^{2} \int_{\Om}\Dcalbf^{\alpha\beta}(\nabla_x\Phi_\beta^0+\vec{E}) \nabla_x\varphi
= 0, 
\end{split}
\end{align}
\begin{equation}
\label{eq_macro_dimless2}
\int_{\Om}\left[\Acalbf\strx{\ub^0}-{\Bcalbf} P^0\right]:\strx{\vb} \dV 
+ \sum\limits_{\beta=1}^{2} \int_{\Om}\Ccalbf^\alpha\Phi_\beta^0:\strx{\vb} \dV = \int_{\Om}\fb\cdot\vb \dV,
\end{equation}
for all $\varphi\in H_\#^1(\Om),q\in L^2(\Om),\vb\in \Hb^1_{\# 0}(\Om)$ with $L$-periodic boundary conditions.

One may recall that for the derivation of the homogenized macroscopic model \Eq{eq_macro_dimless}, we did assume $L$-periodic boundary conditions on the outer boundary $\partial\Om$, see also Remark~\ref{rem:L-per}. This assumption simplified the homogenization process, as it eliminated the necessity to deal with the upscaling of various boundary conditions on $\partial\Om$. {To extend the applicability of the derived model \Eq{eq_macro_dimless}-\Eq{eq_macro_dimless2} to other than periodic problems, we may use the differential form of the problem obtained by per-partes derivation:}{}
%\begin{remark}[Dealing with $L$-periodicity]\label{rem:boundary}
%	By virtue of derivation by parts, we may obtain the differential form of the problem \Eq{eq_macro_dimless}-\Eq{eq_macro_dimless2}. Then it is possible to forgo the assumption of $L$-periodicity and split the boundary $\partial\Om$ into parts, such that for any variable $\vphi$ we may impose the classical Dirichlet type condition on the part of the boundary $\partial_\vphi\Om$ and the Neumann type condition on the part of the boundary $\partial_\vphi^N\Om = \partial\Omega\setminus\partial_\vphi\Om$. 	
%\end{remark}
Find $P^0,\ub^0$ and $\Phi_\alpha^0,\alpha=1,2,$ such that
\begin{align}\begin{split}
\label{eq_macro_piezo_strong}
\Mcal \pdt P^0 -\Bcalbf : \strx{\pdt\ub^0}-\Ncal^\beta (\pdt{\Phi_\beta^0}-\pdt\Psi^\ext)+\nabla_x\cdot \wb^0&=0\\
\sum\limits_{\beta=1}^{2} \left[(\Qcalbf^{\alpha\beta}+{\Scalbf}^{\alpha\beta})\pdt{\Phi_\beta^0} +(\hat{\Qcalbf}^{\alpha\beta}+\hat{\Scalbf}^{\alpha\beta})\pdt{\Psi^\ext} \right]
+\nabla_x\cdot\jb^0&= 0, \\
-\nabla_x\cdot\sigmabf^0 &=\fb,
\end{split}
\end{align}
for  $\alpha=1,2$, where we distinguish the fluid seepage velocity $\wb^0$, the ionic diffusion fluxes $\jb_\alpha^0, \alpha=1,2$, and the porous body stress $\sigmabf^0$, which are defined by
\begin{equation}
\begin{split}
\label{eq_macro_fluxes_piezo}
\wb^0&=\sum_{\beta=1}^{2} \Jcalbf_\beta  (\nabla_x\Phi_\beta^0+\mathbf{E})-\Kcalbf (\nabla_x P^0 -\fb)  \inOm, \\
\sigmabf^0&=\Acalbf e_x(\ub^0)-P^0\hat{\Bcalbf}+\sum_{\beta}^{2}\Ccalbf^\beta\Phi_\beta^0  \inOm,\\
\jb_\alpha^0&=\sum_{\beta=1}^{2} \Dcalbf_{\alpha\beta}\left( \nabla_x\Phi_\beta^0+\mathbf{E} \right)-\Lcalbf_\alpha (\nabla_x P^0-\fb) \inOm \quad \alpha=1,2. 
\end{split}
\end{equation}
%The total velocity $\wb^0$ can be split into three components; pressure driven velocity $\wb_p$ and potential driven velocity $\wb_{\Phi_\alpha},\alpha=1,2$, such that
%\begin{equation}
%\begin{split}
%\label{eq_macro_w_piezo}
%\wb_p&=\Kcal (\nabla_x P^\eff -\fb)  \inOm, \\
%\wb_{\Phi_\alpha}&=\Jcalbf^\alpha (\nabla_x\Phi_\alpha^0+\mathbf{E}),
%\alpha=1,2 \inOm. 
%\end{split}
%\end{equation}

The system \Eq{eq_macro_piezo_strong} needs to be completed by a suitable set of boundary conditions. For any variable $a$, we may impose the Dirichlet type condition on boundary $\partial_a\Om$ and the Neumann type condition on boundary $\partial_a^N\Om = \partial\Omega\setminus\partial_a\Om$ as follows %% . Then, the Neumann type boundary conditions for system \Eq{eq_macro_elkin_diff11}-\Eq{eq_macro_elkin_diff12} read
the Neumann type condition on boundary $\partial_a^N\Om$ read
\begin{align}
%\begin{split}
\nb\cdot\sigmabf^0&=\bar{\gb},\textrm{ on }\partial_u^N\Om,\quad &\ub^0&=\bar{\ub},\textrm{ on }\partial_u\Om,\nonumber\\
\nb\cdot\jb_\alpha^0&=\bar{j}_\alpha,\textrm{ on }\partial_{\Phi_\alpha}^N\Om,\quad &\Phi_\alpha^0&=\bar{\Phi}_\alpha,\textrm{ on }\partial_{\Phi_\alpha}\Om, \alpha=1,2,\label{eq_neumann_piezo}\\
\nb\cdot\wb^0&=\bar{w},\textrm{ on }\partial_P^N\Om,\quad &P^0&=\bar{P}, \textrm{ on }\partial_P\Om.\nonumber
%\end{split}
\end{align}

\subsection{Reconstruction of macroscopic solution at the microscale}\label{sec:rec}
One of the most remarkable advantages of the chosen homogenization method is the ability to reconstruct macroscopic fields on the microscopic scale. After computing the solution to the macroscopic problem \Eq{eq_macro_dimless}-\Eq{eq_macro_dimless2}, in other words, the global (dimensionless) response $\left\{\ub^0, P^0, \Phi_\beta^0\right\}$, it is possible to reconstruct the associated microscopic quantities. This process is also called down-scaling, which is in contrast to the up-scaling process that leads to the macroscopic model, and we will describe it in the following text.

\par We consider a given finite scale $\veps_0>0$ corresponding to the real size of the microstructure, which will enable us to form the following remark.

\begin{remark}[Partition of cell $Y$]
For a given $\veps_0>0$, we introduce the rescaled cell $Y^{\veps_0}=\veps_0Y$. Further, we introduce its local copies $Y^{K,\veps_0}$ labeled by index $K$. We consider the domain $\Om$ generated as a lattice of non-overlapping copies $Y^{K,\veps_0}$ with $\{x^K\}_K$ being the set of their centers (in macroscopic coordinates). Thus, the partitioning $Y_{\veps_0(\Om)}$ of $\Om$ is defined as

\begin{displaymath}
Y_{\veps_0}(\Om)=\bigcup_K \bar{Y}^{K,\veps_0},\quad Y^{K,\veps_0}=Y^{\veps_0}+x^K,
\end{displaymath}

For any global position $x\in Y^{K,\veps_0}$, the local "microscopic" coordinates 
\begin{equation}\label{eq_rem_par}
y=(x-x^K)/\veps_0,
\end{equation}
are introduced, such that $y\in Y$, \cite{rohan2021multiscale}.
\end{remark}

For a defined partitioning $Y_{\veps_0}(\Om)$ of the domain $\Om$, we will recover the macroscopic fields in $\Om$
using a nonlocal reconstruction procedure, which is based on the so-called folding approach.

The folding method of reconstruction lies in using the so-called folding operator, which can be considered the inverse operation to the periodic unfolding. This approach enables us to "fold" the macroscopic responses $(\ub^0(x), P^0(x), \Phi_\beta^0(x))$ using the scale separation splits related to the periodic lattice.

The folding operator $\Fcal^{\veps_0}$ folds any macroscopic field $f^0(x)$ onto the partitioning $Y_{\veps_0}(\Om)$ so that the local microscopic fields are given by the so-called folding mapping \cite{rohan2017darcy}, such that
\begin{equation}
\Fcal^{\veps_0}(\hat{x}):(\ub^0, P^0, \Phi_\beta^0)\rightarrow(\ub^\rec,P^\rec,\wb^\rec,\Phi_\beta^\rec)(y),\qquad y\in Y_{\veps_0}(\Om).
\end{equation}
The folding operator $\Fcal^{\veps_0}$ combines the corrector base functions defined in $Y$ with the interpolated macroscopic responses transformed to the local zoomed RVE $Y_{\veps_0}(x)$ by the interpolation operator  $\Qcal^{\veps_0}$. The $\Qcal^{\veps_0}$ is defined by an $Q_1$  interpolation scheme in the sense of the FEM approximation over partitioning $Y_{\veps_0}(\Om)$ of $\Om$.  

Note that the folding approach requires smoothness of gradients $\nablax f^0(x)$. The smoothness can be achieved by introducing a projection $\Pi_\veps\left[g\right]$ of a given function $g(x)$ into the space of piecewise polynomials defined over the partition $Y_{\veps_0}(\Om)$.

To apply the folding approach, we first introduce the smoothed gradients $\Pi_{\veps_0}\left[\partial_k^x P^0\right]$, $\Pi_{\veps_0}\left[e_{ij}^x \ub^0\right]$ and $\Pi_{\veps_0}\left[\partial_k^x\Phi^0_\alpha\right]$ for $\alpha=1,2$.  By introducing them into the split forms \Eq{eq_split_piezo1}--\eqref{eq_split_piezo2}, for any $x\in\Om$ and $y\in Y$ assigned to x by virtue of \Eq{eq_rem_par}, the two-scale microscopic fields can be reconstructed using the following formulae:

\begin{equation}
\begin{split}\label{eq_rec_elkin3}
{\Phi}_\alpha^\mic(x,y,t)&= \theta^{0,k}(y)\left(f_k-\Pi_{\veps_0}\left[\partial_k^x P^0\right] \right)(x)+\sum\limits_{\beta=1}^{2}\theta^{\beta,k}(y)\left(\Pi_{\veps_0}\left[\partial_k^x \Phi^0_\beta\right] +\mathrm{E}_k\right)(x),\\
P^\mic(x,y,t)&=\pi^{0,k}(y)\left(f_k-\Pi_{\veps_0}\left[\partial_k^x P^0\right] \right)(x) +\sum\limits_{\beta=1}^{2}\pi^{\beta,k}(y)\left(\Pi_{\veps_0}\left[\partial_k^x \Phi^0_\beta\right] +\mathrm{E}_k\right)(x),
\\
\ub^\mic(x,y,t)&=  \widetilde{\wb}^{ij}(y)\Pi_{\veps_0}\left[e_{ij}^x \ub^0\right]-P^0(x)\wb^P(y)+\sum\limits_{\beta=1}^{2}\wb^\beta(y)z_\beta\Phi_\beta^0(x),
\end{split}
\end{equation}

%\begin{itemize}
%\item Microscopic fields $\wb^\rec, P^{\textrm{mic},\veps}, \Phi_\beta^{\textrm{mic},\veps}$ related to the decoupled electrokinetic system:
%\begin{align}
%P^\rec=P^0
%\end{align}
%\item Microscopic displacement field $\ub^{\textrm{mic},\veps},\Psi_s^{\rec,\veps},\Psi_f^{\rec,\veps}$(in case of deformable porous medium): 
%\end{itemize}
where by $\tilde{\sqcup}$ we denote the extensions of quantities from $Y_s$ to entire $Y$. Now, the fields that are relevant to the microscopic heterogeneity are evaluated using
\begin{align}\label{eq_rec_piezo}
\begin{split}
P^\rec(x,t)&=P^0(x,t)+\veps_0P^\mic(x,y,t)\\
\ub^\rec(x,t)&=\ub^0(x,t)+\veps_0\ub^\mic(x,y,t)\\
\Phi_\alpha^\rec(x,t)&=\Phi_\alpha^0(x,t)+\veps_0{\Phi}_\alpha^\mic(x,y,t), \alpha=1,2.
\end{split}
\end{align}
In addition, by the virtue of the \Eq{eq_split_piezo1}$_1$,\eqref{eq_split_piezo1}$_1$ and \eqref{eq_split_piezo1}$_2$, we obtain a reconstruction of velocity and potential fields
\begin{equation}\label{eq_rec_wpot}
\begin{split}
\wb^\rec(x,y,t)&=  \sum\limits_{\beta=1}^{2}\ombf^{\beta,k}(y)\left(\Pi_{\veps_0}\left[\partial_k^x \Phi^0_\beta\right] +\mathrm{E}_k\right)(x)
+\ombf^{0,k}(y)\left(f_k-\Pi_{\veps_0}\left[\partial_k^x P^0\right] \right)(x),\\
\Psi_f^\rec(x,y,t)&=\sum\limits_{\beta=1}^{2}\varpi^\beta(y)\Phi_\beta^0(x) ,\\
\Psi_s^\rec(x,y,t)&=  \eta^{ij}(y)\Pi_{\veps_0}\left[e_{ij}^x \ub^0\right]-P^0(x)\eta^P(y)+\sum\limits_{\beta=1}^{2}\eta^\beta(y)z_\beta\Phi_\beta^0(x) ,
\end{split}
\end{equation}
The reconstruction of the macroscopic solution on a microscopic scale is useful for determining the influence of macroscopic fluxes. 
%po tomto Zkontrolovano grammarly

\section{Numerical modeling}\label{sec_numeric}
This section aims to illustrate the behavior of the derived two-scale homogenized model described in the previous text. For this purpose, we present a numerical simulation of the ionic transport through a porous medium occupying a simple-shaped macroscopic domain with a periodic microstructure.

The two-scale homogenized model was implemented in \emph{SfePy}, a software for solving problems with coupled partial differential equations (PDEs) in weak forms through the finite element method (FEM) for 2D and 3D problems, \cite{sfepy}. \emph{SfePy} is based on the \textit{Python} programming language and its packages \emph{NumPy} and \emph{SciPy}, \cite{scipy}.

\subsection{Semi-discretized macroscopic problem}
In what follows, we introduce the semi-discretized form of macroscopic problem used for finite element (FE) modeling. The computational time $T$ is discretized into the $n$ equidistant time levels by the fixed time step $\Delta t$. The column vectors $\Pbm^n,\Phibf^n_{\alpha}, \alpha=1,2,$ and $\ubm^n$ represent all degrees of freedom of FE mesh nodes associated with partitioning of the macroscopic domain $\Om$ at $n-$th time level and by $\qbm,\vbm$ and $\phibf$ we refer to column vectors of their respective test functions. We present the following approximations of the terms involved in the macroscopic problem \Eq{eq_macro_dimless}-\Eq{eq_macro_dimless2}:

\begin{equation}
\begin{array}{lcllcl}
\qbm^T\KK\Pbm^n & \approx & \int_{\Om}\Kcalbf \nabla_x P^n\nabla_xq dV, &\quad \qbm^T\JJ^\alpha\Phibf^n_\alpha& \approx &\int_{\Om} \Jcalbf^\alpha \nabla_x\Phi^n_\alpha\nabla_xq \dV, \\
\phibf^T\LL^\alpha\Pbm^n& \approx &\int_{\Om}\Lcalbf^\alpha \nabla_x P^n\nabla_x\testphi \dV,  &\quad \phibf^T\DD^{\alpha\beta}\Phibf^n_\beta& \approx & \int_{\Om}\Dcalbf^{\alpha\beta} \nabla_x\Phi^n_\beta\nabla_x\testphi \dV,\\
\qbm^T\MM^\alpha\Pbm^n & \approx & \int_{\Om}\Mcalbf^\alpha P^n\nabla_x \testp\dV, &\quad \vbm^T\mathbb{B}\Pbm^n& \approx & \int_{\Om}\Bcalbf P^n:e_x(\mathbf{\vb}) \dV,\\
\vbm^T\CC^\alpha\Phibf^n_\alpha & \approx & \int_{\Om}\Ccalbf^\alpha\Phi^n_\alpha:e_x(\mathbf{\vb}) \dV, &\quad \vbm^T\mathbb{A}\ubm^n& \approx & \int_{\Om}\Acalbf e_x(\ub^n):e_x(\mathbf{\vb})\dV,\\
\phibf^T\QQ^{\alpha\beta}\Phibf_\beta^n & \approx & \int_{\Om}(\Qcalbf^{\alpha\beta}+\Scalbf^{\alpha\beta})\Phi_{\beta}^n \testphi,&\quad \qbm^T\NN^\alpha\Phibf_\alpha^n & \approx & \int_{\Om}\Ncalbf^\alpha \Phi_\alpha^n\nabla_x q,\nonumber
\end{array}
\end{equation}
with time derivatives approximated by forward finite differences. Then, the  discretized problem for $\fb=\mathbf{0}, \vec{E}=\mathbf{0}$ reads

\footnotesize
\begin{equation}\label{eq_disc_piezo}
\left[\begin{array}{cccc}
\Delta t\KK+\MM & -\Delta t\JJ^1+\NN & -\Delta t\JJ^2+\NN & -\mathbb{B}\\
\Delta t\LL^1 & {\QQ}^{11}-\Delta t\DD^{11} & {\QQ}^{12}-\Delta t\DD^{12} & \mathbf{0}\\
\Delta t\LL^2 & {\QQ}^{21}-\Delta t\DD^{21} &{\QQ}^{22}-\Delta t\DD^{22}& \mathbf{0}\\
\mathbb{B} &  -\CC^1 & -\CC^2 & \mathbb{A}
\end{array}\right]
\left[\begin{array}{c}
\Pbm^n\\
\Phibf_{1}^n\\
\Phibf_{2}^n\\
\ubm^n
\end{array}\right] =
\left[\begin{array}{c}
\MM \Pbm^{n-1}\\
\left(\sum\limits_{\beta=1}^2\QQ^{1\beta}+\NN^1\right) \Phibf_{1}^{n-1}\\
\left(\sum\limits_{\beta=1}^2{\QQ}^{2\beta}+\NN^2\right)\Phibf_{2}^{n-1}\\
\mathbb{B} \ubm^{n-1}
\end{array}\right].
\end{equation}
\normalsize
The initial values of $\Pbm^0,\Phibf^0_{\alpha}, \alpha=1,2,$ and $\ubm^0$ are taken from a solution of steady state problem.

\subsection{Numerical implementation}
\par Presented computational examples do not consider volume forces and also disregard effects of an external electric field; thus, we put $\fb=\mathbf{0}$ and $\vec{E}=\mathbf{0}$. All used values of the quantities and parameters are in Tab.~\ref{tab_constants}. All the computations are performed for the given pore size $\ell_\mic$.
The numerical implementation of the presented homogenized model consists of a few subproblems on microscopic and macroscopic geometry. Thus, numerical simulation consists of the following steps:
\begin{enumerate}
	\item\label{item1} Solve the potential distribution in equilibrium $\Psi^\eq$ on cell $Y$ as a solution of \Eq{eq_adim_PB0}.
	\item Compute concentrations $c_\beta^\eq,\beta=1,2$ from \Eq{eq_equilibrium_conc}.
	\item Compute corrector functions $(\ombf^{P,k},\pi^{P,k},\theta_\beta^{P,k})$, and $(\ombf^{\alpha,k},\pi^{\alpha,k},\theta_\beta^{\alpha,k})$ where $k=1,\dots,n$, $\alpha,\beta=1,2$, related to electrokinetic system as a solution of local problems \Eq{eq_lp_el0}.
	\item Compute effective coefficients relevant to the electrokinetic system from \Eq{eq_ec_el_p1} and \Eq{eq_ec_el_p2}.
	\item Compute corrector functions  $\varpi^{\alpha}, \alpha=1,2$ related to the potential perturbation as a solution of local problems \Eq{eq_lp_pot}.
	\item Compute corrector functions  $(\wb^{ij}, \wb^P, \wb^\alpha)$, $\alpha=1,2$, $i,j=1,\dots,n$, related to the displacement perturbation as a solution of local problems \Eq{eq_lp_disp_p1}-\Eq{eq_lp_disp_p3}.
	\item\label{item7} Compute effective coefficients relevant to the Biot poroelasticity from \Eq{eq_ec_piezo_disp_1}.
	\item Compute solution to the macroscopic homogenized system of equations \Eq{eq_macro_dimless}-\Eq{eq_macro_dimless2}.
\end{enumerate}
The steps \ref{item1} through \ref{item7} are in more detail explored in our previous work, \cite{turjani2018}.

\begin{table}[t]\centering
	\begin{tabular}{llll}
		\hline\noalign{\smallskip}
		Symbol	& Quantity &Value&Unit\\
		\noalign{\smallskip}\hline\noalign{\smallskip}
		$\textrm{e}$&Electron charge&$1.6\times 10^{-19}$&	C	       \\
		$\textrm{k}_B$&Boltzmann constant&$1.38\times 10^{-23}$&	J/K\\  
		$\textrm{T}$&Absolute temperature&$298$&	K\\	       
		$\kappa$&Dielectric constant&$6.93\times 10^{-10}$& C/(mV)	\\	        
		$\eta_f$&Dynamic viscosity of fluid&$1\times 10^{-3}$&kg/(ms)\\
		$D_1$&Diffusivity of 1st ionic species&$13.33\times 10^{-10}$&m$^2$/s\\
		$D_2$&Diffusivity of 2nd ionic species&$20.32\times 10^{-10}$&m$^2$/s\\
		$\ell_\mic$&Characteristic pore size&$1.0\times 10^{-7}$&m\\
		$L_c$&Characteristic macrostructure size&$1.0\times 10^{-5}$&m\\
		$c_c$&Characteristic concentration&$(6.02\times 10^{24},6.02\times 10^{26})$& particles/m$^3$\\
		$\Sigma_c$&Surface charge density&$-0.129$&C/m$^2$\\
		$\lambda_D$&Debye length $\sqrt{\kappa\textrm{k}_B}\textrm{T}/(\textrm{e}^2 c_c)$&$0.042$&nm\\
		$g_c$&Characteristic piezoelectric coupling &$18.4559\times 10^{-4}\ ^\dag$&	C/m$^2$      \\
		\noalign{\smallskip}\hline
	\end{tabular}
	\caption{The electrochemical parameters characterizing the two-component electrolyte solution. Source: \cite{allaire2015ion}}
	\label{tab_constants}
\end{table}
\begin{table}[!t]\centering
	\begin{tabular}{llll}
		\hline
		Symbol	& Quantity &Value&Unit\\
		\hline
		$g_{14}$&Piezoelectric coupling &$18.4559\times 10^{-4}\ ^\dag$&	C/m$^2$      \\
		$d_{11}$&Dielectric permitivity&$88.54\times 10^{-12}$&	C$^2$/N$\cdot$ m$^2$\\
		$d_{33}$&Dielectric permitivity&$106.25\times 10^{-12}$& C$^2$/N$\cdot$ m$^2$\\ 
		$E_{1}$& Young modulus&13.90$\times 10^{9}$& Pa\\
		$E_{3}$& Young modulus&22.21$\times 10^{9}$& Pa\\
		$G_{14}$& Shear modulus&3.18$\times 10^{9}$& Pa\\
		$\nu_{12}$& Poisson's ratio&0.2537& -\\
		$\nu_{31}$& Poisson's ratio&0.3002& -\\
		$g_c$&Characteristic piezoelectric coupling &$18.4559\times 10^{-4}\ ^\dag$&	C/m$^2$ \\
		$ E_c$&Characteristic elastic modulus&$22.21\times 10^{9}$&Pa\\ 
		\hline
	\end{tabular}
	\caption{The mechanical and piezoelectric parameters characterizing the collagen-hydroxyapatite matrix of cortical bone tissue. Sources: \cite{silva2001collagen},\cite{fotiadis1999wave} and \cite{predoi2008human}
		\\ \footnotesize{$\dag$ Computed from the piezoelectric coefficient of strain-charge form.}	
	}
	\label{tab:ident_piezo}
\end{table}

\subsection{Comparison of poroelastic and poropiezoelectric model}
This section explores and illustrates the properties of the homogenized two-scale models described in the preceding sections. The presented model of electrolyte flux through the poropiezoelectric porous medium (PZPM) aims to extend the model considering only linear elastic porous medium (EPM) as was published in \cite{turjani2018}. Below, we present some selected numerical simulations to compare the properties of the PZPM and EPM homogenized two-scale models. 

It can be shown that the derived macroscopic problem is formally given by \Eq{eq_macro_piezo_strong} for both PEPM and EPM. The difference lies at the microscopic scale, in the expressions of cell problems and, subsequently, the expressions of effective coefficients. The expressions of electrokinetic coefficients \Eq{eq_ec_el_p1} describing PZPM remain the same for EPM, but the Biot-like coefficients \Eq{eq_ec_piezo_disp_1} differ. Thus, we will use the expressions of Biot-like effective coefficients in the form given in \cite{turjani2018}, \ie effective coefficients characterizing the behavior of linear elastic material, to describe the solid matrix of EPM.  

\begin{figure}[h]
	\centering	
	\begin{subfigure}{0.49\linewidth}
		\includegraphics[width=\linewidth]{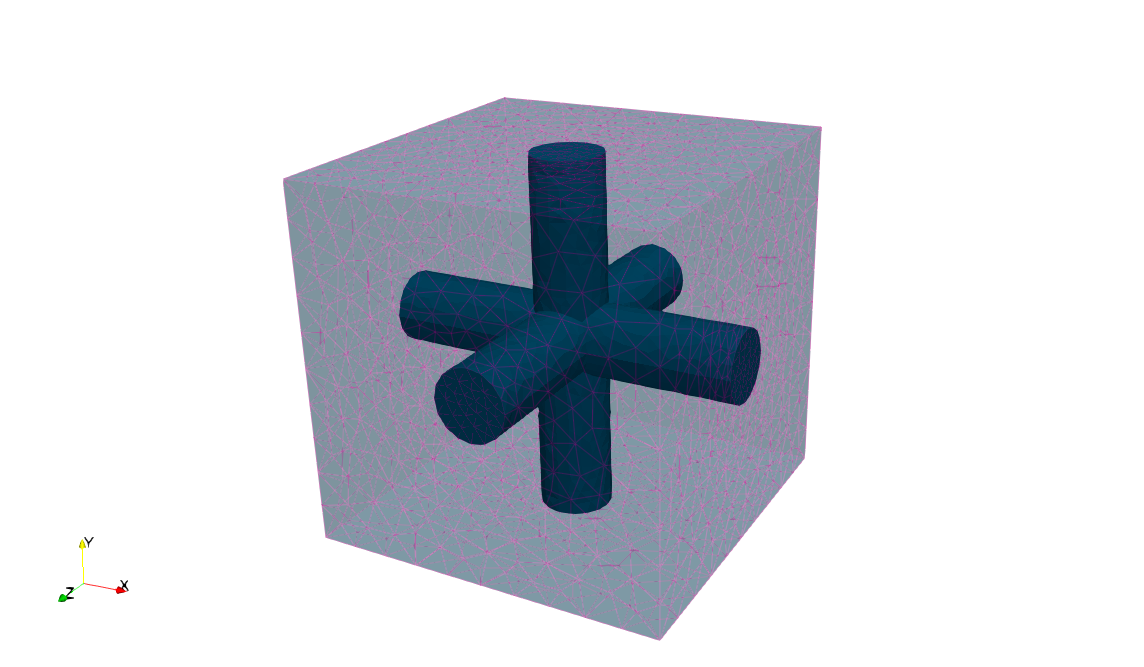}
		\caption{Cupic periodic RVE $Y$}
		\label{RPC}
	\end{subfigure}
	\begin{subfigure}{0.45\linewidth}
		\includegraphics[width=\linewidth]{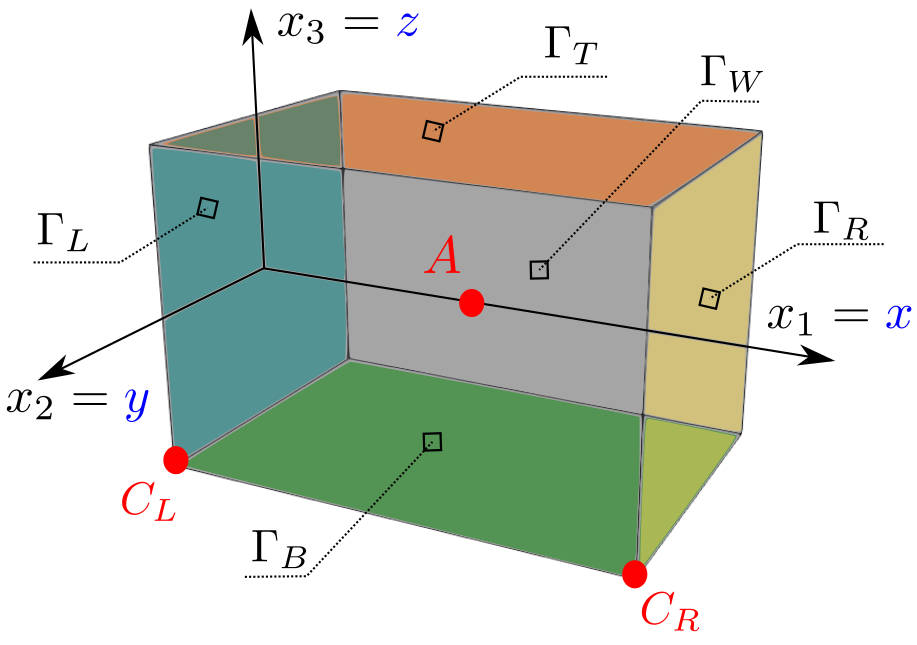}
		\caption{Macroscopic specimen with boundaries $\Gamma_A$}
		\label{fig:specimen}
	\end{subfigure}
	\caption{Geometry and mesh representation of microscopic structure and macroscopic specimen.}
	\label{fig:geo}
\end{figure}
\subsubsection{Description of microscopic level.}
At the microscopic level, the microstructure is represented by cubic periodic RVE $Y$ containing a spherical inclusion laying on the cross-section of three channels in the directions of the three coordinate axes. Fig.~\ref{RPC} depicts the microstructure geometry. The microstructure size is given by {$\veps:=\veps_0=0.05$}, which
determines the scaling of the piezoelectric coupling $\gb$ and dielectric tensor $\db$.

The fluid saturating the channels is characterized by electrochemical constants presented in Tab.~\ref{tab_constants}.

The piezoelectric matrix usually exhibits some degree of anisotropy. We use collagen-hydroxyapatite to constitute the solid matrix material in our numerical tests. Thus, the mechanical and piezoelectric parameters from Tab.~\ref{tab:ident_piezo} characterize the solid matrix of PZPM. 

However, we use only the mechanical parameters in Tab.~\ref{tab_constants} to comprise the stiffness tensor of EPM. 

\subsubsection{The boundary value problem}
The weak formulation of the macroscopic problem \Eq{eq_macro_dimless} was presented for a simple set of boundary conditions (BCs). We need to prescribe some specific BCs according to the boundary $\partial\Omega$ decomposition into segments $\Gamma_A$ for the particular numerical examples reported below. We refer to these segments using an intuitive notation, e.g. $\Gamma_L$ refers to the "left" face, $\Gamma_T$ to the "top" face, etc., see Fig.~\ref{fig:specimen}. To simplify the description of results, we use the notation $(x,y,z)$ for macroscopic coordinates instead of $(x_1,x_2,x_3)$ to denote the components of macroscopic tensors, as illustrated by Fig.~\ref{fig:specimen}.

To easily compare both models, we propose a simple BVP that describes the following situation. The specimen is fixed to prevent displacement $u_1$ and $u_2$ at one corner point $C_L$ and $u_2$ at corner point $C_R$ to prevent rotation. Additionally, the displacement is restrained in $x_3$-direction on $\Gamma_B$ so that $u_3=0$. At the top of the specimen, \ie at $\Gamma_T$, the cyclic compression is applied, which is realized through the traction on $\Gamma_T$ given as a function
\begin{equation}\label{eq_cyc_compress}
\bar{h}(t)=\bar{h}_a(\cos(\bar{h}_ft)-1),\quad t\in<0,T>,
\end{equation}
with amplitude $\bar{h}_a=0.1/\sigma_c$ and frequency $\bar{h}_f=0.25$. Additionally, the walls are considered non-permeable. Thus, the BVP is defined by \Eq{eq_macro_piezo_strong} and by the BCs of Neumann and Dirichlet types that are listed in Tab.~\ref{tab:bc}. Initial conditions are taken from the steady-state solution (\ie for $t=0$ and neglecting all time derivatives) of the BVP.

\begin{table}[h]\centering
	\begin{tabular}{llll}
		\hline
		Boundary part & $\mathbf{u}^0$ & $\mathbf{\Phi}^\alpha$ & P\\
		\hline
		$\Gamma_T$& $\nb\cdot\sigmabf_s=\bar{h}(t)$&$\nb\cdot\jb_\alpha=0$&$\nb\cdot\wb=0$\\
		$\Gamma_B$& $u_3=0$&$\nb\cdot\jb_\alpha=0$&$\quad\nb\cdot\wb=0$\\
		$\Gamma_W$& $\nb\cdot\sigmabf_s=0$&$\nb\cdot\jb_\alpha=0$&$\nb\cdot\wb=0$\\
		$\Gamma_L$& $\nb\cdot\sigmabf_s=0$&$\nb\cdot\jb_\alpha=0$&$ P=\bar{P}$\\
		$\Gamma_R$& $\nb\cdot\sigmabf_s=0$&$\nb\cdot\jb_\alpha=0$&$\nb\cdot\wb=\bar{w}$\\
		$C_L$& $u_1=0$ and $u_2=0$& &\\
		$C_R$& $u_2=0$& &\\
		\hline
	\end{tabular}
	\caption{The Dirichlet and Neumann conditions for the particular
		fields involved in the macroscopic model. The conditions are given for all $t\in ]0,T[$, where $\sigmabf_s,\wb$ and $\jb_\alpha, \alpha=1,2$, are given by \Eq{eq_macro_fluxes_piezo} and $\bar{h}(t)$ is given by \Eq{eq_cyc_compress}}	
	\label{tab:bc}
\end{table}

Below, we report on numerical solutions of a boundary value problem (BVP) for both the EPM and PZPM models.

\begin{remark}[Dimensionalized quantities]\label{rem:dimension}
	There are benefits to presenting the homogenized model in its non-dimensional form, especially in the context of numerical simulations. In order to interpret the results of such simulation, it may be more beneficial to return computed quantities to their physical dimensions. That may be achieved easily by recalling the adimensional choices made in \sref{sec:dimless}. That gives us the following macroscopic dimensionalized quantities denoted by $\sqcup^\eff$
	\begin{displaymath}
	P^\eff=p_cP^0,\quad \Phi_\alpha^\eff=\Psi_c\Phi_\alpha^0,\quad \ub^\eff=u_c\ub^0,\quad \wb^\eff=v_c\wb^0.
	\end{displaymath}
\end{remark}

\begin{figure}[h]		\centering
	\begin{subfigure}[b]{0.49\linewidth}
		\centering
		\includegraphics[width=0.99\linewidth]{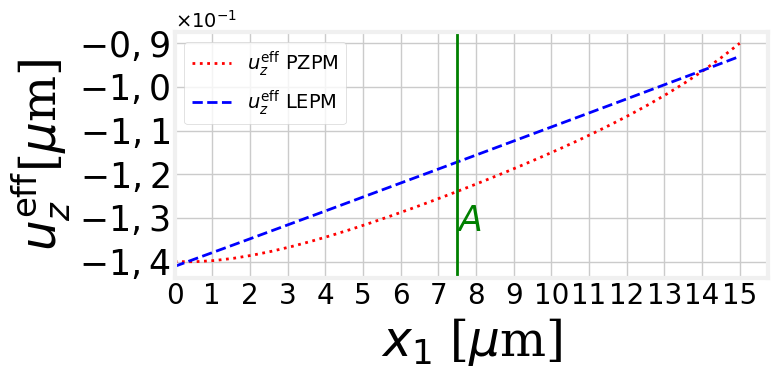}
		%		\caption{}
		%		\label{fig:puI}
	\end{subfigure}
	\begin{subfigure}[b]{0.49\linewidth}
		\centering
		\includegraphics[width=0.99\linewidth]{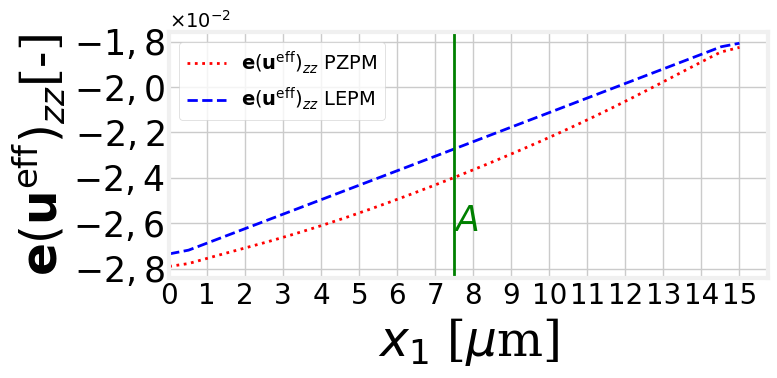}
		%		\caption{}
		%		\label{fig:puI}
	\end{subfigure}
	\begin{subfigure}[b]{0.49\linewidth}
		\centering
		\includegraphics[width=0.99\linewidth]{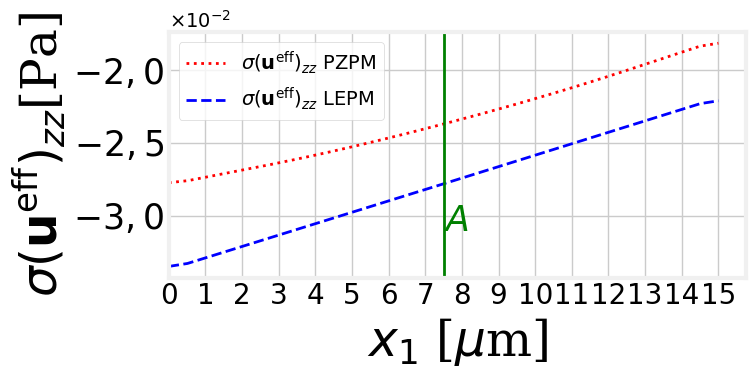}
		%		\caption{}
		%		\label{fig:puI}
	\end{subfigure}
	\begin{subfigure}[b]{0.49\linewidth}	\centering
		\includegraphics[width= 0.99\linewidth]{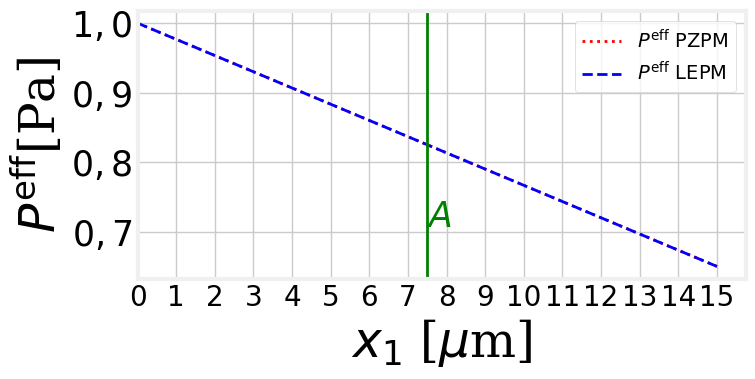}
		%		\caption{}
		%		\label{fig:phiI}
	\end{subfigure}
	\begin{subfigure}[b]{0.49\linewidth}
		\centering
		\includegraphics[width=0.99\linewidth]{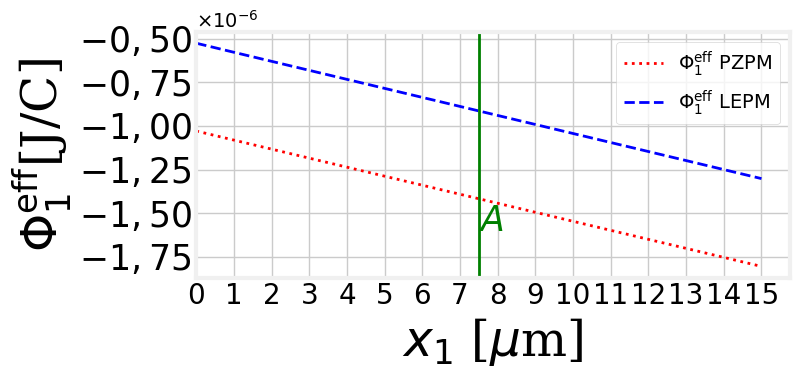}
		%		\caption{}
		%		\label{fig:puII}
	\end{subfigure}
	\begin{subfigure}[b]{0.49\linewidth}	\centering
		\includegraphics[width=0.99\linewidth]{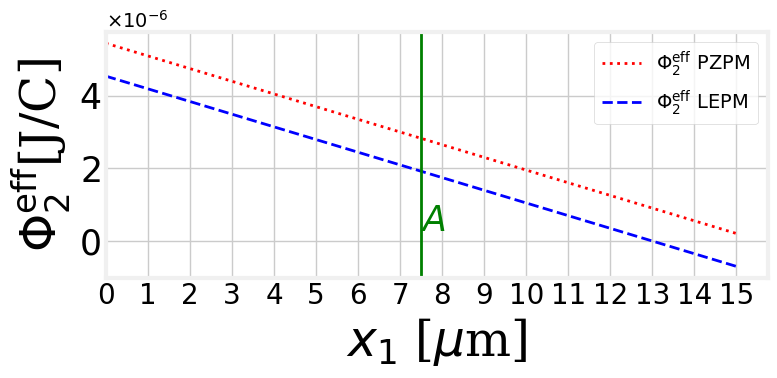}
		%		\caption{}
		%		\label{fig:phiII}
	\end{subfigure}
	\caption{{Comparison of PZPM and EPM models: Steady-state distribution of macroscopic fields $(p^\eff,\ub^\eff,\Phi^\eff_\alpha),\alpha=1,2,$ at $t=0$ along the $x_1$-axis. The  pozition of point A within the macroscopic specimen is denoted by green vertical line.}}
	\label{fig:graphs_x}
\end{figure}
\begin{figure}[h]		\centering
	\begin{subfigure}[b]{0.49\linewidth}
		\centering
		\includegraphics[width=0.99\linewidth]{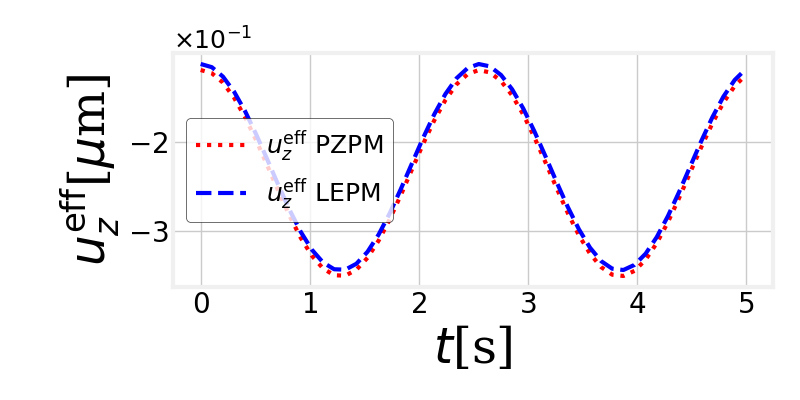}
		%		\caption{}
		%		\label{fig:puI}
	\end{subfigure}
	\begin{subfigure}[b]{0.49\linewidth}
		\centering
		\includegraphics[width=0.99\linewidth]{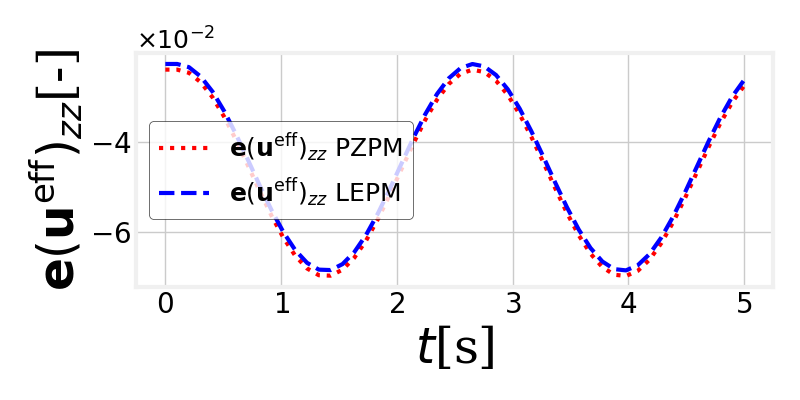}
		%		\caption{}
		%		\label{fig:puI}
	\end{subfigure}
	\begin{subfigure}[b]{0.49\linewidth}
		\centering
		\includegraphics[width=0.99\linewidth]{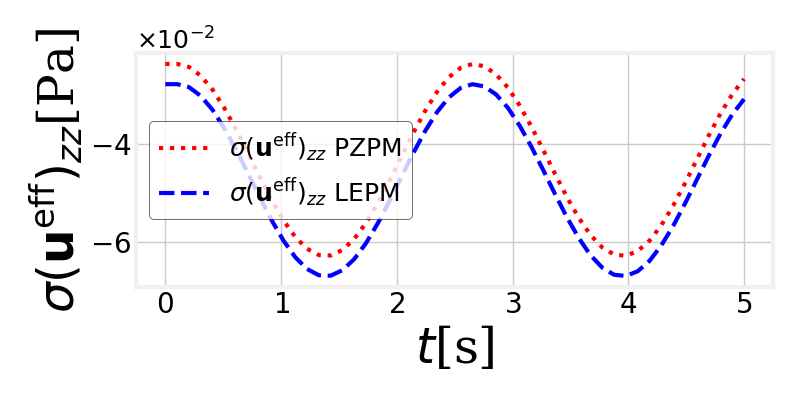}
		%		\caption{}
		%		\label{fig:puI}
	\end{subfigure}
	\begin{subfigure}[b]{0.49\linewidth}	\centering
		\includegraphics[width= 0.99\linewidth]{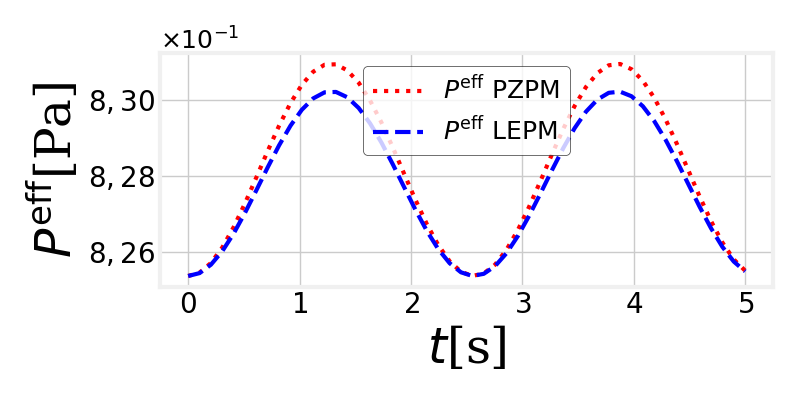}
		%		\caption{}
		%		\label{fig:phiI}
	\end{subfigure}
	\begin{subfigure}[b]{0.49\linewidth}
		\centering
		\includegraphics[width=0.99\linewidth]{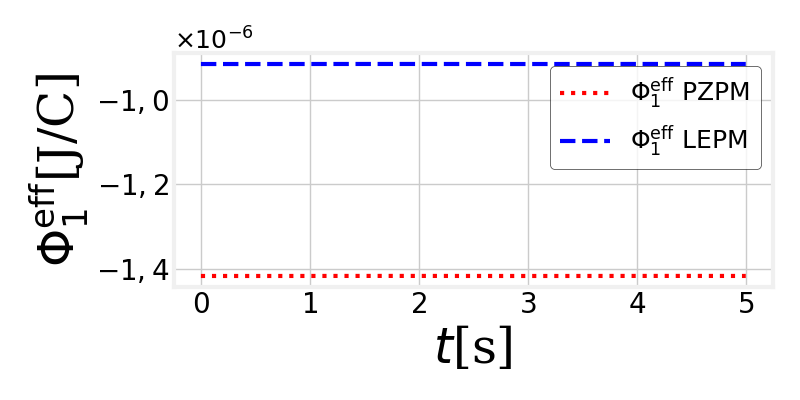}
		%		\caption{}
		%		\label{fig:puII}
	\end{subfigure}
	\begin{subfigure}[b]{0.49\linewidth}	\centering
		\includegraphics[width=0.99\linewidth]{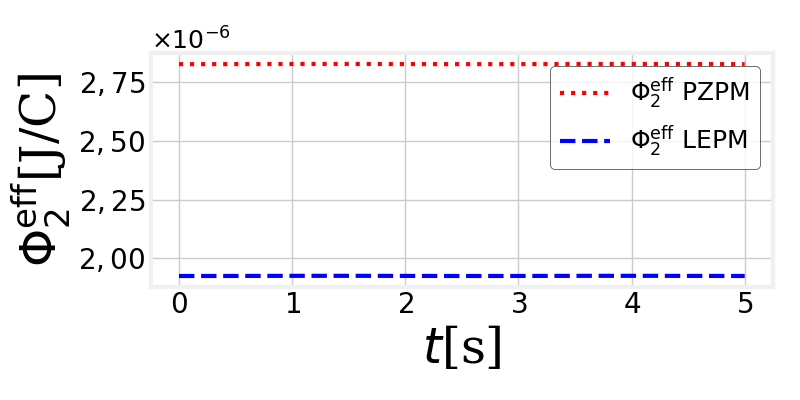}
		%		\caption{}
		%		\label{fig:phiII}
	\end{subfigure}
	\caption{{Comparison of PZPM and EPM models: Evolution of macroscopic fields $(p^\eff,\ub^\eff,\Phi^\eff_\alpha),\alpha=1,2,$ at point $A$, $t\in[0,5]$s.}}
	\label{fig:graphs_time}
\end{figure}

\begin{figure}[p]		\centering
	\begin{center}
		\begin{tabular}{
				m{0.03\linewidth}|
				>{\centering\arraybackslash}m{0.45\linewidth}|
				>{\centering\arraybackslash}m{0.45\linewidth}}
			& PZPM &EPM \\
			\hline
			\rotatebox[origin=c]{90}{$u^\eff_z$}
			&		\includegraphics[width=0.99\linewidth]{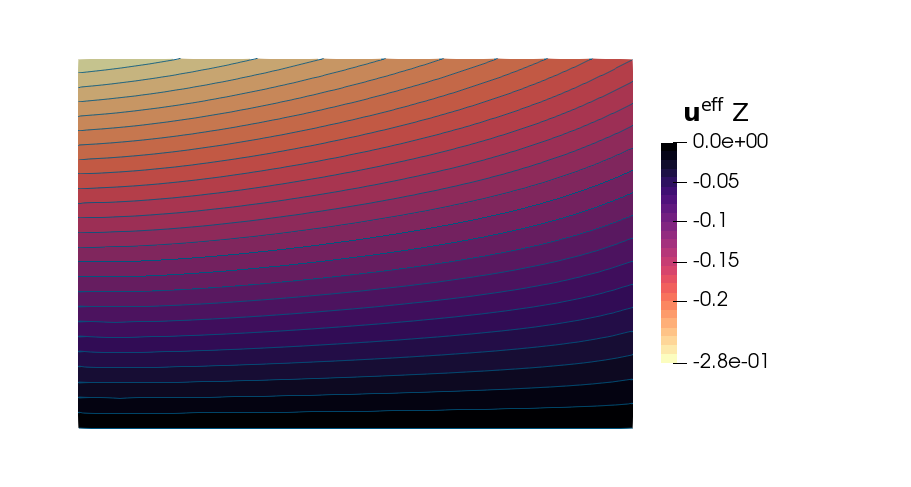}	&\includegraphics[width=0.99\linewidth]{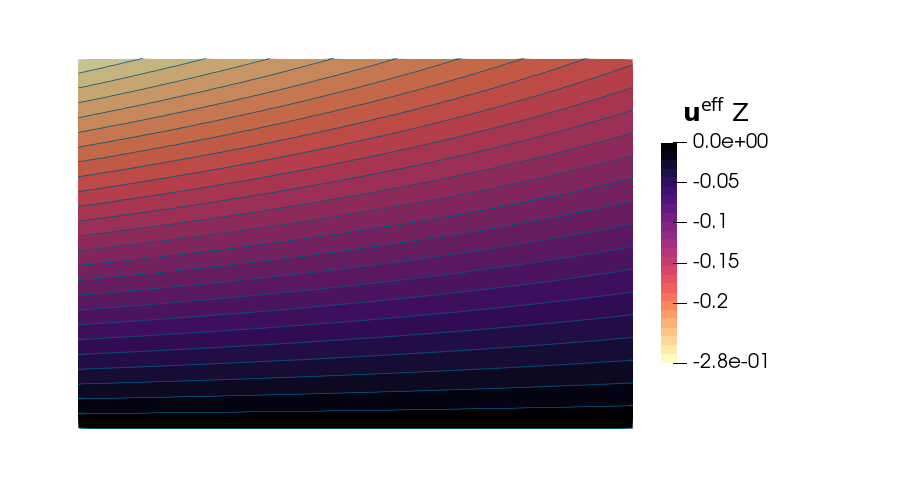}\\
			\hline
			\rotatebox[origin=c]{90}{$e(\ub^\eff)_{xx}$}
			&\includegraphics[width=0.99\linewidth]{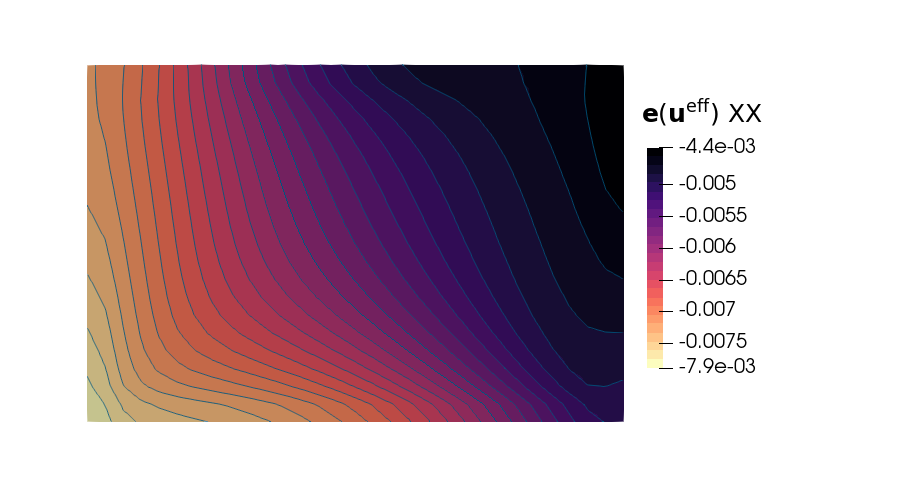}	&\includegraphics[width=0.99\linewidth]{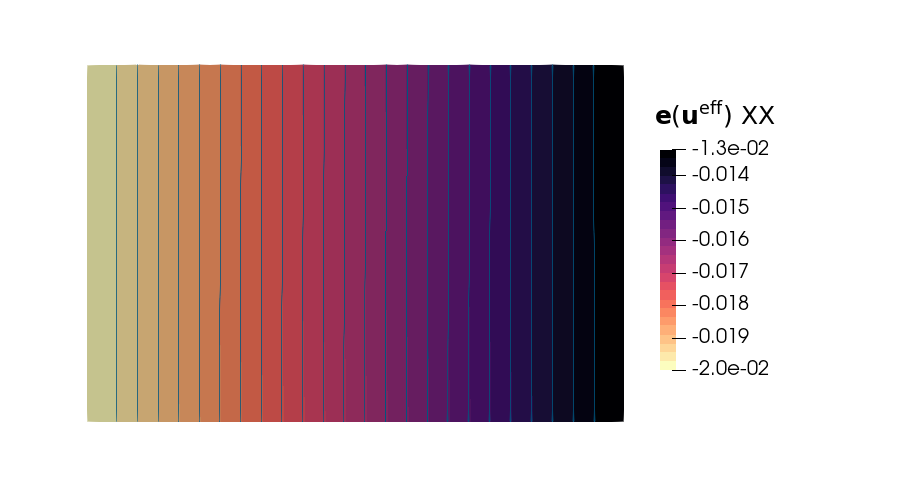}\\
			\hline
			\rotatebox[origin=c]{90}{$e(\ub^\eff)_{yy}$}
			&\includegraphics[width=0.99\linewidth]{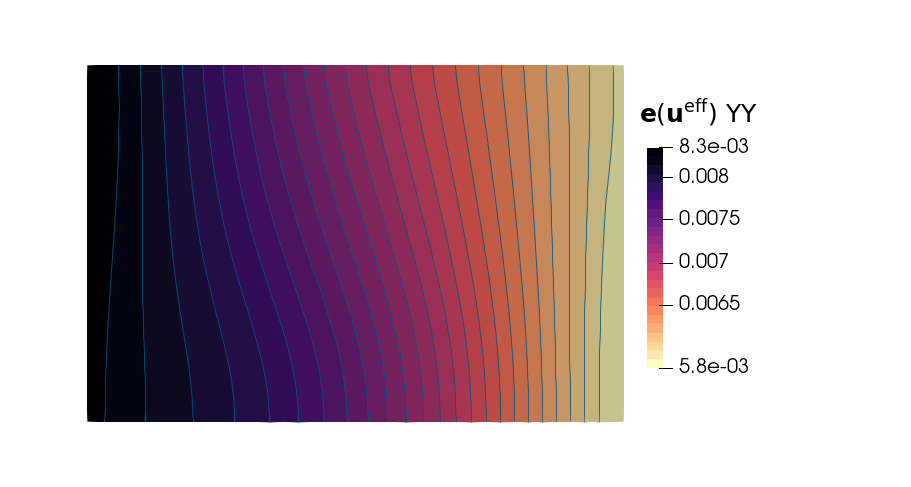}	&\includegraphics[width=0.99\linewidth]{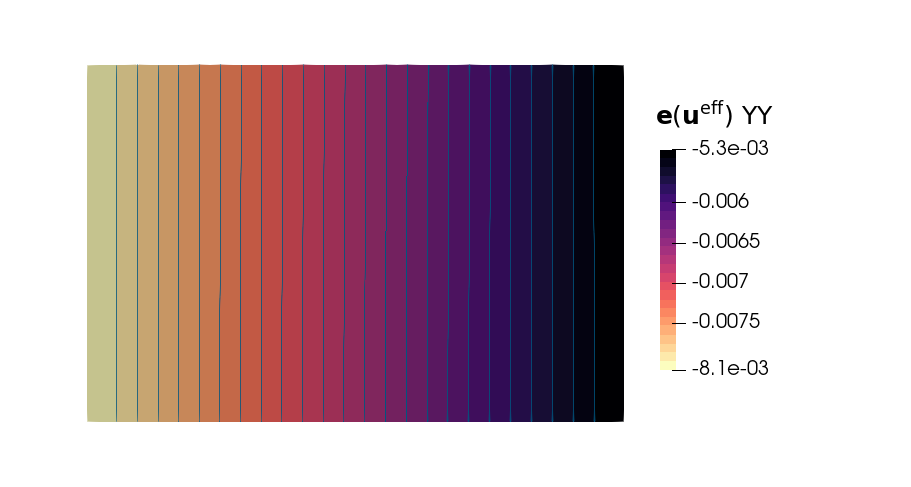}\\
			\hline
			\rotatebox[origin=c]{90}{$e(\ub^\eff)_{zz}$}
			&\includegraphics[width=0.99\linewidth]{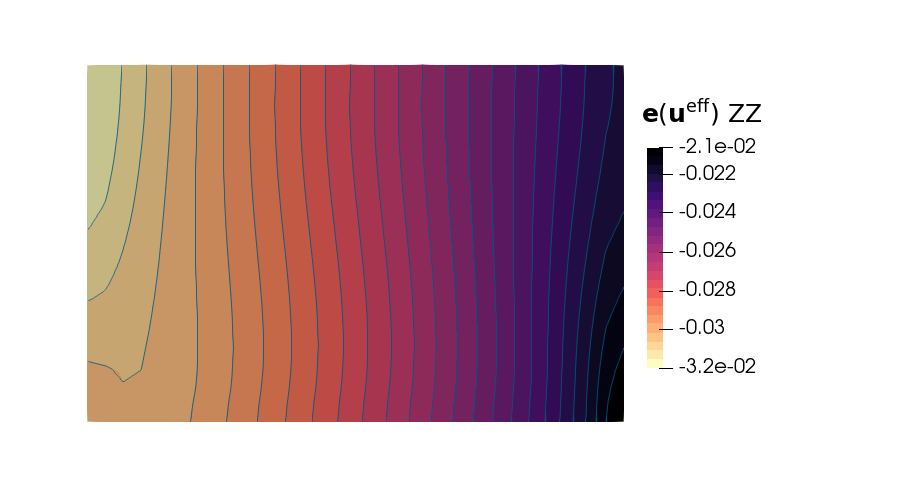}	&\includegraphics[width=0.99\linewidth]{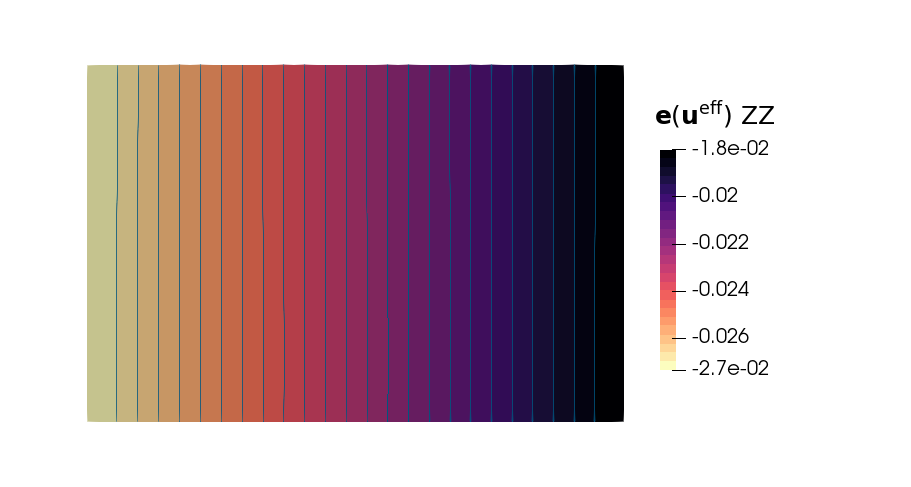}\\
			\hline
			\rotatebox[origin=c]{90}{$\sigma(\ub^\eff)_{zz}$}
			&\includegraphics[width=0.99\linewidth]{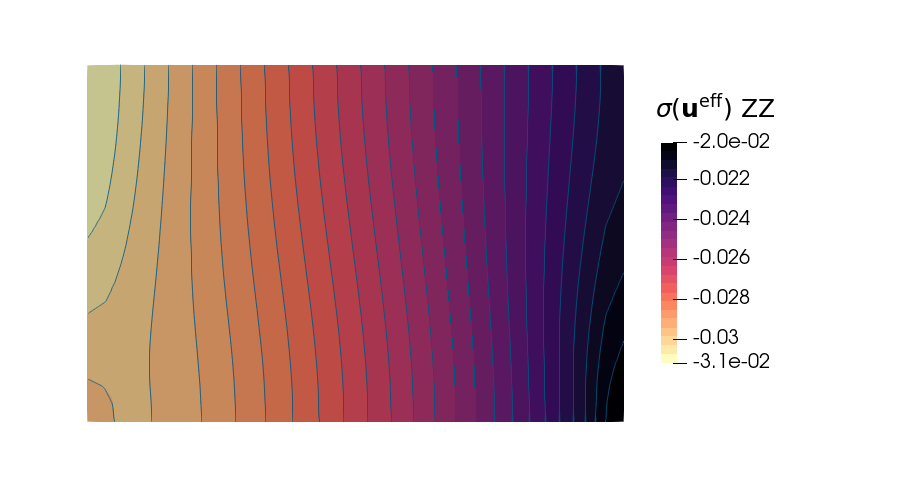}	&\includegraphics[width=0.99\linewidth]{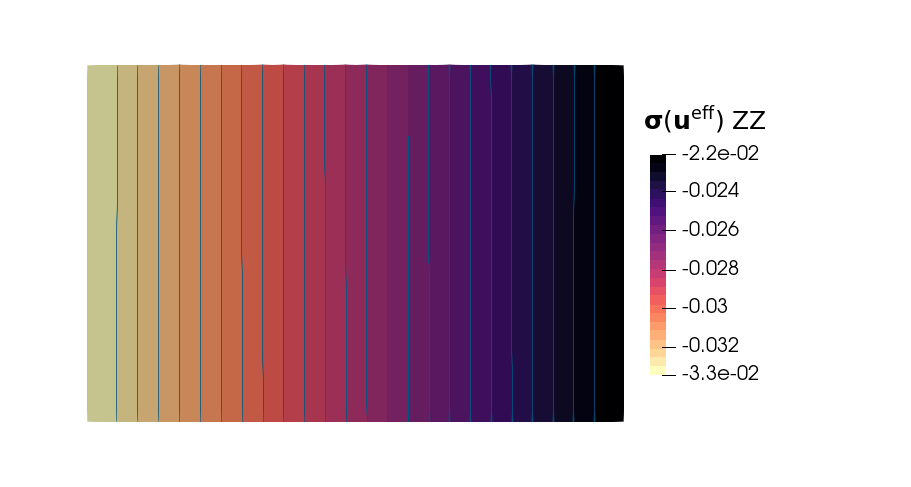}\\
		\end{tabular}
	\end{center}
	\caption{{Comparizon of steady-state solution of BVP for PZPM and EPM model: Steady-state distribution of components of macroscopic displacement $u_z^\eff$[$\mu$m] and components of Cauchy strain $e(\ub^\eff)_{xx},e(\ub^\eff)_{yy}$ and $e(\ub^\eff)_{zz}$[-] and Cauchy stress $\sigma(\ub^\eff)_{zz}$[Pa] in the $xz$-slice of macroscopic specimen.}}
	\label{fig:solution1}
\end{figure}

\begin{figure}[t]		\centering
	\begin{subfigure}[b]{0.49\linewidth}
		\centering
		\includegraphics[width=0.99\linewidth]{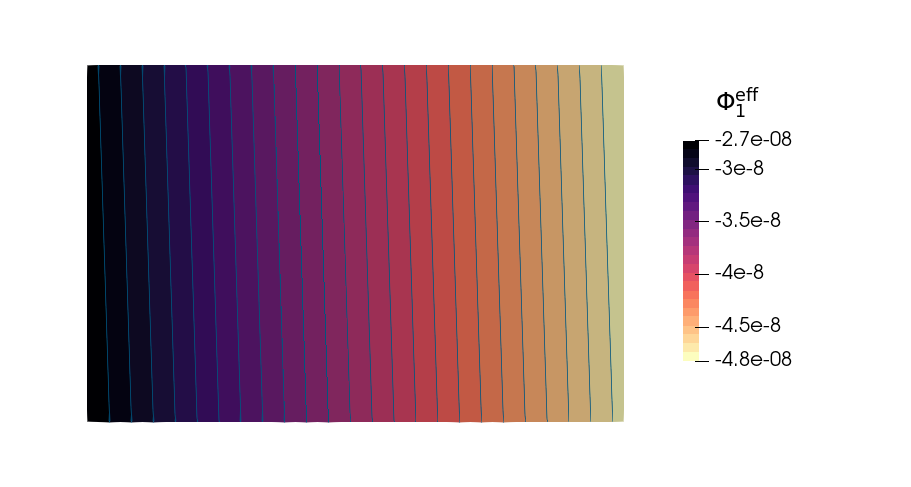}
		%		\caption{}
		%		\label{fig:puII}
	\end{subfigure}
	\begin{subfigure}[b]{0.49\linewidth}	\centering
		\includegraphics[width=0.99\linewidth]{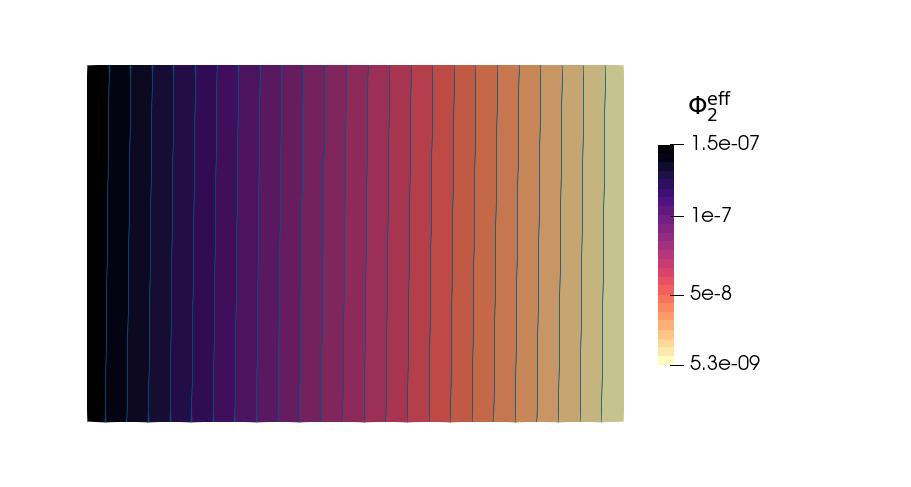}
		%		\caption{}
		%		\label{fig:phiII}
	\end{subfigure}
	\begin{subfigure}[b]{0.49\linewidth}	\centering
	\includegraphics[width= 0.99\linewidth]{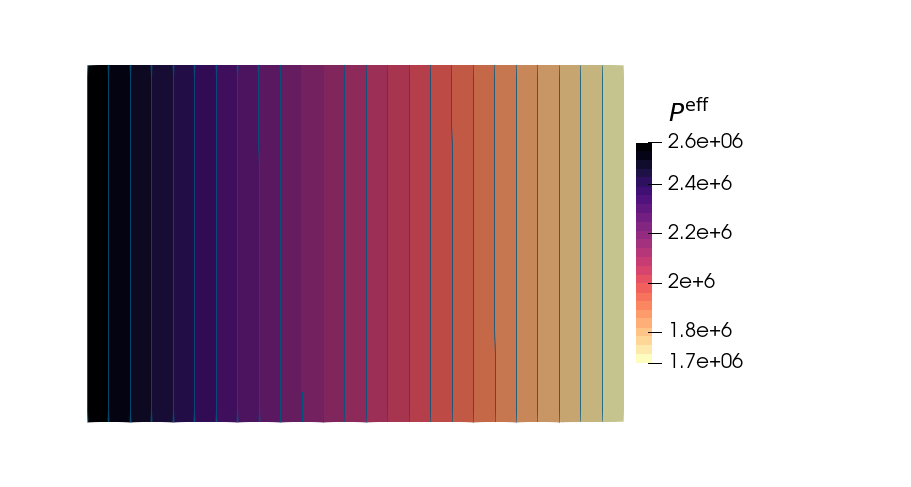}
	%		\caption{}
	%		\label{fig:phiI}
	\end{subfigure}
	\begin{subfigure}[b]{0.49\linewidth}	\centering
	\includegraphics[width= 0.99\linewidth]{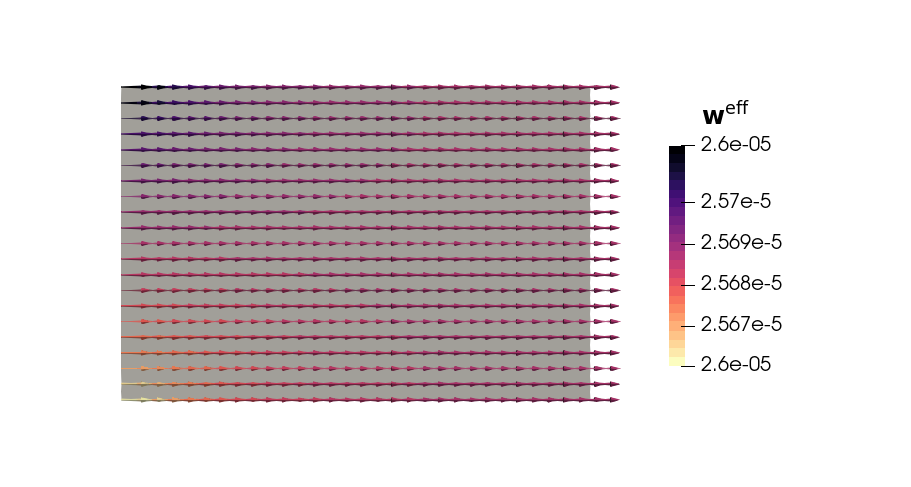}
	%		\caption{}
	%		\label{fig:phiI}
	\end{subfigure}
	\caption{{Solution steady-state solution of BVP for PZPM  model: Steady-state distribution of macroscopic fields $(P^\eff,\wb^\eff,\Phi^\eff_\alpha),\alpha=1,2,$ in the $xz$-slice of macroscopic specimen.}}
	\label{fig:solution2}
\end{figure}

\subsubsection{Poroelastic vs poropiezoelectric model}
The presented numerical example BVP can serve to compare the differences between solutions of the presented model of electrolyte flux through the poropiezoelectric porous medium (PZPM) with the model considering only linear elastic porous medium (EPM) as was published in \cite{turjani2018}. 
It can be shown that the derived macroscopic problem describing steady-state is formally given by \Eq{eq_macro_piezo_strong} for both PZPM and EPM. However, in the case of EPM, we will use the expressions of Biot-like effective coefficients in the form given in \cite{turjani2018}, \ie effective coefficients characterizing the behavior of linear elastic material, instead of coefficients \Eq{eq_ec_piezo_disp_1}.  

In our comparison, we use the microstructure and material properties from Tab.~\ref{tab_constants} and \ref{tab:ident_piezo}. Following the same procedure, we compute the macroscopic solution of BVP for both models describing EPM and PZPM, respectively. 

Even though the macroscopic specimen is not loaded in the steady state, we observe the non-zero displacement $\ub^\eff$. This is caused by the so-called swelling of the specimen, which occurs in some water-saturated porous media, \eg swelling clays \cite{moyne2002electro}. The swelling of a specimen was also observed in our previous work \cite{turjani2018}. The Cauchy stress and Cauchy strain are computed from the resulting macroscopic displacement field $\ub^\eff$. 

We visualize the distribution of resulting macroscopic fields along the $x$-axis of the macroscopic specimen for both models, see Fig.~\ref{fig:graphs_x}, where the vertical green line denotes the position of point A. In the case of the vector fields, we show their component that corresponds to the direction of loading, \ie components $u^\eff_z, e_{zz}(\ub^\eff)$ and $\sigma_{zz}(\ub^\eff)$.

The results show both a quantitative and qualitative difference between the PZPM and EPM models across all macroscopic fields except for pressure field $P^\eff$. However, this is most likely caused by the choice of Dirichlet boundary conditions prescribed to the pressure $P^\eff$. The macroscopic quantities obtained from the solution of the EPM model exhibit linear distribution along the $x$-axis as is to be expected. The PZPM model seems a better choice to describe the materials with nonlinear behavior.

The  spatial distribution of displacement component $u^\eff_z$ and consequently components of strain $e_{zz}(\ub^\eff)$ and stress $\sigma_{zz}(\ub^\eff)$ of both models is depicted in Fig.~\ref{fig:solution1}. There, we see a visible qualitative difference in the distribution of $\ub^\eff$ in the $xz$-slices of the macroscopic specimen. Once again, the steady-state Cauchy stress $\sigma_{zz}(\ub^\eff)$ and strain  $e_{zz}(\ub^\eff)$ resulting from the EPM model appears to be linear in comparison to these quantities resulting from the PZPM model. 

Both the potentials $\Phi_\alpha^\eff, \alpha=1,2,$ and the pressure $P^\eff$ shows only quantitative difference between models PZPM and EPM as can be seen in Fig.~\ref{fig:graphs_x}. Thus, we visualize their spatial distribution only for model PZPM in Fig.~\ref{fig:solution2} for completeness.

The difference between the macroscopic fields is more pronounced when computing the non-steady solution of BVP. Once again, we solved \Eq{eq_macro_piezo_strong} for both PZPM and EPM, where to describe EPM, we used the expressions of Biot-like effective coefficients from \cite{turjani2018}. The evolution of the macroscopic solution at point A of the macroscopic specimen (see Fig.~\ref{fig:boundaries}) is depicted in Fig.~\ref{fig:graphs_time}. The cyclic compression (given by function \eqref{eq_ramp}) acting on the boundary $\Gamma_T$ influences the time evolution of
all resulting macroscopic fields. 

%To better visualize the difference between the results of PZPM and EPM, we also provide the visualization of the absolute difference between macroscopic fields of PZPM and EPM models, see \TODO{Suppl.~\ref{}}{dát do supplementary?}, where we observe the absolute difference of components of macroscopic fields growing and receding with the cyclic compression.

The main cause for the difference between results from PZPM and LEMP models is  the additional piezoelectric terms included in Biot-like coefficients \Eq{eq_ec_piezo_disp_1}. Because the PZPM model includes the piezoelectric terms in Biot-like coefficients \Eq{eq_ec_piezo_disp_1}, the changes in the distribution of ions will have a greater effect on the electro-mechanical responses of the specimen. The existing difference between the resulting potentials of EPM and PZPM models leads us to the conclusion that even in the case of weakly piezoelectric materials, the piezoelectric effect's contribution should be considered.

\section{Application to the modeling of cortical bone porous tissue}\label{sec:app}
The applications of the presented homogenized model can be found in a range of scientific fields, the most significant being geoscience, energetics, and biomechanics. Our interest lies in the modeling of the cortical bone porous structure, where the transport of the ions in the proximity of charged collagen-apatite matrix has shown to play a role in bone regrowth and remodeling. Upon stress, bone tissue generates an electrical potential that directly influences the activity of bone cells. With this in mind, the model of ionic transport in PEPM seems suitable for the modeling of cortical bone tissue.   
%that was described in Sec.~\ref{Sec:piezo} 

\subsection{Cortical bone porous structure}
The cortical bone is a strictly hierarchical system with a complicated porous structure on different scale levels. From the macroscopic point of view, the cortical bone tissue comprises a system of approximately cylindrical sub-units called osteons. Each osteon has a radius of approximately 100-150 $\mu$m, \cite{yoon}, with a hollow canal in its center. It is called the Haversian canal and contains blood vessels and nerves. The rest of the space is occupied by bone fluid. The walls of the Haversian canal are covered by bone cells. Behind this bone cell layer, the walls of the Haversian canal are perforated by a network of small interconnected channels known as canaliculi \cite{yoon}. The canaliculi network connects the Haversian canal and lacunae, small ellipsoidal cavities containing one bone-creating cell,\ie  an osteocyte. The lacunar-canalicular network (further referred to by LCN) is saturated by bone fluid that transports nutrients and information about mechanical loading. 
For the purpose of numerical modeling, we simplify the bone structure into two structural levels and propose their geometry representation, as follows:
\begin{itemize}
	\item The microscopic level represents the LCN  filled with bone fluid. The cubic RVE $Y$ represents three channels with a single ellipsoidal lacuna. The three channels have a cross-sectional area corresponding to the sum of cross-sectional areas of all the canaliculi in the given direction to preserve the flow rate between lacunae, see Fig.~\ref{Fig:Y_CLN}, \cite{beno}.
	\item The macroscopic level is represented by a single osteon which has an approximately cylindrical shape with a hollow canal in its center, see Fig.~\ref{Fig:Mac_geom},  \cite{GAUTHIER2019526}
\end{itemize}

\begin{figure}[t]
	\begin{subfigure}[t]{0.49\linewidth}
		\centering
		\includegraphics[width=0.7\linewidth]{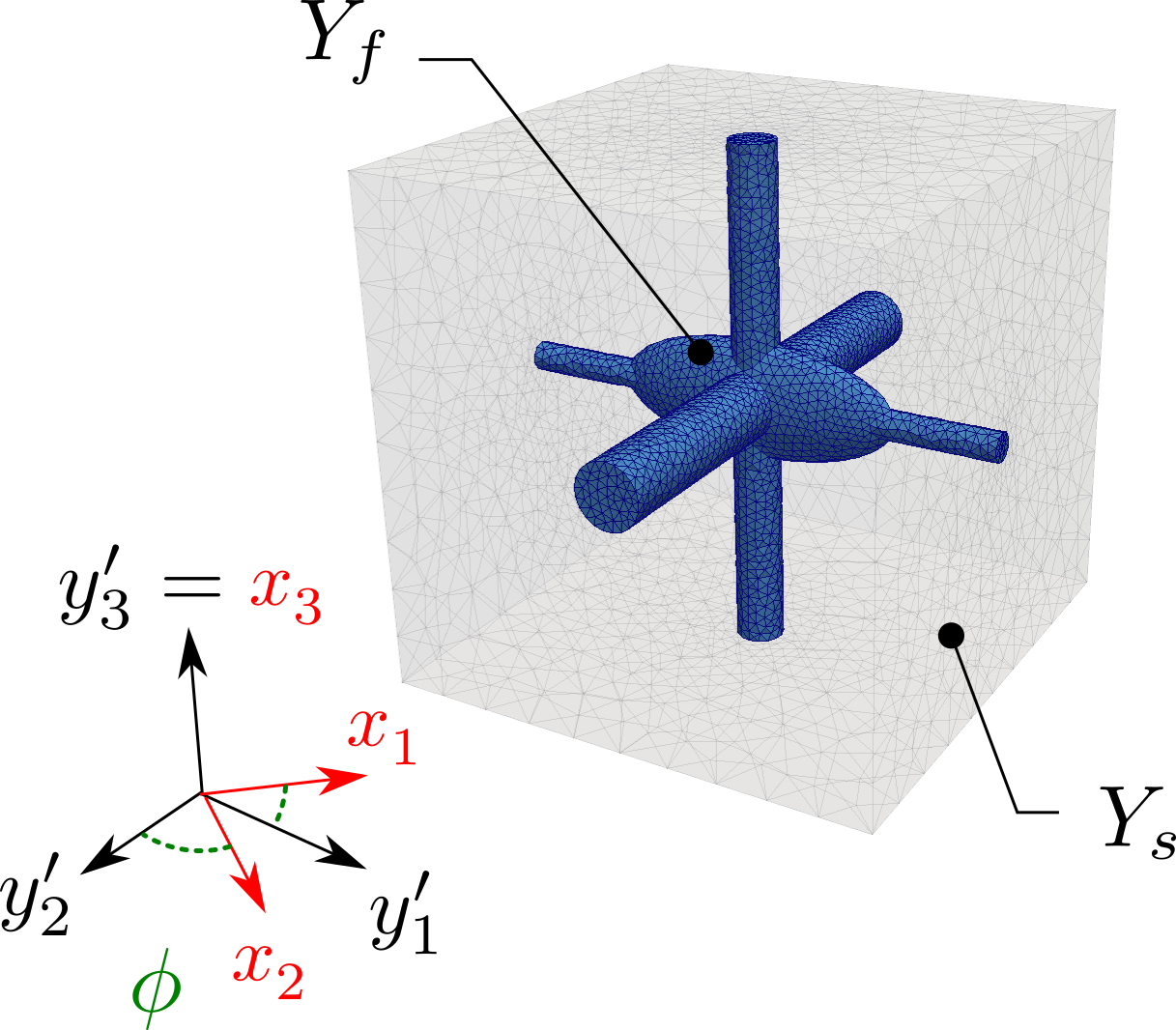}
		\caption{Mesh representation of RVE Y}
		\label{Fig:Y_CLN}
	\end{subfigure}
	\begin{subfigure}[t]{0.49\linewidth}
		\centering
		\includegraphics[width=0.7\linewidth]{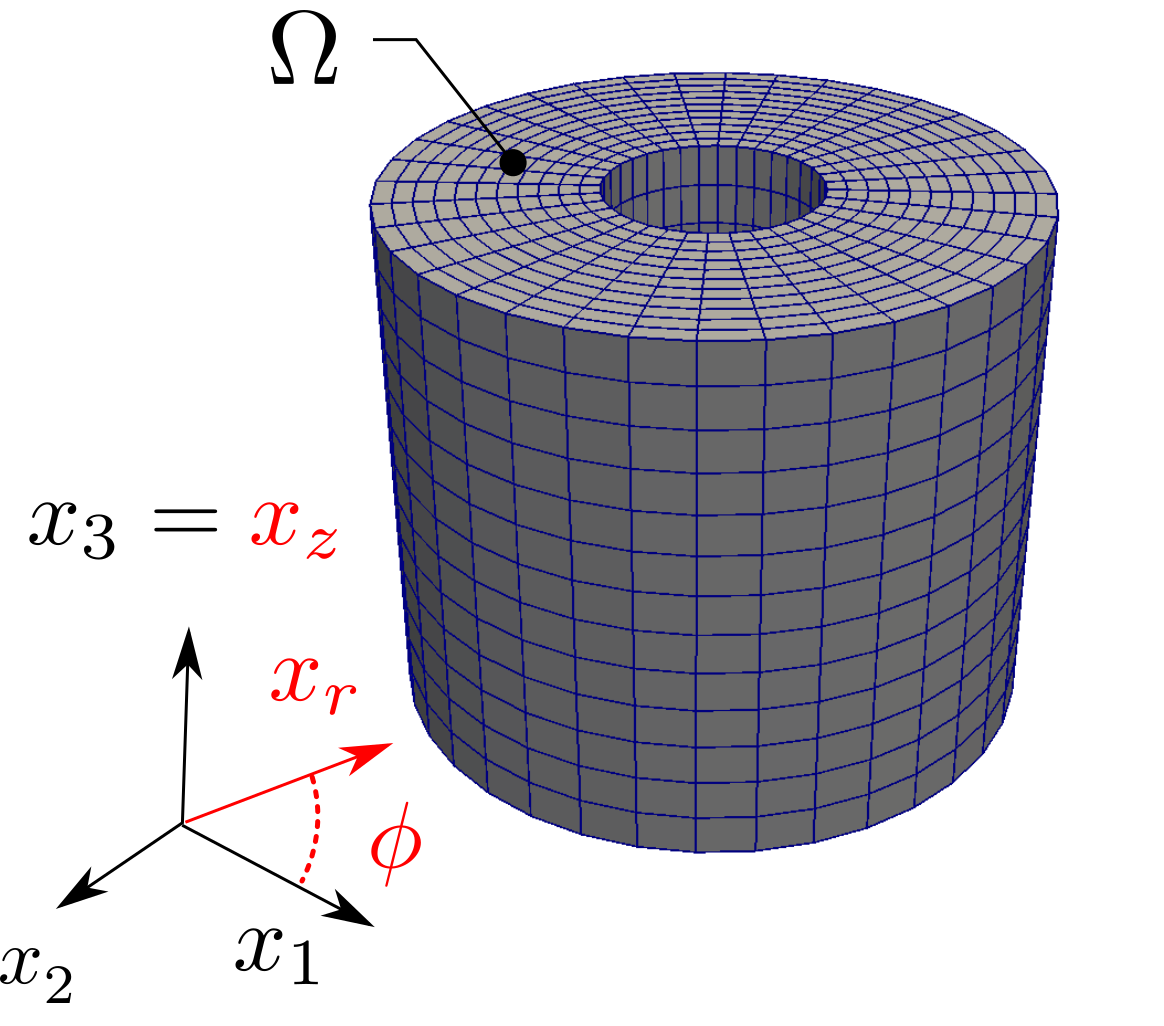}
		\caption{Mesh representation of single bone osteon}
		\label{Fig:Mac_geom}
	\end{subfigure}
	\caption{Mesh representation of micro- and macroscopic structure made within software \textit{GMSH}.}
	\label{fig:meshes}
\end{figure}

\begin{paragraph}{Microstructure orientation in the osteon}
	%\label{rem:orient}
	In the context of
	the osteonal structure, the microporosity has orthotropic properties aligned with a cylindrical coordinate system
	introduced for a central canal of the osteon. Therefore, in \Eq{eq_macro_piezo_strong}, we consider all the macroscopic tensors rotated according to such local coordinate system. In Fig.~\ref{Fig:Y_CLN}, the RVE $Y$
	is defined in a local coordinate system $\yb^\prime=(y_1^\prime,y_2^\prime,y_3^\prime)$ labeled by prime. Thus,
	within the osteon, the orthotropy axes arising from the homogenization vary, as illustrated in Fig.~\ref{Fig:Y_CLN}.
\end{paragraph}
\subsection{Numerical modeling of single bone osteon}\label{sec:num_bone}
In this part of the text, we use all the knowledge about osteonal geometry, microstructure, and material properties to provide a computational model of a single bone osteon.

We present the FE mesh of both micro- and macroscopic geometry in Fig.~\ref{fig:meshes}. At the microscopic level, the mesh represents cubic periodic RVE $Y$ containing a single lacuna with channels representing the collection of canaliculi leading in the directions of the three coordinate axes. It consists of 65546 tetrahedron elements with 11656 vertices. The computational mesh represents a single bone osteon with the Haversian canal at the macroscopic scale. It comprises 14896 hexahedron elements with 16800 vertices. Then the macroscopic problem to solve has 10106544 DOFs. 

Both meshes were made in the open-source software \textit{GMSH}, which provides a wide variety of meshing algorithms and is suitable for modeling periodic meshes,  \cite{geuzaine2009gmsh}. 
The microstructure size is given by $\veps:=\veps_0=0.03$ which
determines the influence of the piezoelectric coupling $\gb$ and dielectric tensor $\db$. 

To describe the material properties at the microscopic structure we use the values from Tab.~\ref{tab:ident_piezo} that characterize collagen-hydroxyapatite matrix as a transversally isotropic material with weak piezoelectric properties. Electrochemical properties of bone fluid were taken from Tab.~\ref{tab_constants}.

Once again, to compute the homogenized coefficients and to simulate processes on the macroscopic scale, we used the implementation made in the \textit{SfePy} software. Then, to respect the orientation of LCN in osteon, the computed effective coefficients are  circumferentially rotated around the $x_z$-axis of the central canal.

% For more informations about implementation and discretization we refer to Chapter~\ref{sec_numeric} and especially its parts Sec.~\ref{Sec:num_algorithm} and Sec.~\ref{sec:discretized}.

\begin{figure}[!t]
	\centering
	\includegraphics[width=\linewidth]{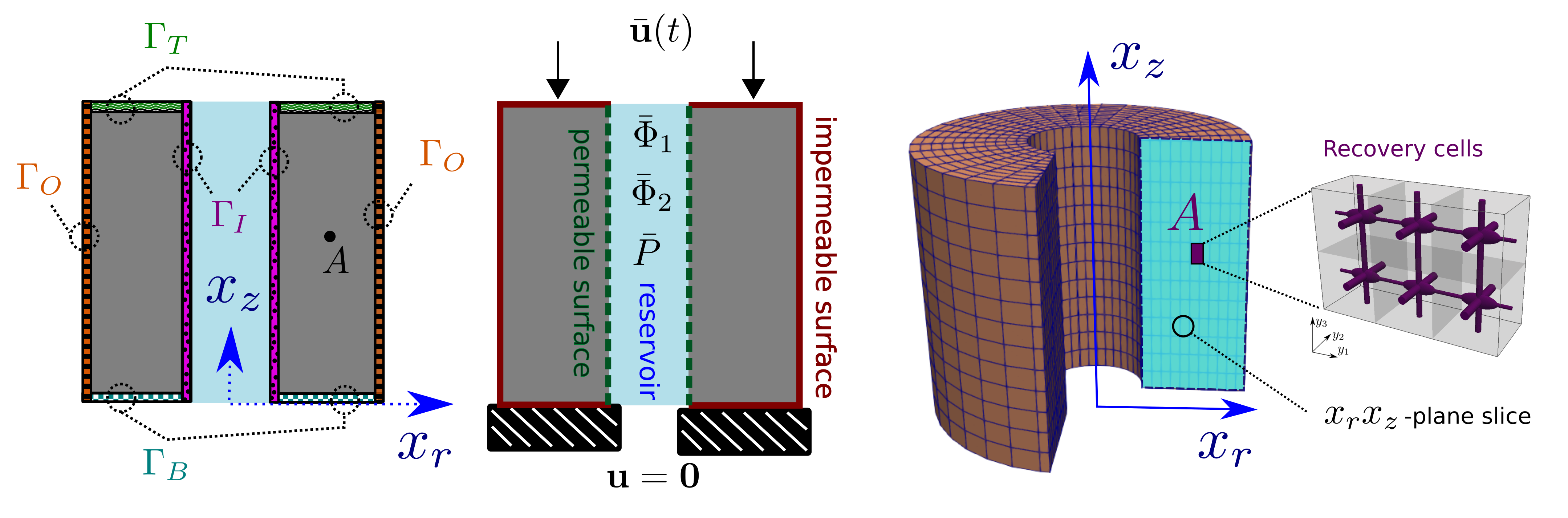}
	\caption{Left: Definition of boundaries on the macroscopic specimen; Middle: Boundary conditions defining BVP; Right: The slice through $x_zx_r$-plane of macroscopic specimen and placement of point A. The point A lies inside the recovery region that encompasses six copies of cell $Y$.}
	\label{fig:boundaries}
\end{figure}

\paragraph{Boundary conditions}
For simplicity, we denote the inner and outer osteonal wall and its top and bottom base by $\Gamma_\textrm{I}, \Gamma_\textrm{O}, \Gamma_\textrm{T}$, and $\Gamma_\textrm{B}$, respectively. These boundaries are shown in Fig.~\ref{fig:boundaries}.

On the outer wall of osteon $\Gamma_\textrm{O}$ there is the so-called cement surface, through which only a few canaliculi can cross. Thus, as an idealization, it is reasonable to consider the outer wall to be impermeable with no fluid flow and no ionic exchanges, see \cite{remond2008interstitial}. In addition, we also consider the top and bottom bases of osteon $\Gamma_\textrm{T}$ and $\Gamma_\textrm{B}$ to be impermeable. 

Even though we model the osteon as a hollow cylindrical body, the influence of the fluid in the Haversian canal has to be reflected by the boundary conditions. When compared with the porosity of the LCN network, the Haversian canal is large enough to enable the fluid inside to relax, so that the pressure can be assumed to be constant at the osteonal inner wall $\Gamma_\textrm{I}$. The ionic potentials are considered constant at the inner osteonal wall $\Gamma_\textrm{I}$. This leads to a prescribed Dirichlet condition on the boundary $\Gamma_\textrm{I}$.
In addition, we also consider the inner wall to be stress-free, see \cite{nguyen2010poroelastic}.

With this assumptions, we define the boundary value problem (BVP) to demonstrate the macroscopic behavior of our effective model. The initial values of effective quantities $(P,\Phi_\alpha^\eff, \ub^\eff), \alpha=1,2,$ were taken from its steady state solution for $\fb=\mathbf{0}, \Eb=\mathbf{0}$. The computational time was taken $t\in ]0,T[$, where $T=1$s and time step $\Delta t\approx0.05$s.

\subsubsection{The boundary value problem}
The BVP describes a situation, where we consider the inner osteonal wall $\Gamma_I$ to be non-permeable. At the top of the osteon, \ie at $\Gamma_T$, the gradual compression is applied which is realized through the boundary condition on $u_{3}(t)$ given as a ramp-and-
hold function
\begin{equation}\label{eq_ramp}
\bar{u}(t)=\begin{cases}
-\hat{u}t&\textrm{for } 0\leq t<t_r,\\
-\hat{u}t_r&\textrm{for } t_r\leq t<T,\\
\end{cases}
\end{equation}
where $t_r=0.45$s and $\hat{u}=0.1$.

The BVP is defined by \Eq{eq_macro_piezo_strong} and by the boundary conditions of Neumann and Dirichlet type. The boundary conditions are applied to the parts of macroscopic specimen boundary described by Fig.~\ref{fig:boundaries}   and are listet in Tab.~\ref{tab:bc2}. Initial conditions are taken from the steady-state solution (\ie  for $t=0$ and neglecting all time derivatives) of the BVP. 

\begin{table}[h]\centering
	\begin{tabular}{llll}
		\hline
		Boundary part & $\mathbf{u}^0$ & $\mathbf{\Phi}^\alpha$ & P\\
		\hline
		$\Gamma_T$& $u_{1}=u_{2}=0,\quad u_{3}(t)=\bar{u}(t)$&$\nb\cdot\jb_\alpha=0$&$\nb\cdot\wb=0$\\
		$\Gamma_B$& $\ub(t)=\mathbf{0}$&$\nb\cdot\jb_\alpha=0$&$\quad\nb\cdot\wb=0$\\
		$\Gamma_O$& $\nb\cdot\sigmabf_s=0$&$\nb\cdot\jb_\alpha=0$&$\nb\cdot\wb=0$\\
		$\Gamma_I$& $\nb\cdot\sigmabf_s=0$&$\nb\cdot\jb_\alpha=0$&$ P=\bar{P}$\\
		\hline
	\end{tabular}
	\caption{The Dirichlet and Neumann conditions for the particular
		fields involved in the macroscopic model. The conditions are given for all $t\in ]0,T[$, where $\sigmabf_s,\wb$ and $\jb_\alpha, \alpha=1,2$, are given by \Eq{eq_macro_fluxes_piezo} and $\bar{u}(t)$ is given by \Eq{eq_ramp}. The prescribed values of boundary conditions are $\bar{P}=0.1,\bar{j}_1=-0.01,\bar{j}_2=0.01$}	
	\label{tab:bc2}
\end{table}

\begin{figure}[p]		\centering
	\begin{center}
		\begin{tabular}{ 
				m{0.03\linewidth}|
				>{\centering\arraybackslash}m{0.17\linewidth}
				>{\centering\arraybackslash}m{0.17\linewidth} >{\centering\arraybackslash}m{0.17\linewidth} >{\centering\arraybackslash}m{0.25\linewidth}}
			&$t=0$ &$t=t_1$ & $t=t_r$&$t=T$\\
			\hline
			\rotatebox[origin=c]{90}{$p^\eff$}&
			\includegraphics[height=4.7cm]{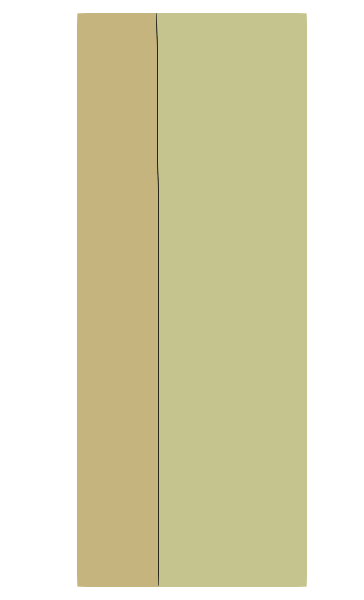}			
			&
			\includegraphics[height=4.7cm]{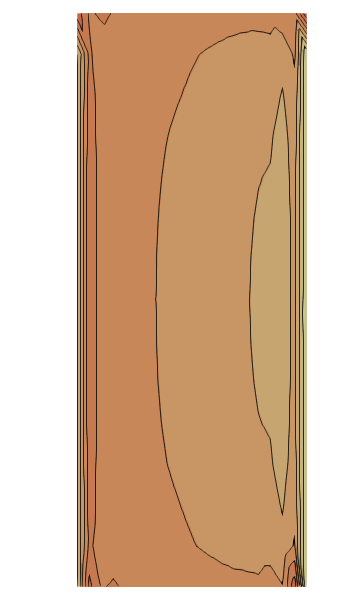}			
			&\includegraphics[height=4.7cm]{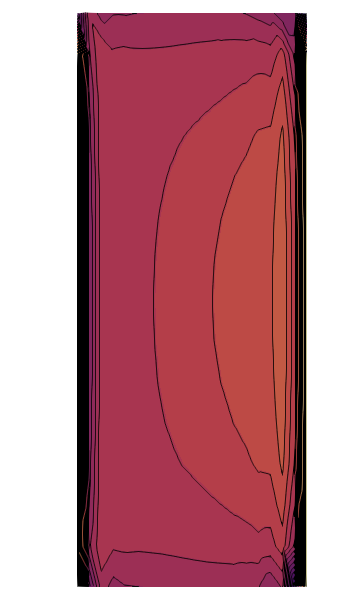}
			&\includegraphics[height=4.7cm]{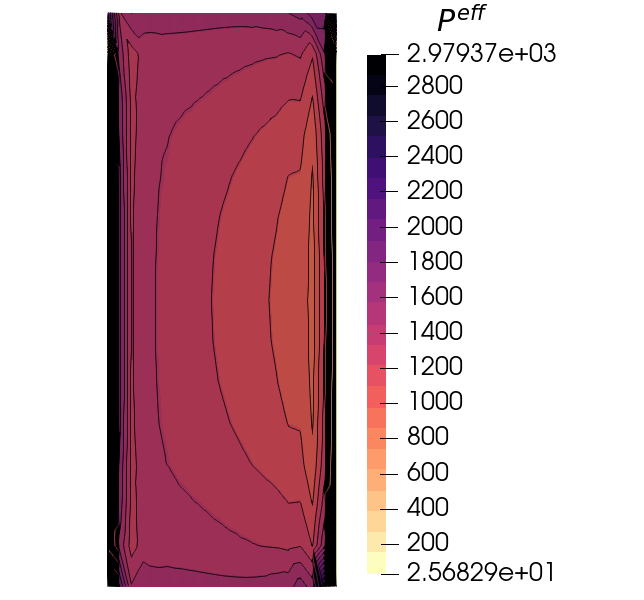}\\
			\hline
			\rotatebox[origin=c]{90}{$\wb^\eff$}&
			\includegraphics[height=4.7cm]{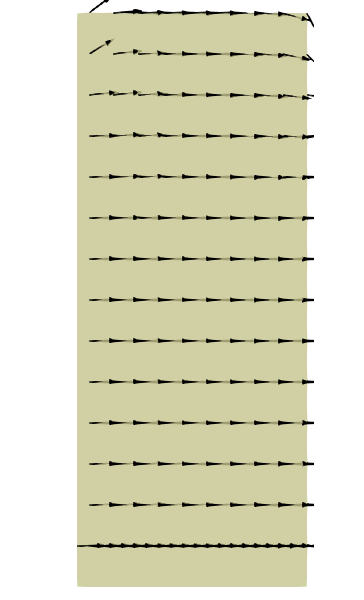}			
			&
			\includegraphics[height=4.7cm]{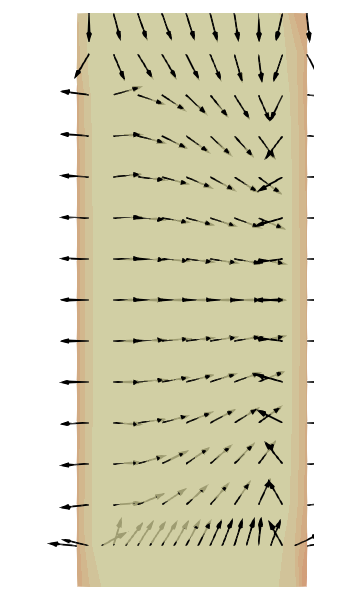}
			&\includegraphics[height=4.7cm]{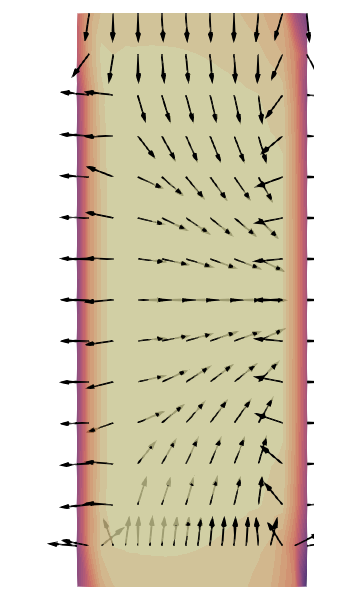}
			&\includegraphics[height=4.7cm]{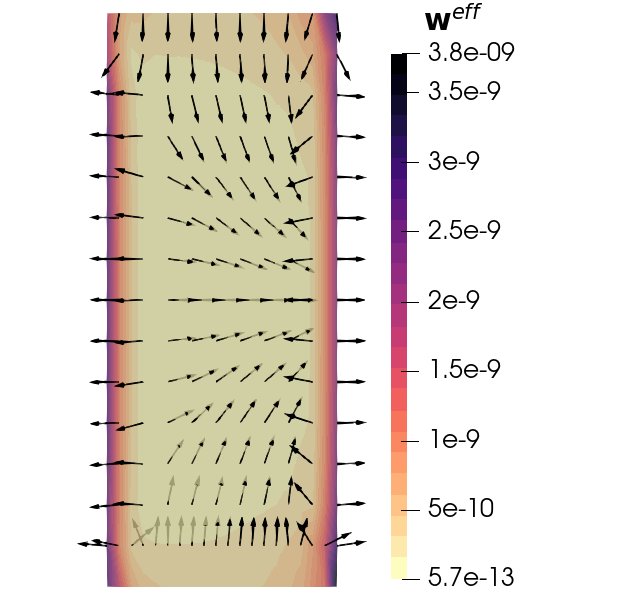}\\
			\hline
			\rotatebox[origin=c]{90}{$\Phi_1^\eff$}&
			\includegraphics[height=4.7cm]{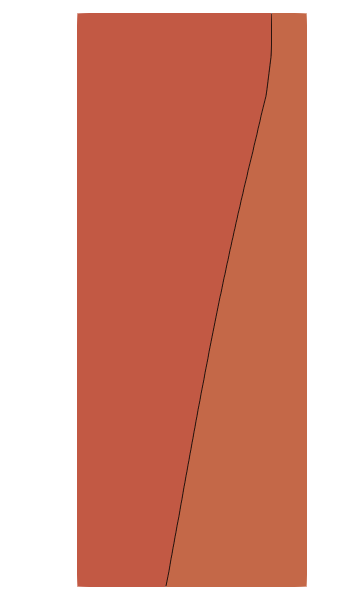}			
			&
			\includegraphics[height=4.7cm]{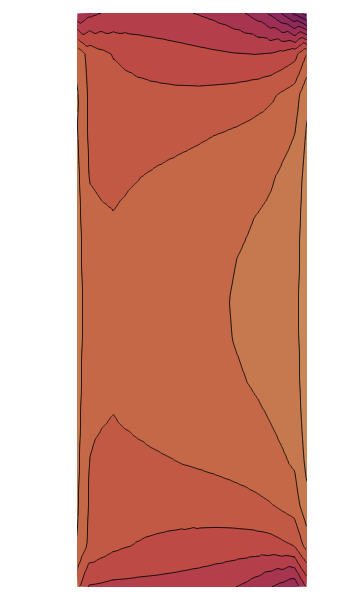}			
			&\includegraphics[height=4.7cm]{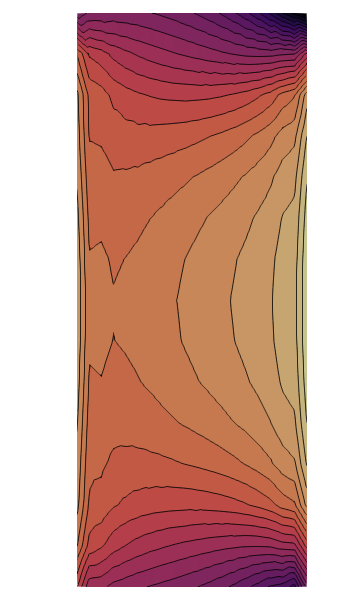}
			&\includegraphics[height=4.7cm]{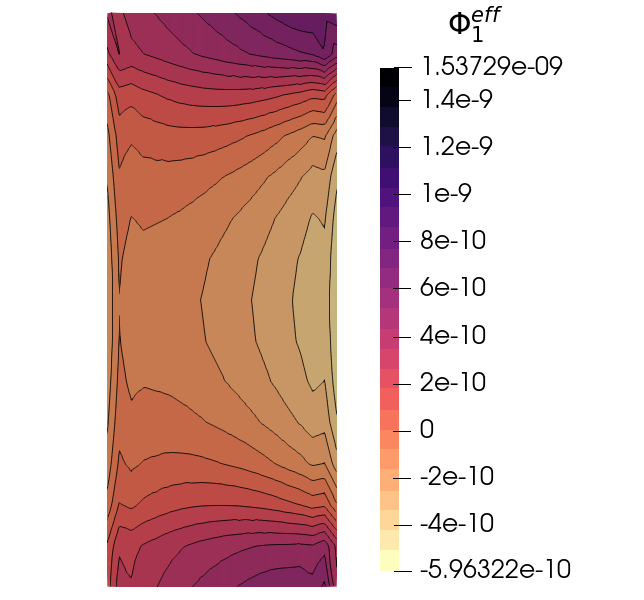}\\
			\hline
			\rotatebox[origin=c]{90}{$\Phi_2^\eff$}&
			\includegraphics[height=4.7cm]{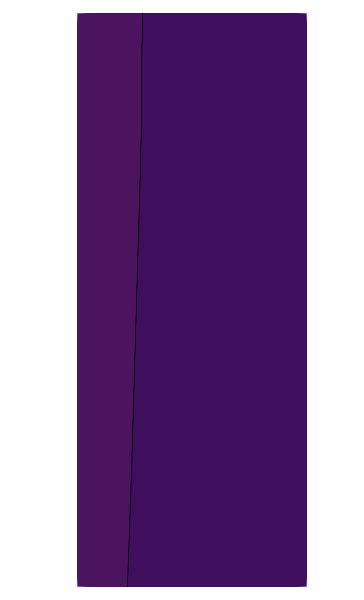}			
			&
			\includegraphics[height=4.7cm]{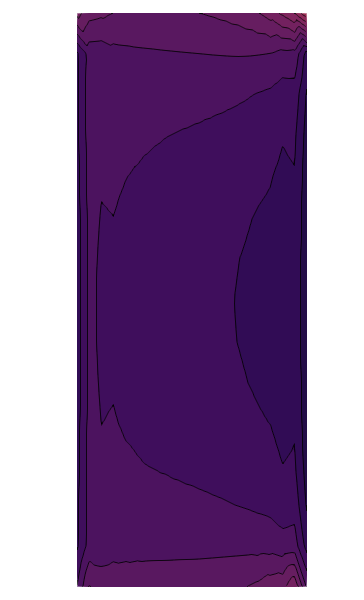}
			&\includegraphics[height=4.7cm]{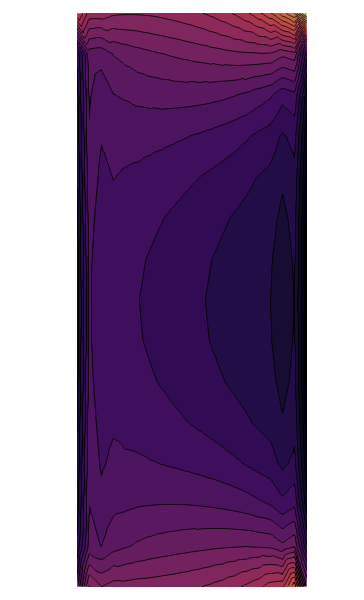}
			&\includegraphics[height=4.7cm]{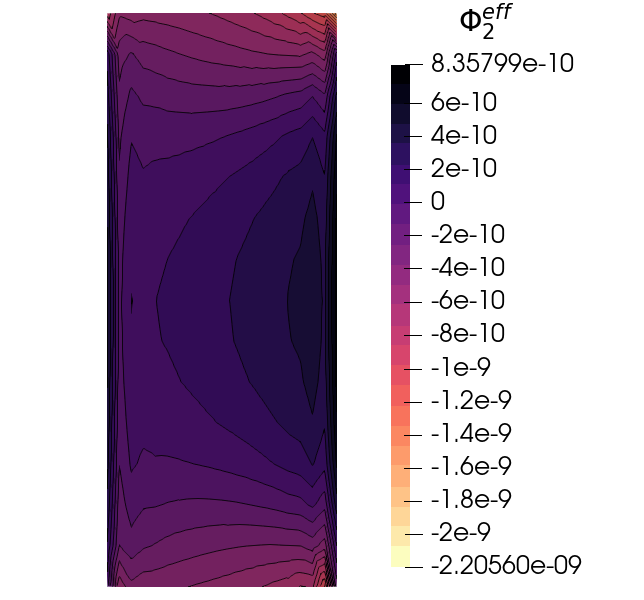}
		\end{tabular}
	\end{center}
	\caption{Solution of the BVP: Evolution of macroscopic pressure $p^\eff$[Pa], velocity $\wb^\eff$[m/s] and  potentials $\Phi_1^\eff$[J/C] and $\Phi_2^\eff)$[J/C] in the slice through $x_rx_z$-plane of macroscopic specimen. The distribution of the macroscopic fields is shown at time steps  $t\in\{t_0,t_1,t_r,T\}=\{0,0.15,0.45,1\}$s, where $t_r$ is time included in the definition of the ramp-and-hold function \eqref{eq_ramp}.}
	\label{fig:macro_solution_BPV_IV_phi}
\end{figure}
\begin{figure}[h]		\centering
	\begin{center}
		\begin{tabular}{
				m{0.03\linewidth}|
				>{\centering\arraybackslash}m{0.25\linewidth}|
				>{\centering\arraybackslash}m{0.17\linewidth} >{\centering\arraybackslash}m{0.17\linewidth} >{\centering\arraybackslash}m{0.25\linewidth}}
			&$t=0$ &$t=t_1$ & $t=t_r$&$t=T$\\
			\hline
			\rotatebox[origin=c]{90}{$u^\eff_z$}
			&\includegraphics[height=4.7cm]{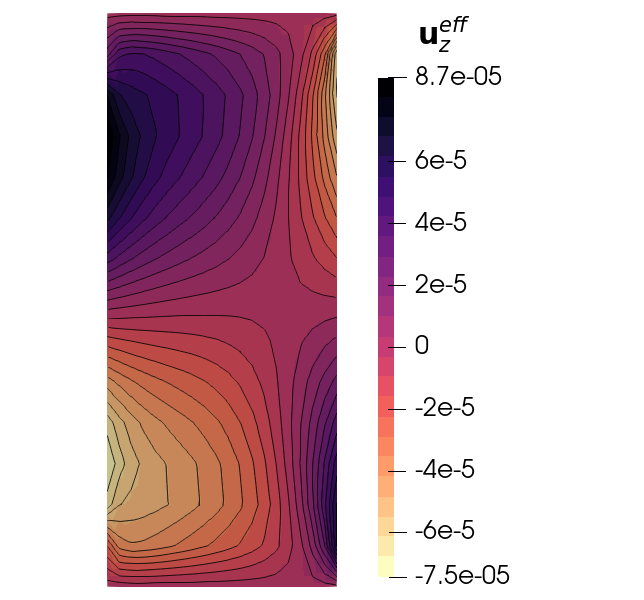}&			
			\includegraphics[height=4.7cm]{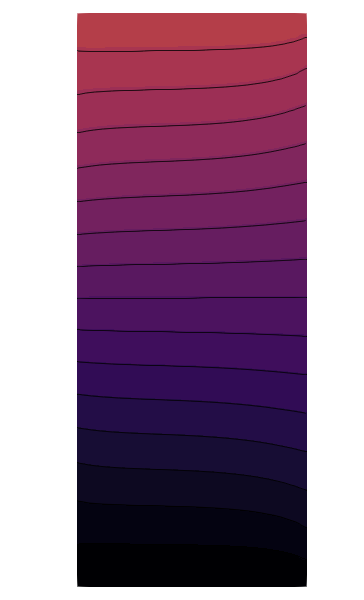}			
			&\includegraphics[height=4.7cm]{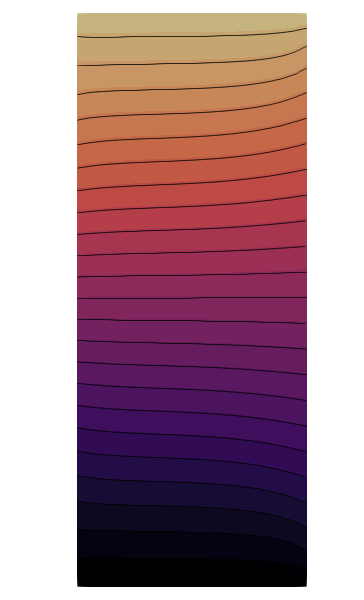}
			&\includegraphics[height=4.7cm]{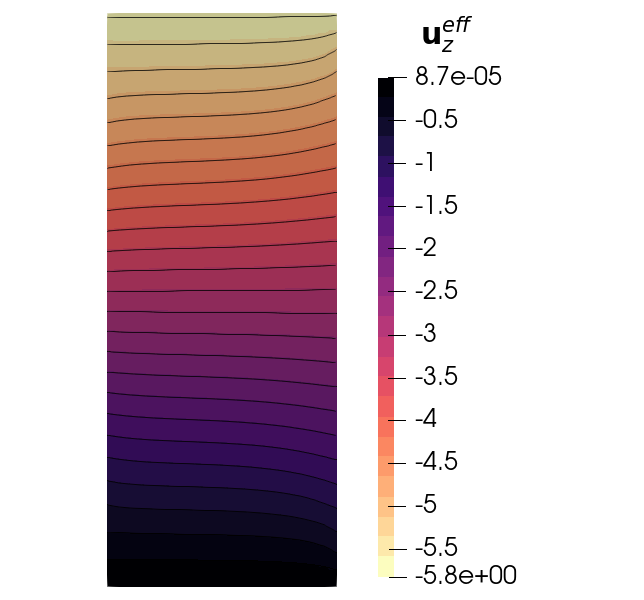}\\
			\hline
			\rotatebox[origin=c]{90}{$e(\ub^\eff)_zz$}
			&\includegraphics[height=4.7cm]{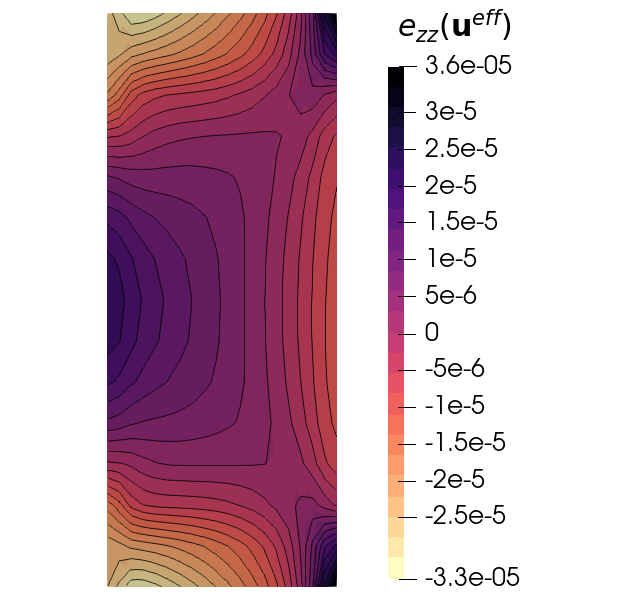}&			
			\includegraphics[height=4.7cm]{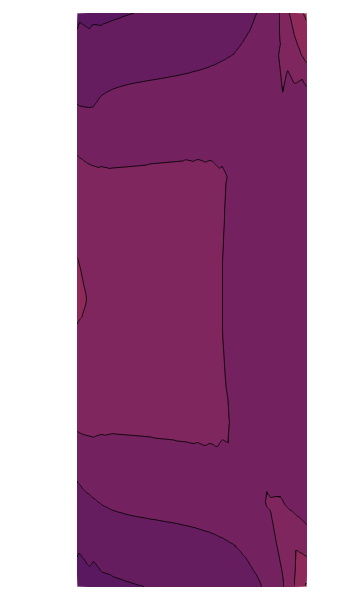}
			&\includegraphics[height=4.7cm]{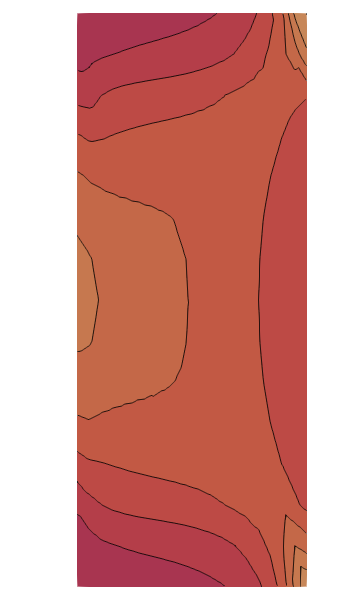}
			&\includegraphics[height=4.7cm]{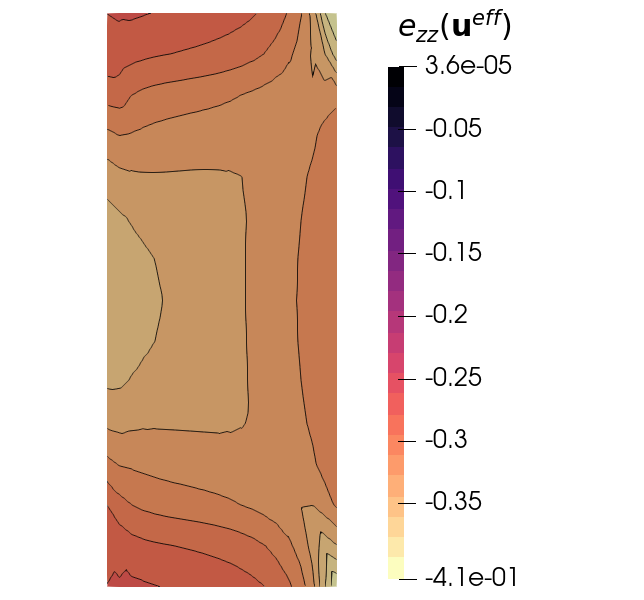}
		\end{tabular}
	\end{center}
	\caption{ Solution of the BVP: Evolution of magnitude of component of macroscopic field $u_z^\eff$[m] and the component of strain $e(\ub^\eff)_{zz}$[-] in the slice through $x_rx_z$-plane of macroscopic specimen.  Left: The steady state distribution of the macroscopic fields; Right: The distribution of the macroscopic fields is shown at time steps  $t\in\{t_1,t_r,T\}=\{0.15,0.45,1\}$s, where $t_r$ is time included in the definition of the ramp-and-hold function \eqref{eq_ramp}.}
	\label{fig:macro_solution_BPV_IV_pu}
\end{figure}

We solve the BVP of the presented model of electrolyte flux through the poropiezoelectric porous medium (PZPM) via numerical simulation. To better illustrate the evolution of the spatial distribution of resulting macroscopic fields $(p^\eff,\ub^\eff,\Phi^\eff_\alpha)$, $\alpha=1,2,$ in the macroscopic specimen, we provide the figures depicting them in a slice through $x_rx_z$-plane. This is possible because due to the circumferential orientation of the microstructure in the macroscopic specimen, all the presented results are axially symmetric. Additionally,  we also computed macroscopic velocity field $\wb^\eff$ and Cauchy strain $\str{\ub^\eff}$ from the solutions of BVP.
The Figs.~\ref{fig:macro_solution_BPV_IV_phi} and \ref{fig:macro_solution_BPV_IV_pu}  depict the resulting macroscopic fields at the selected time steps $t=T=1$s. We observe a visible deformation of the macroscopic specimen, which is mainly due to the increasing displacement on the boundary $\Gamma_T$ but is partly caused by the swelling of the specimen. That is visible in the left part of Fig.~\ref{fig:macro_solution_BPV_IV_pu}, where for $t=0$, \ie steady state solution, the deformation of the specimen is caused purely by swelling.

In this case, the distribution of displacement $\ub^\eff$ directly influences all the other macroscopic quantities, showing a direct connection between specimen deformation and ion distribution, (represented by ionic potentials $\Phi_\alpha, \alpha=1,2$ here).

\subsubsection{Reconstruction of macroscopic solution}
One of the most remarkable advantages of the homogenization method is the possibility to reconstruct the solution at the microscopic scale. Once the macroscopic BVP has been solved, the macroscopic functions $p^\eff, \ub^\eff$ and $\Phi_\alpha^\eff, \alpha=1,2,$ can be used to reconstruct the local responses at the microscopic level. This process is also called downscaling in contrast to the upscaling process leading to the macroscopic model and was briefly explained in \sref{sec:rec}. The reconstructed fields $\ub^{\mic},\wb^{\mic}, P^{\mic},  \Phi_\alpha^{\mic},\alpha=1,2,$ and $\Psi_s^\mic$ are computed from expressions \Eq{eq_rec_elkin3}-\Eq{eq_rec_wpot}.

For a given $\veps_0=0.03$, the macroscopic functions are reconstructed in the recovery region that lies in the proximity of point $A$, see Fig.~\ref{fig:boundaries}. This region  encompasses the block of six periodically repeated copies of the RVE $Y^{\veps_0}=\veps_0 Y$ on the microscopic level. The  fluctuation part of reconstructions $(\Phi_1^\mic, \Phi_2^\mic, P^\mic, \wb^\mic)$ and $(\ub^\mic, \Psi_s^\mic)$ of macroscopic fields $P^\eff, \ub^\eff$ and $\Phi_\alpha^\eff, \alpha=1,2,$ are shown in Figs.~\ref{fig:rec_BPV_IV_phis} and \ref{fig:rec_BPV_IV_up}, respectively. The reconstructed quantities are shown in their steady-state $t=0$ and and the $t=T$.

The microscopic reconstructions follow a similar pattern as the macroscopic solutions from whose they are obtained. We observe a slow increase in both $p^\rec$, $\ub^\rec$, and $\Phi_2^\rec$ and the decrease in $\Phi_1^\rec$. In the case of reconstructed potential of the solid $\Psi_s^\mic$ we observe a decrease, however, as we consider the solid to have a negative charge, this seems to indicate the deformation induced generating of negative potential. The distribution of reconstructed velocity $\wb^\mic$ indicates that the main direction of flow is in the $x_z$-direction. 
%\begin{figure}[h]		\centering
%	\begin{subfigure}[b]{0.49\linewidth}	\centering
%		%		\includegraphics[width=0.99\linewidth]{img/case_I/rec/u_mic_T_croped.png}
%		\includegraphics[width= 0.99\linewidth]{img/case_I/rec/u_rec_T_croped.png}
%	\end{subfigure}
%	\begin{subfigure}[b]{0.49\linewidth}	\centering
%		%%			\includegraphics[width=0.99\linewidth]{img/case_I/rec/P_mic_T_croped.png}
%		\includegraphics[width= 0.99\linewidth]{img/case_I/rec/P_rec_T_croped.png}
%	\end{subfigure}
%	\begin{subfigure}[b]{0.49\linewidth}	\centering
%		\includegraphics[width=\linewidth]{img/case_I/rec/phi1_rec_T_croped.png}
%	\end{subfigure}
%	\begin{subfigure}[b]{0.49\linewidth}	\centering
%		\includegraphics[width= \linewidth]{img/case_I/rec/phi2_rec_T_croped.png}
%	\end{subfigure}
%	\caption{Evolution of total reconstructions $(\ub^\rec,P^\rec,\Phi_1^\rec, \Phi_2^\rec)$ of macroscopic fields obtained as solution of the BVP, $t=T=1$s.}
%	\label{fig:rec_BPV_IV_phis}
%\end{figure}
\begin{figure}[p]		\centering
	\begin{center}
		\begin{tabular}{ m{0.05\linewidth}| >{\centering\arraybackslash}m{0.465\linewidth} >{\centering\arraybackslash}m{0.465\linewidth} }
			&$t=0$ &$t=T$\\
			\hline
			\rotatebox[origin=c]{90}{$\Phi_1^\mic$}
			&\includegraphics[width=\linewidth]{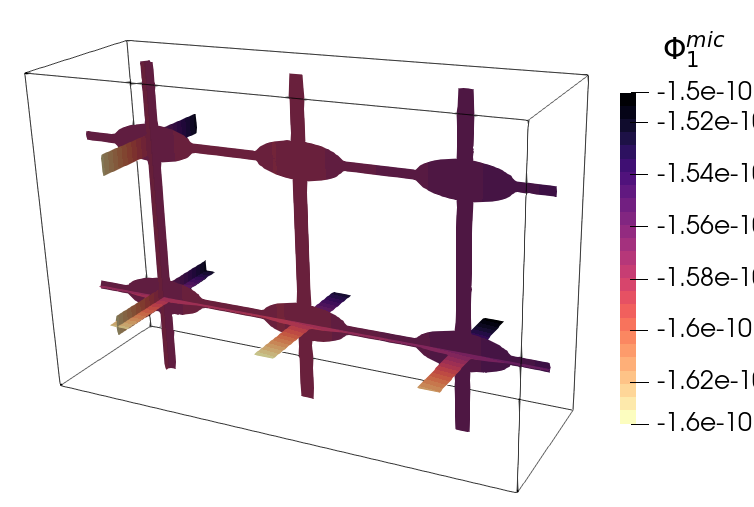}
			&\includegraphics[width= \linewidth]{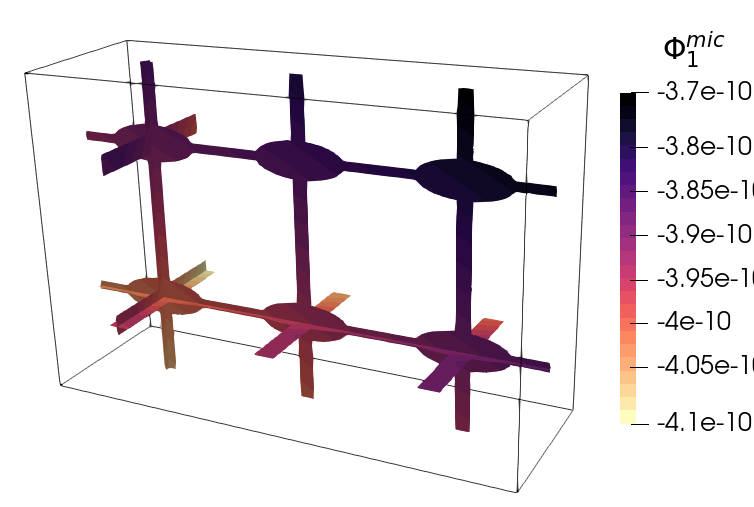}\\
			%			$t=t_3$
			%			&\includegraphics[width=\linewidth]{\figPathMac/rec/phi1_rec_t3_croped.png}
			%			&\includegraphics[width= \linewidth]{\figPathMac/rec/phi2_rec_t3_croped.png}\\			\hline
			\hline\rotatebox[origin=c]{90}{$\Phi_2^\mic$}
			&\includegraphics[width=\linewidth]{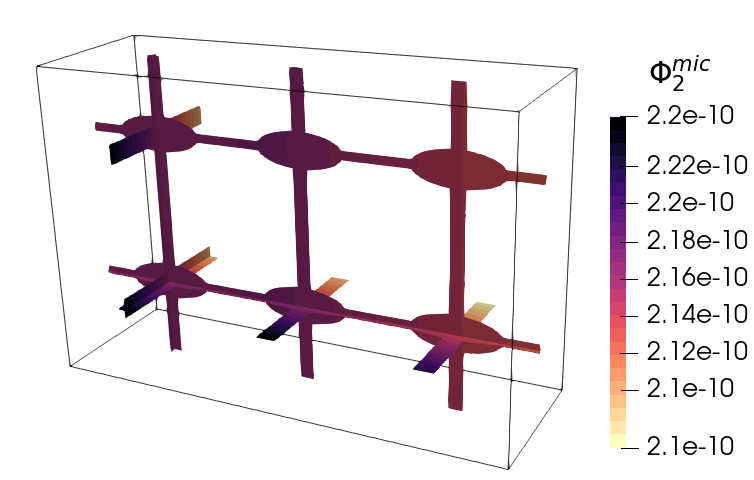}
			&\includegraphics[width= \linewidth]{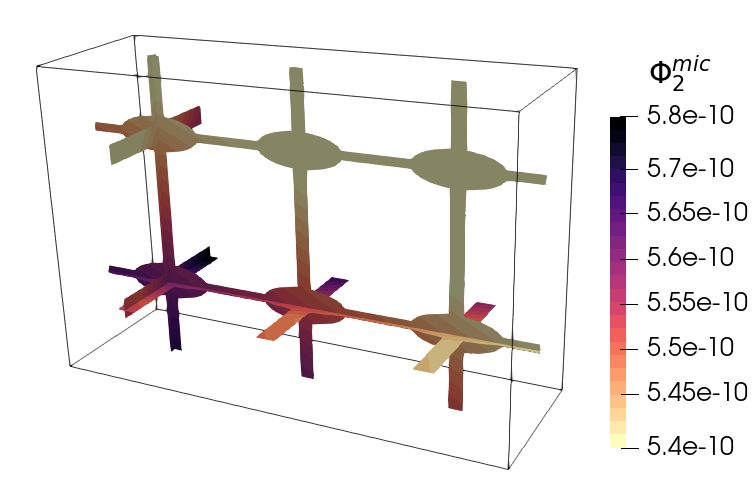}\\
			\hline	\rotatebox[origin=c]{90}{$P^\mic$}
			&\includegraphics[width=\linewidth]{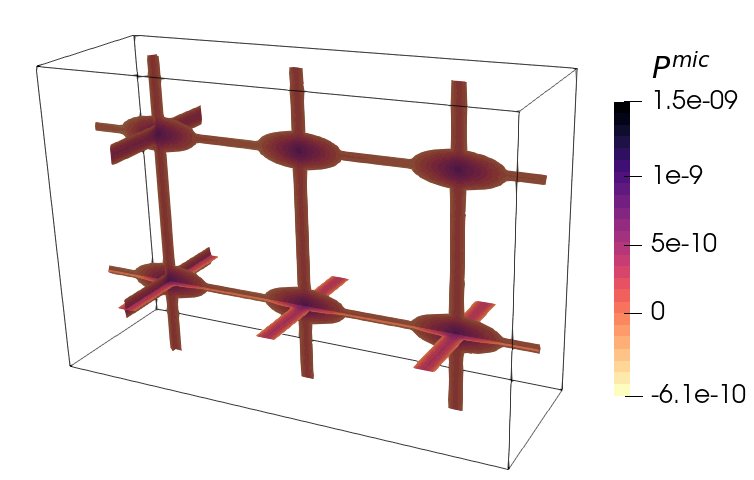}
			&\includegraphics[width= \linewidth]{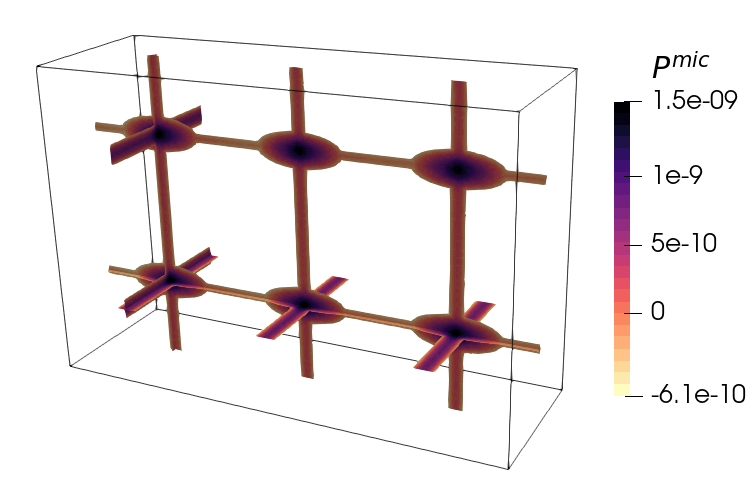}\\
			\hline\rotatebox[origin=c]{90}{$\wb^\mic$}			&\includegraphics[width=0.99\linewidth]{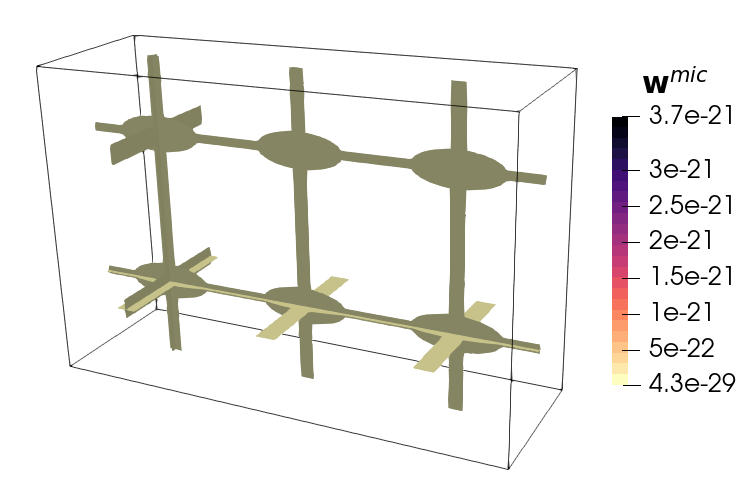}
			&\includegraphics[width= 0.99\linewidth]{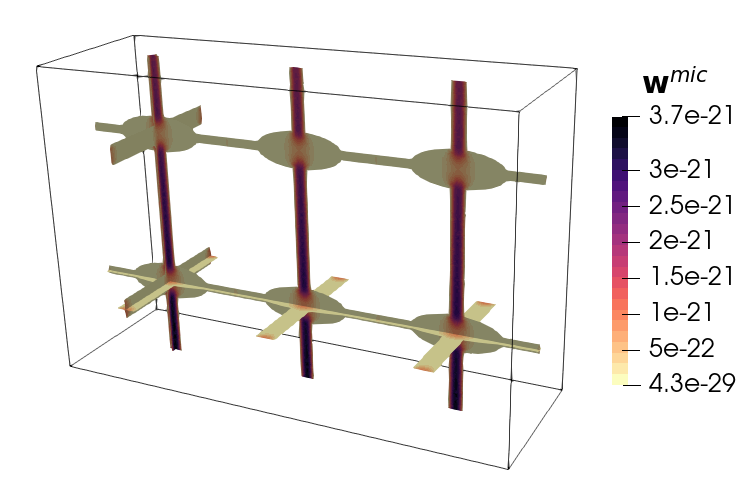}
		\end{tabular}
	\end{center}
	\caption{Reconstruction of BVP solution at point A of microscopic scale: Evolution of fluctuations of macroscopic potential fields $\Phi_1^\mic,\Phi_2^\mic, P^\mic$ and $\wb^\mic$ at times $t\in\{t_0,T\}=\{0,1\}$.}
	\label{fig:rec_BPV_IV_phis}
\end{figure}

%\begin{figure}[h]		\centering
%	\begin{center}
%		\begin{tabular}{ m{0.07\linewidth} m{0.465\linewidth} m{0.465\linewidth} }
%			$t=t_0$
%			&\includegraphics[width=\linewidth]{\figPathMac/rec/phi1_mic_t0_rez_croped.png}
%			&\includegraphics[width= \linewidth]{\figPathMac/rec/phi2_mic_t0_rez_croped.png}\\
%%			$t=t_3$
%%			&\includegraphics[width=\linewidth]{\figPathMac/rec/phi1_rec_t3_croped.png}
%%			&\includegraphics[width= \linewidth]{\figPathMac/rec/phi2_rec_t3_croped.png}\\
%			$t=T$
%			&\includegraphics[width=\linewidth]{\figPathMac/rec/phi1_mic_T_rez_croped.png}
%			&\includegraphics[width= \linewidth]{\figPathMac/rec/phi2_mic_T_rez_croped.png}
%		\end{tabular}
%	\end{center}
%	\caption{Reconstruction of BVP solution at microscopic scale: Evolution of total reconstructions of macroscopic potential fields, $t\in\{t_0,t_r,T\}=\{0,0.45,1\}$. Left: total reconstruction  $\Phi_1^\rec$; Right: total reconstruction $\Phi_2^\rec$}
%	\label{fig:rec_BPV_IV_phis}
%\end{figure}
%

\begin{figure}[!h]	
	\begin{center}
		\begin{tabular}{ m{0.05\linewidth}| >{\centering\arraybackslash}m{0.465\linewidth} >{\centering\arraybackslash}m{0.465\linewidth} }
			&$t=t_0$&$t=T$\\
			\hline\rotatebox[origin=c]{90}{$\ub^\mic$}
			&\includegraphics[width=0.99\linewidth]{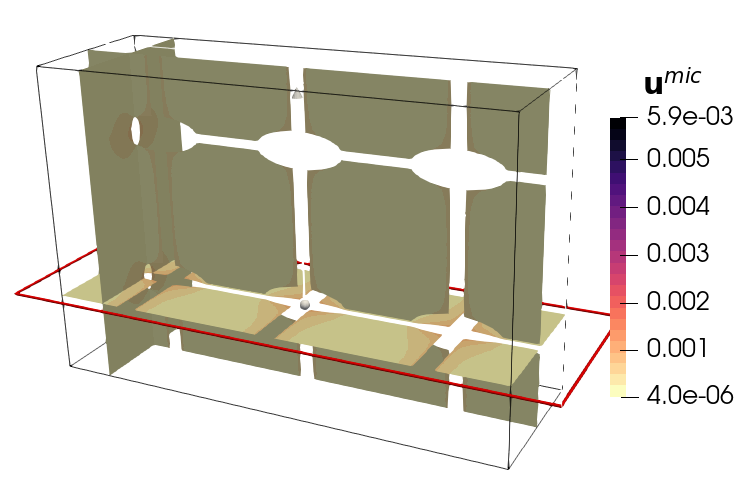}
			&\includegraphics[width= 0.99\linewidth]{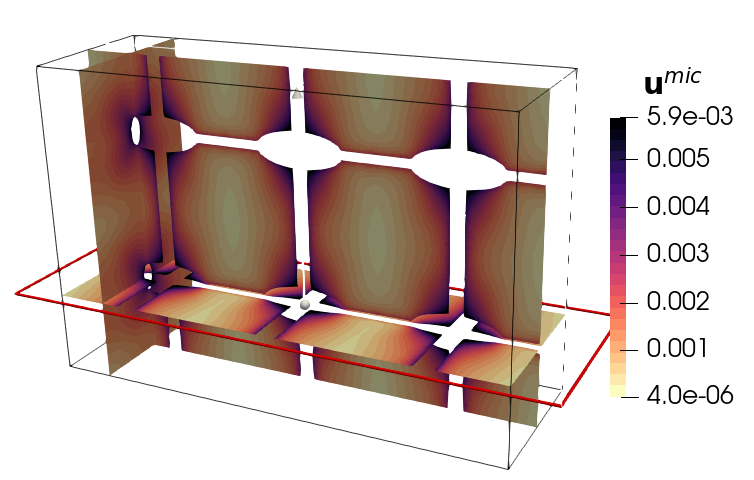}\\
			%			$t=t_r$
			%			&\includegraphics[width=0.99\linewidth]{\figPathMac/rec/u_mic_t3_croped.png}
			%			&\includegraphics[width= 0.99\linewidth]{\figPathMac/rec/psis_mic_t3_croped.png}\\
			\hline\rotatebox[origin=c]{90}{$\Psi_s^\mic$}
			&\includegraphics[width=0.99\linewidth]{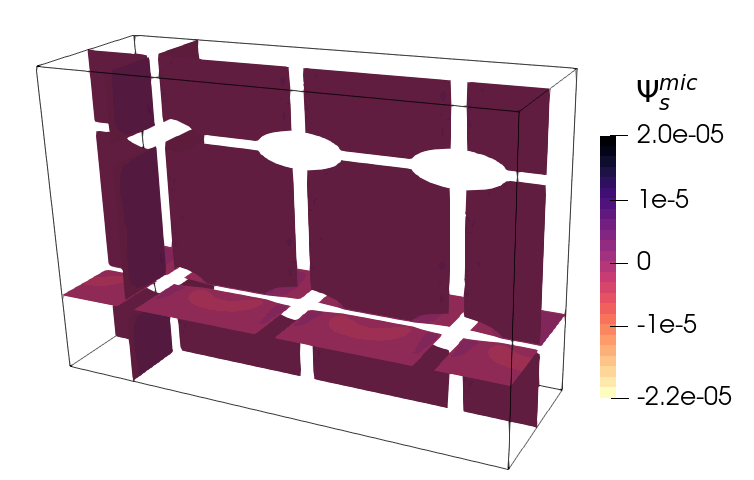}
			&\includegraphics[width= 0.99\linewidth]{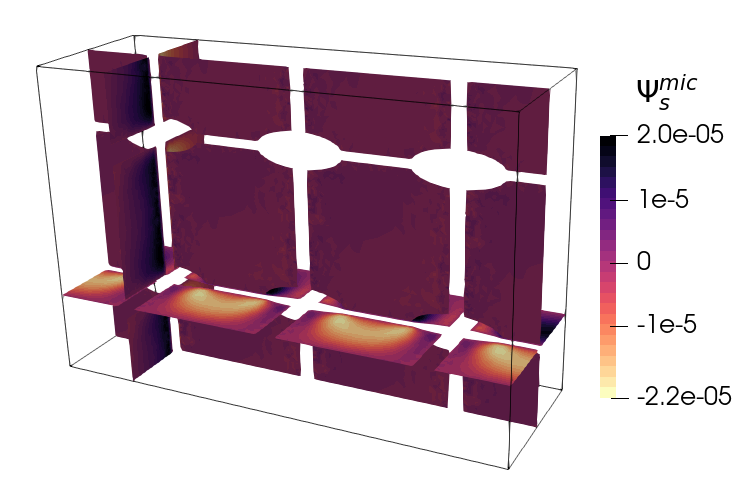}
		\end{tabular}
	\end{center}
	\caption{Reconstruction of BVP solution at poit A of microscopic scale: Evolution of reconstruction of displacement field $\ub^\mic$ and solid potential field $\Psi_s^\mic$, $t\in\{t_0,T\}=\{0,1\}$}
	\label{fig:rec_BPV_IV_up}
\end{figure}

\section{Conclusion}\label{sec_conclusion}
We presented the quasi-static model of the weakly piezoelectric porous media saturated by a two-component electrolyte and its two-scale homogenization. It extends our previous work \cite{turjani2018} that presents only the stationary case of the poroelastic medium, thus delivering a model with weaker coupling between the electromechanical processes. Indeed, the incorporation of piezoelectricity in the solid leads to stronger coupling between mechanics, fluid flow, and ionic transport. The presented model also allows ionic exchanges between phases on their interface. However, due to the scaling of a Damk\"{o}hler number, the effect of the ionic exchanges remains microscopic. 

The two-scale homogenization is performed on the dimensionless linearized mathematical model obtained for the equilibrium reference state, which operates under the assumptions of zero fluxes (fluid and solid velocities and electroneutrality in the bulk of the electrolyte). As the principal result, we derived the fully coupled quasi-static two-scale model, which enables us to study the influence of the skeleton piezoelectric property on the electrochemical processes in the pore fluid under various loading conditions. Values of the homogenized coefficients depend on the microstructure size given by parameter $\ell_\mic$ which reflects the scale-dependent effect associated with the electric double layer and viscous flow velocity profile in the pores.

The resulting two-scale model was implemented in our in-house developed and open-source software \emph{SfePy}. The reported computational examples show the behavior of the macroscopic homogenized model and illustrate the coupling between electrochemical and mechanical phenomena. The software can also reconstruct macroscopic solutions at the chosen region of the macroscopic specimen and, thus, provide more detailed information about their distribution at the microscale. The obtained results can contribute to a better understanding of the processes undergoing at the microscopic level.

The presented numerical simulations compare the derived model of electrolyte flow in a weakly piezoelectric porous medium (PZPM) with the previously published model of the same phenomenon in the linear elastic porous medium (EPM), \cite{turjani2018}. The comparison clearly shows the impact of the piezoelectric skeleton, resulting in remarkable differences in spatial distributions and evolutions of macroscopic fields, especially in the displacement and computed strain and stress fields, influencing the ionic transport. In particular, the modified Biot coefficients \Eq{eq_ec_piezo_disp_1} involved in the PZPM model lead to significant changes in the ions distributions and amplify the overall electromechanical responses of the specimen when compared to corresponding responses of the EPM model.

The presented model and its computational implementation have the potential for further extensions, such as the skeleton flexoelectric property. Modeling a nonlinear problem would bring a whole set of new mathematical and numerical challenges and significantly increase the computational requirements. In the case of non-steady flow, the fluid-structure interaction will be more involved, thus leading to a strong coupling between the fluid flow, ionic concentrations, and deformation, \cf \cite{Rohan-Lukes-PZarXive2023}. The scale decoupling procedure will become more complicated and will lead to fading memory effects of the macroscopic responses, as the homogenized coefficients will serve for time convolution kernels, \cite{Auriault1993,rohan-etal-jmps2012-bone}. Other interesting applications can be pursued, such as modeling energy storages, \eg \cite{Werner-2021}.

Notably, the presented two-scale model provides a basis for identifying material parameters of the solid skeleton in the microstructure. It can contribute to exploring complex physiological processes in tissues, such as cortical bone tissue, and motivate systematic modelling-based research in biomaterial design.

\paragraph{Acknowledgment} This research was supported by project GA~22-00863K  of
the Scientific Foundation of the Czech Republic.

%%----------------------- JANA:

%The existing research in the field of bone piezoelectricity and its effect was comprehensively compiled in the paper \cite{Zhang2023a}.

%The recent research shows, that the knowledge about the mechanoelectric microenviroment is usefull for development of scafold for bone tissue regeneration strategies, \cite{Silva2022}. The result of this research prove that the effective cell growth in the electromechanically responsive piezoelectric scafolds is enhanced. Indeed, the correct viscoelastic and piezoelectric stimuli simulating the cues from bone collagen fibers show to be crucial for development of artificial bone substitute materials for regeneration of bone defects, \cite{Zhang2023b}. 

%%-----------------------------

\appendix
%\chapter{Appendices}
%\input{appendix_dimless}

\section{Equilibrium solution}\label{sec_equilib}
 The equilibrium displacement $\ub^{\eq,\veps}(x)$ and potential $\Psi_s^{\eq,\veps}(x)$ are obtained as a solution  of the following dimensionless problem: find $\ub^\eq$ and $\Psi_s^\eq$ such that
\begin{equation}\label{eq_piezo_eq_bc}
\begin{aligned}
-\nabla\cdot(\Ab\str{\ub^\eq}-\veps M_g\bar\gb^{T}\nabla\Psi_s^\eq)&=0,\inom{s}\\
-\nabla\cdot(\veps\bar\gb\str{\ub^\eq}+\veps^{2}M_\Psi\bar\db\nabla\Psi_s^\eq)&=0\inom{s},\\
(\Ab\str\ub-\veps M_g\bar\gb\nabla\Psi^\eq_s)\cdot\nb&=C_p\Sigma\ongamma.
\end{aligned}\end{equation}
Following the approach by \cite{moyne2003} shows, that displacement and potential of the solid are of type 
\begin{equation}
\ub^\eqe=\veps\ub_\pi^1\left(\frac{x}{\veps}\right),\quad \Psi_s^\eqe=\Psi_s^\eq\left(\frac{x}{\veps}\right)\quad x\in\Omega^\veps_s,
\end{equation}
where both $\ub_\pi^1$ and $\Psi_s^\eq$ are $Y$-periodic functions defined as a solution to
\begin{equation}\label{eq_equlibrium_upsi}
\begin{aligned}
\nabla_y \cdot (\Ab\stry{\ub_\pi^1}-M_g\bar\gb^{T}\nabla_y \Psi_s^0)=&0, \inYs,\\
\nabla_y \cdot (\bar\gb\stry{\ub_\pi^1}-M_\Psi\bar\db^{T}\nabla_y \Psi_s^0)=&0, \inYs\\
(\bar\gb\stry{\ub_\pi^1}-M_\Psi\bar\db^{T}\nabla_y \Psi_s^0)\cdot \nb=&0, \ongammay\\
(\Ab\stry{\ub_\pi^1}-M_g\bar\gb^{T}\nabla_y \Psi_s^0)\cdot \nb=\ &\\
=-\sum\limits_{\beta=1}^2 c_\beta^\eq \exp\left({-z_\beta\Psi_f^\eq}\right)\Ib \cdot\nb +& \gamma^{-1}\left(\nablay\Psi_f^\eq\otimes\nablay\Psi_f^\eq-\frac{1}{2}|\nabla_y\Psi_f^\eq|^2\Ib\right)\cdot\nb, \ongammay.
\end{aligned}\end{equation}

Similarly to the equilibrium solution of the Poisson-Boltzmann problem, see \cite{allaire2013asymptotic}, the Neumann boundary condition \Eq{eq_piezo_eq_bc} appears only in the equilibrium solution and it can by  eliminated from the problem of perturbations during linearization. 

\section{Unfolding homogenization}\label{sec_A}
The unfolding homogenization method is based on the properties of unfolding operator $\Tuft$  which is similar to the dilatation operator.  
By virtue of the coordinate decomposition into "coarse" and "fine" parts, any function $\psi=\psi(x)$ can be unfolded into a function of $x$ and $y$. The convergence results in the unfolded
domains $\Om \times Y$ can be found in  \cite{cioranescu2008periodic}. By virtue of its definition, for specific subsequences of $\veps$, domain $\Om = ]0,L[^d$ contains    the ``entire'' periods $\veps Y$, thus
\begin{equation}\label{eq:3}\nonumber
\begin{split}
\hat \Om^\veps & = \mbox{interior} \bigcup_{\zeta \in \Xi^\veps} Y_\zeta^\veps\;, \quad Y_\zeta^\veps= \veps (\ol{Y} + \zeta )\\
\mbox{ where } \Xi^\veps & = \{\zeta \in \ZZ^3\,|\; \veps (\ol{Y} + \zeta) \subset \Om\}\;.
\end{split}
\end{equation}

For all $z \in \RR^3$,  let $[z]$ be the unique integer such that $z- [z] \in Y$.
We may write $z = [z]+\{z\}$ for all $z\in \RR^3$, so that
for all $\veps >0$, we get the unique decomposition
\begin{equation}\label{eq:3a}
x = \veps\left ( \left [\frac{x}{\veps}\right ] + \left \{\frac{x}{\veps}\right \}\right) =  \xi + \veps y
\quad \forall x \in \RR^3\;,\quad \xi = \veps\left [\frac{x}{\veps}\right ]\;.
\end{equation}
Based on this decomposition, the periodic unfolding operator
$\Tuft:  L^2(\Om;\RR) \rightarrow L^2(\Om \times Y;\RR)$ is defined as follows: for
any function $v \in L^1(\Om;\RR)$, extended to $L^1(\RR^3;\RR)$ by zero outside $\Om$,
i.e. $v=0$ in $\RR^3 \setminus \Om$,
\begin{equation}\nonumber
\Tuf{v}(x,y) =
\left \{
\begin{array}{ll}
v\left( \veps \displaystyle  \left [\frac{x}{\veps}\right ] + \veps y \right)\;,
\quad &  x \in \hat\Om^\veps, y \in Y\;, \\
0 & \mbox{ otherwise }. \\
\end{array}
\right .
\end{equation}
Unfolding operator $\Tuft$  has the following three important properties: For all functions $\psi$ and $\chi$:
\begin{eqnarray}
(i) & & \Tuf{\psi(x)\chi(x)}=\Tuf{\psi(x)}\Tuf{\chi(x)}, \\
(ii) & & \int\limits_\Om{\psi(x)} \dx = \int\limits_\Om \frac{1}{|Y|}\int\limits_Y\Tuf{\psi}(x,y) \dx  \dy = \intOmY\Tuf{\psi}(x,y),\\ \label{uo_3}
(iii) & & \Tuf{\nabla_x\psi(x)}=\frac{1}{\varepsilon}\nabla_y(\Tuf{\psi}(x,y)).
\end{eqnarray}
By $\Mcal_Y(\cdot)$ we denote the average operator over $Y$, if $\Tuf{w^\veps}\rightharpoonup \hat w$ weakly in $L^p(\Om\times Y)$, then ${w^\veps}\rightharpoonup \Mcal_Y(\hat w)$
weakly in $L^p(\Om)$.
For any $D\subset Y$,
$\intYsmall_{D} = \frac{1}{|Y|}\int_{D}$; the analogical notation is employed for any $A\subset Z$, thus $\intYsmall_{A} = \frac{1}{|Z|}\int_{A}$. Further, for any $D\subset Y$, $\Vb^1(D)$ is the Sobolev space $\Wb^{1,2}(Y) = \Hdb(Y)$ of vector-valued Y-periodic functions (indicated by the subscript $\#$).

\par Importantly, the unfolding operator also transforms the integration in domain $\Om$ to $\Om\times Y$, so that standard means of the weak convergence in Lebesgue spaces $L^q(\Om\times Y)$ can be employed. For more details see  \cite{cioranescu2008periodic}.
%\clearpage
\bibliographystyle{unsrt} 

%In our case, the unfolded equations of the weak formulation are obtained using the unfolding operator $\mathcal{T}_\veps$ defined in  \ref{sec_A}, see \cite{cioranescu2008periodic}.
%Due to the a~priori estimates on the solutions \Eq{eq_weak_piezo1}-\Eq{eq_weak_piezo3}, we obtained convergence result by the unfolding. As these convergence results for $\veps\rightarrow 0$ and any fixed time $t>0$ do not differ from the previously explored steady state case presented in our work, \cite{turjani2018}, we only explain the main differences.

\section{Convergences}\label{appendix_conv}
The unfolded equations of the weak form given by \Eq{eq_weak_piezo1}--\eqref{eq_weak_piezo3} are obtained using the unfolding operator $\mathcal{T}_\veps$ defined in  \ref{sec_A}, see \cite{cioranescu2008periodic}.
Due to the a~priori estimates on the solutions of \Eq{eq_weak_piezo1}--\eqref{eq_weak_piezo3}, according to 
\cite{allaire2015ion}, the following convergence result for $\veps\rightarrow 0$ and any fixed time $t>0$ can be proved:
There exist limit fields $(\wb^0, P^0)\in L^2(\Omega;H^1_\#(Y_f)^d)\times L^2(\Omega)$, $\left\{\Phi^0_\alpha,\Phi^1_\alpha\right\}_{\alpha=1,2}\in (H^1(\Omega)\times L^2(\Omega;H^1_\#(Y_f)))^2$, $\Psi^0\in L^2(\Omega;H^1_\#(Y_f))$, $\ub^0\in H^1_{\#}(\Om_s^\veps)^d$ and $\ub^1\in L^2(\Omega; H^1_{\#}(\Om_s^\veps)^d)$ such that the following convergences hold
\begin{align}
\Tuf{\dwbe} 	\rightharpoonup &\wb^0&\qquad\textrm{w. in }L^2(\Omega\times Y_f),\nonumber\\
\veps\Tuf{\nabla\dwbe} 	\rightharpoonup &\nabla_y\wb^0&\qquad\textrm{w. in }L^2(\Omega\times Y_f),\nonumber\\
\Tuf{\Pe} 	\rightarrow & P^0&\qquad \textrm{ s. in }L^2(\Om),\nonumber\\
\Tuf{\nabla\Pe} 	\rightharpoonup & \nabla_xP^0+\nabla_yP^1&\qquad \textrm{w. in }L^2(\Omega\times Y_f),\nonumber\\
\Tuf{\left\{\dPhije\right\}} 	\rightarrow &\left\{\Phi^0_\beta\right\} &\qquad\textrm{ s. in }L^2(\Om),\nonumber\\
\Tuf{\left\{\nabla\dPhije\right\}} 	\rightharpoonup &\left\{\nabla_x\Phi^0_\beta+\nabla_y\Phi^1_\beta\right\}&\qquad\textrm{w. in }L^2(\Omega\times Y_f),\label{converg_piezo}\\
\Tuf{\dPsies}\rightharpoonup&\Psi^0&\qquad\textrm{w. in }L^2(\Omega\times Y_f),\nonumber\\
\Tuf{\veps\nabla\dPsie}\rightharpoonup&\nabla_y\Psi^0 &\qquad\textrm{w. in }L^2(\Omega\times Y_f),\nonumber\\
\Tuf{\dube}\rightharpoonup&\ub^0&\qquad\textrm{w. in }L^2(\Omega\times Y_s),\nonumber\\
\Tuf{\nabla\dube}\rightharpoonup&\nablax\ub^0+\nablay\ub^1&\qquad\textrm{w. in }L^2(\Omega\times Y_s).\nonumber
\end{align}

\end{document}